\documentclass[lettersize,onecolumn]{IEEEtran}
\IEEEoverridecommandlockouts
\usepackage[utf8]{inputenc}
\usepackage{stackengine}
\usepackage{graphicx}
\usepackage{amssymb}
\usepackage{amsmath}
\usepackage{mathrsfs}
\usepackage{ulem}
\usepackage{epsf,epsfig}
\usepackage{cite}
\usepackage{color,xcolor}
\usepackage{dsfont}
\usepackage{bbm}
\usepackage{amsthm}
\usepackage{physics}
\usepackage{tcolorbox}
\usepackage{wrapfig}
\usepackage{url}
\usepackage{graphicx}
\usepackage{stmaryrd}
\usepackage{hyperref}
\usepackage{eurosym}
\usepackage{enumitem}
\usepackage{nopageno}
\usepackage{fancyhdr}
\usepackage{mathtools}
\usepackage{tabularx}
\usepackage{array} 
\usepackage{multirow}

\def\complex{\mathbb{C}}

\def\integers{\mathbb{Z}}

\def\Expectation{\mathbb{E}}

\newcommand{\msout}[1]{\text{\sout{\ensuremath{#1}}}}

\definecolor{darkgreen}{rgb}{0,0.5,0}
\definecolor{darkmagenta}{rgb}{.5,0,.5}
\definecolor{darkblue}{rgb}{0,0,0.5}

\def\CalA{\mathcal{A}}

\def\CalC{\mathcal{C}}
\def\CalD{\mathcal{D}}

\def\CalF{\mathcal{F}}

\def\CalH{\mathcal{H}}

\def\CalJ{\mathcal{J}}

\def\CalM{\mathcal{M}}
\def\CalN{\mathcal{N}}

\def\CalP{\mathcal{P}}
\def\CalQ{\mathcal{Q}}

\def\CalT{\mathcal{T}}
\def\CalU{\mathcal{U}}
\def\CalV{\mathcal{V}}
\def\CalW{\mathcal{W}}
\def\CalX{\mathcal{X}}
\def\CalY{\mathcal{Y}}

\def\ulineD{\underline{D}}

\def\ulineG{\underline{G}}

\def\ulineM{\underline{M}}

\def\ulineQ{\underline{Q}}
\def\ulineR{\underline{R}}

\def\ulineU{\underline{U}}
\def\ulineV{\underline{V}}
\def\ulineW{\underline{W}}

\def\ulineY{\underline{Y}}
\def\ulineZ{\underline{Z}}

\def\ulinea{\underline{a}}

\def\ulined{\underline{d}}
\def\ulinee{\underline{e}}

\def\ulinem{\underline{m}}

\def\ulineq{\underline{q}}

\def\ulineu{\underline{u}}
\def\ulinev{\underline{v}}

\def\ulinex{\underline{x}}

\def\ulinez{\underline{z}}

\def\CalA{\mathcal{A}}

\def\CalC{\mathcal{C}}
\def\CalD{\mathcal{D}}

\def\CalF{\mathcal{F}}

\def\CalH{\mathcal{H}}

\def\CalJ{\mathcal{J}}

\def\CalM{\mathcal{M}}
\def\CalN{\mathcal{N}}

\def\CalP{\mathcal{P}}
\def\CalQ{\mathcal{Q}}

\def\CalT{\mathcal{T}}
\def\CalU{\mathcal{U}}
\def\CalV{\mathcal{V}}
\def\CalW{\mathcal{W}}
\def\CalX{\mathcal{X}}
\def\CalY{\mathcal{Y}}

\def\ulineCalM{\underline{\mathcal{M}}}

\def\ulineCalQ{\underline{\mathcal{Q}}}

\def\ulineCalU{\underline{\mathcal{U}}}
\def\ulineCalV{\underline{\mathcal{V}}}

\def\ulineCalX{\underline{\mathcal{X}}}

\def\parsec{\par\noindent}
\def\med{\medskip\parsec}

\def\define{\mathrel{\ensurestackMath{\stackon[1pt]{=}{\scriptstyle\Delta}}}}

\def\olineB{\overline{B}}

\def\olineG{\overline{G}}

\def\olineZ{\overline{Z}}

\def\ScrC{\mathscr{C}}
\def\ScrD{\mathscr{D}}

\def\ScrQ{\mathscr{Q}}
\def\ScrR{\mathscr{R}}

\def\sfb{\mathsf{b}}

\def\sfs{\mathsf{s}}

\def\tildez{\tilde{z}}

\newtheorem{Notation}{Notation}
\newtheorem{remark}{Remark}
\newtheorem{theorem}{Theorem}

\newtheorem{definition}{Definition}
\newtheorem{lemma}{Lemma}
\newtheorem{example}{Example}
\newtheorem{proposition}{Proposition}
\newtheorem{fact}{Fact}
\newtheorem{note}{Note}

\def\span{\mbox{span}}

\newcommand{\colBlue}[1]{\textcolor{blue}{#1}}
\newcommand{\colRed}[1]{\textcolor{red}{#1}}
\newcommand{\colMagenta}[1]{\textcolor{magenta}{#1}}
\newcommand{\colViolet}[1]{\textcolor{violet}{#1}}

\usepackage{float}
\usepackage{accents}

\usepackage{caption}

\mathchardef\mhyphen="2D
\def\olinekappa{\overline{\kappa}}
\def\3To1BC{$3-$to$-1$}

\def\dbrackthree{\llbracket 3 \rrbracket}

\def\SemiPrivateRVSet{\CalU}

\def\Prime{\upsilon}

\def\fieldpij{\CalF_{\Prime_{j}}}

\def\bias{\mathtt{b}}
\def\Bias{\mathtt{B}}

\usepackage{xcolor}
\allowdisplaybreaks
\begin{document}

\sloppy
\title{An Achievable Rate Region for  $3-$User
 Classical Quantum Broadcast Channel via Coset Codes
}

\author{
\IEEEauthorblockN{Fatma Gouiaa and Arun Padakandla\\
\vspace{-0.15in}
}
}
\maketitle

\thispagestyle{empty}

\begin{abstract}
\color{black} We undertake a Shannon theoretic study of the problem of communicating statistically independent bit streams over a 3-user classical quantum broadcast channel ($3-$CQBC) and focus on characterizing inner bounds to its capacity region. We propose a coding strategy based on coset codes possessing algebraic closure properties. Elevating Sen's technique of tilting, smoothing, and augmentation - originally designed only for IID codes - we design new POVMs that can simultaneously decode into a combination of unstructured IID and coset codes to efficiently decode univariate and bivariate interferences respectively. Analyzing the information-theoretic performance of the proposed coding strategy we characterize a new inner bound to the capacity region of the $3-$CQBC that subsumes all currently known bounds and is proven to be strictly larger for identified examples.
\end{abstract}

\section{Introduction}
\label{Sec:Introduction}

{Even while information theory's connections \cite{199806TIT_Ver} with mathematical physics \cite{201610IC_CoeFriSpe,201003NJP_BarBarCla,199210WPC_Lan}, information geometry \cite{BkInfoGeomAmari_2016,198503IS_Cam}, computer science \cite{BkChaitinAlgInfTh_1987,1991MMTFE_Yam,BkJacquetSzpanskowski_2015,BkKrichevsky_1994} have blossomed, the problem of characterizing the information carrying capacity of a channel has had a large influence \cite{197109ISIT_Ahl,197905TIT_Mar,198101TIT_HanKob,197303PPI_Hol,199801TIT_Hol,199707PhyRev_SchWes,201512TIT_SavWil,202102SAD_Sen} on its agenda. In solving this problem via \textit{random coding} in the context of a classical point-to-point channel, Shannon \cite{194807BSTJ_Sha} invented an ingenious and powerful technique. It is indeed remarkable that the average error probability of a code whose codewords are \textit{independent and identically distributed} (IID) with a product distribution falls exponentially for any rate below capacity. Note that codewords of highly probable codes within Shannon's ensemble possess empirical properties, but are otherwise \textit{bereft of any additional structural properties}. Indeed, we can only state that {individual} codewords are \textit{typical} \cite{BkNITElGamalKim_2011} with high probability, while unable to make any assertions on their structural relationships or their behaviour when jointly operated via multi-variate functions. Shannon's recipe of employing unstructured IID codes in characterizing capacity regions has pervaded through all of classical \cite{197905TIT_Mar,197201TIT_Cov,198101TIT_HanKob,197109ISIT_Ahl,BkNITElGamalKim_2011} and quantum information theory \cite{200107TIT_Win,201206TIT_FawHaySavSenWil,201512TIT_SavWil,202103SAD_Sen,202102SAD_Sen} and forms the defacto approach in any information theoretic study, be it the one-shot or the asymptotic regime. The problems of characterizing the capacity regions of the general quantum interference and quantum broadcast channels remaining open, it is natural to inquire whether \textit{unstructured IID codes} are optimal for multi-terminal (network) quantum communication?}

{The broad import of our findings, as we will elaborate in the sequel, establishes that the conventional approach of unstructured IID codes is sub-optimal for network quantum communication. In essence, we prove that multi-terminal quantum channels (q-channels), particularly those with three or more terminals, possess higher degrees of freedom that \textbf{cannot} be exploited via unstructured IID codes owing to their lack of additional structural properties. Therefore, to characterize the capacity regions of multi-terminal q-channels and identifying codes that can achieve the same, deliberate attempt must be made to incorporate richer structural properties, in addition to individual empirical ones. Moreover, to harness the full communication capabilities of multi-terminal q-channels, these codes must be coupled with newer encoding maps and decoding POVMs that exploit their structural relationships. Towards stating our specific findings, we begin by describing the q-channel and the problem of interest.}

\begin{figure}
\vspace{-0.15in}
\centerline{\includegraphics[scale=0.55]{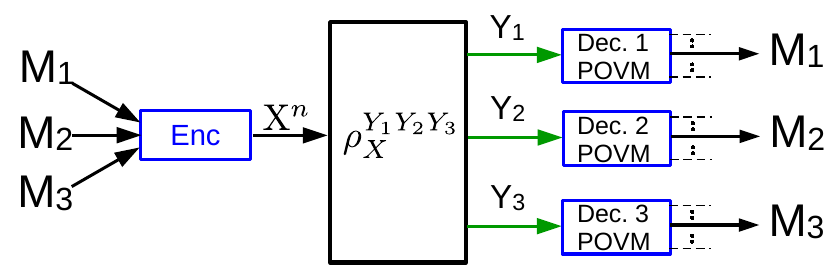}}
\caption{Three independent messages to be communicated over a $3-$CQBC.}
\label{Fig:3CQBC}
\vspace{-0.2in}
\end{figure}

{Consider a $3-$user classical-quantum broadcast channel ($3-$CQBC) with three distributed receivers (Rxs) depicted in Fig.~\ref{Fig:3CQBC}. The transmitter (Tx) needs to communicate three statistically independent bit streams in an information-theoretically reliably sense \cite{BkWilde_2017} to the three Rxs. The problem of interest is to characterize the capacity region of the $3-$CQBC. Our specific focus is on designing coding strategies, analyzing their performance and deriving \underline{inner bounds} to the capacity region of the $3-$CQBC.}

What is the current known best coding strategy for communicating over a $3-$CQBC? Marton's strategy \cite{197905TIT_Mar} - a combination of Cover's superposition coding \cite{197201TIT_Cov} and precoding \cite{197905TIT_Mar} via binning - designed in the context of a $2-$user classical broadcast channel (BC) \cite[Ch.~8]{BkNITElGamalKim_2011} is the current known best \cite{201512TIT_SavWil,201110TIT_YarHayDev} for any quantum BC with any number of Rxs. Recognizing this, Savov and Wilde \cite{201512TIT_SavWil} design appropriate decoding POVMs to adopt Marton's strategy for communicating over a general $2$ Rx CQBC ($2-$CQBC). Identifying necessary analytical tools such as the \textit{overcounting} to analyze its performance, they \cite{201512TIT_SavWil} prove achievability of an inner bound which is a natural CQ generalization of Marton's inner bound \cite{197905TIT_Mar}. The latter is indeed the current known largest inner bound to the capacity region of a general $2-$CQBC. A natural generalization of their inner bound \cite{201512TIT_SavWil} to the three Rx case yields the current known largest for a $3-$CQBC, which we henceforth refer to as the unstructured inner bound. A characterization of this latter inner bound can be obtained by replacing the information quantities in \cite[Thm.~1]{201804TIT_PadPra} with appropriate Holevo informations.

{Taking a cue from \cite{201804TIT_PadPra} and building on our prior works \cite{202206ISIT_Pad3CQBC,202607ISIT_GouPad3CQBC}
we step beyond unstructured IID code strategies to design a coding strategy based on jointly designed coset codes. As we discuss in the sequel, communication over $3-$CQBC requires design of multiple codes. In an unstructured IID code strategy, the multiple codes possess no inter-relationships. Moreover, codewords of any code possess no intra-relationships. Here, we design a coding strategy wherein codes are chosen to be \textit{cosets} of a \textit{linear} code and therefore algebraically closed. Moreover, \textit{specific} pairs of codes are picked as cosets of a \textit{common linear code} endowing them with inter-relationships (Rem.~\ref{Rem:InterAndIntraStructure}). See Fig.~\ref{Fig:3CQBCStepIICodeStructure}. We exploit these joint structural relationships by designing new \textit{simultaneous} decoding POVMs that can efficiently recover bivariate functions. Identifying appropriate analytical tools to characterize information theoretic performance of the designed coding strategy, we derive a new inner bound (Thms.~\ref{Thm:3CQBCStepIInnerBound}, \ref{Thm:Step2}) to the capacity region of the $3-$CQBC that subsumes the current known largest - the unstructured inner bound (Rem.~\ref{Rem:afterstep3}). Moreover, we identify Ex.~\ref{Ex:3CQBCRoleOfCosetCds} for which the derived inner bound is proven (Lem.~\ref{Lem:CosetCdsInnBndStrictlyLargerForEx1}) to be strictly larger than the latter.}

{We encounter three challenges in design and analysis of the above coding strategy for a general $3-$CQBC. Firstly, since each Rx must decode into several codebooks, we are required to design and analyze POVMs that can \textit{simultaneously decode}. In particular, the advantages of jointly designed coset codes crucially relies on the Rx decoding the bivariate function, as against to the pair of its input arguments. This will require decoding into `functions of codebooks' - the codeword collection obtained by applying the bivariate function on each pair of codewords. Moreover, a general coding strategy must involve two layers - one comprising of unstructured IID code to enable Rxs decode univariate interference components and the other of jointly designed coset codes. Rxs must therefore decode \textit{simultaneously} into a \textit{combination of both coset and unstructured IID codes}. Design and analysis of simultaneously decoding POVMs even into the well studied unstructured IID codes has proven to be a difficult challenge. Indeed, this challenge remained open for over a decade since Winter's recognition \cite[Sec.~VIII]{200107TIT_Win} of the same, spurring several alternate approaches \cite{201206TIT_FawHaySavSenWil,201207ISIT_Sen}. Recently, {Sen has invented \textit{tilting, smoothing, and augmentation} (TSA)} \cite{202103SAD_Sen,202102SAD_Sen} to design and analyze a quantum simultaneous decoder. Sen's technique \cite{202103SAD_Sen,202102SAD_Sen} is remarkably powerful and accomplishes simultaneous decoding even in the more general one-shot regime. However, Sen's study is confined to decoding into unstructured IID codes. Our first challenge is to design and analyze a quantum decoder that can \textit{simultaneously decode} into `functions of codebooks', jointly designed coset \textit{and} unstructured IID codes to recover bivariate functions efficiently.}

{The second challenge pertains to analyzing binning in a $3-$user quantum BC. Since binning entails exponentially many codewords for each message, its analysis over a q-channel results in a positive exponent appearing in the error probability. This has resulted in mis-steps \cite[Sec.~I and Note therein]{201512TIT_SavWil}. Savov and Wilde \cite{201512TIT_SavWil} propose an \textit{overcounting} technique to suppress this positive exponent. While this overcounting technique works for the $2-$CQBC, it leaves the challenge unresolved in the $3-$user case, as elaborated in \cite[Sec.~3]{202508SalHayHay}. We also note here that the analysis steps presented in \cite{202508SalHayHay} needs a clearer mathematical justification. See \cite{202607ISIT_GouPadBinning}. Since binning is an integral part of any general coding strategy involving coset codes, we are required to identify a robust technique to analyze the same that is compatible with the new quantum simultaneous decoder that we design. It is important to note that the above two challenges are specific to the non-commutative nature of the $3-$CQBC and hence not encountered in \cite{201804TIT_PadPra,202206ISIT_Pad3CQBC}. Thirdly, note that codewords of a random coset code are uniformly distributed and only pairwise independent. Achieving inner bounds corresponding to arbitrary distributions over generic $3-$CQBC cannot proceed as in the case of unstructured IID codes, since the latter can be chosen with respect to any product distribution.}

{We overcome the first challenge by elevating Sen's TSA \cite{202103SAD_Sen,202102SAD_Sen} to incorporate efficient bivariate function decoding. Specifically, we first resort back to appropriate conditional typical projectors that can decode bivariate functions. Next, as elaborated in proof of Thm.~\ref{Thm:3to1CQBC}, we identify new auxiliary spaces and new tilting maps to decode the bivariate function and the unstructured codeword simultaneously. To address the second challenge,} we put forth an alternate approach to analyze binning over any q-channel. Recognizing the connection of binning to soft covering \cite[Lemma 19]{CuffPhDThesis}, we employ a stochastic encoder that enables us decouple the binning and channel coding analysis in a way that works for any scenario involving binning over classical or q-channel. Our ability to incorporate and analyze all these different elements - coset codes, simultaneous decoding, binning - in a single coding scenario in a rigorous information-theoretic way constitutes an important element of our work.

Our findings build on considerable prior work. The problem of characterizing bit carrying capacity of network (multi-terminal) q-channels gathered momentum following Holevo \cite{199801TIT_Hol}, Schumacher and Westmoreland's \cite{199707PhyRev_SchWes} characterization of the CQ capacity of a point-to-point (PTP) q-channel. In their study of a more general CQ PTP q-channel, Hayashi and Nagaoka identified powerful analytical tools \cite{200307TIT_HayNag} that form a central part of CQ capacity studies. Winter's study of the CQ capacity of a quantum MAC \cite[Sec.~VIII]{200107TIT_Win} revealed challenges of analyzing a quantum simultaneous decoder. The study of quantum broadcast channel (QBC) was first initiated by Dupuis, Hayden and Li \cite{201005TIT_DupHayLi} with the additional resource of entanglement. This was furthered by Yard, Hayden and Devetak \cite{201110TIT_YarHayDev} who studied the use of QBC to communicate both bits and qubits. Savov and Wilde \cite{201512TIT_SavWil} proved achievability of the CQ analogue of Marton's inner bound for the $2-$CQBC. Recently, Salek, Hayden and Hayashi undertake information theoretic study \cite{202508SalHayHay} of multi-level QBCs. Sen proved achievability of Han-Kobayashi's inner bound \cite{198101TIT_HanKob} in the asymptotic regime via sequential decoding. Inventing the techniques of TSA \cite{202103SAD_Sen}, Sen proved achievability of inner bounds \cite{202102SAD_Sen} for several two-terminal CQ channels in the more general one-shot regime. We should note that all of the above works employ unstructured IID codes. In regards to binning and our approach to overcome the challenge over q-channels, we note that \cite{201604TIT_SonCufPoo}, \cite{201411TIT_YasAreGoh} and \cite{201504TIT_WatKuzTan} have developed analogous techniques in the context of source coding problems.

We highlight the significance of our findings. Characterizing the capacity region of a BC is regarded as a fundamental problem \cite{BkNITElGamalKim_2011} both in classical \cite{199810TITI_Cov} and quantum information theory. Indeed, this problem has continued to receive attention and its pursuit has yielded new coding strategies that are applicable in other channel scenarios. Secondly, the use of unstructured IID codes is ubiquitous in information theoretic studies. The restriction of earlier studies \cite{201512TIT_SavWil,201206TIT_FawHaySavSenWil} to two Rx channels belie an underlying assumption that optimal coding strategies for two Rx channels can be naturally extended to yield optimal strategies for channels with three or more Rxs and that the latter do not possess any newer aspects that require more sophisticated coding strategies. By combining challenging elements such as simultaneous decoding, coset codes and new decoding POVMs, characterizing new achievable rate regions that subsume the current known largest and strictly enlarging the same for examples, our findings put to rest such an assumption and establish that multi-terminal channels admit newer coding strategies.

\med\textbf{Organization}: Following preliminaries - notation and problem statement - in Sec.\ref{Subsec:NotationProblemStatement}, we propose a technique based on likelihood encoder and analysis via soft covering to analyze binning in the context of a $2-$CQBC in Sec.~\ref{Subsec:2CQBC} that we shall employ in proofs of subsequent theorems. In Sec.~\ref{SubSec:NeedForCosetCodes}, considering a specific example, we explain how coset codes possessing algebraic closure properties can outperform unstructured IID codes for communicating over $3-$CQBCs. Following this discussion, in Sec.~\ref{SubSec:StructureOfAGeneralCodingStrategy} we outline the structure of a general coding strategy for communicating over a general $3-$CQBC. We design and analyze this coding strategy in three pedagogical steps. Step I, presented in Sec.~\ref{Sec:3to1CQBC}, enables one of the Rxs to decode bivariate interference. In Step II presented in Sec.~\ref{SubSec:Step2}, a full suite of coset codes is employed to enable each Rx decode bivariate interference. We combine the strategy of Step II with unstructured IID codes in Step III to design and analyze a general coding strategy that enables each Rx decode univariate and bivariate interferences efficiently. The characterized inner bound therein (Thm.~\ref{Thm:3CQBCStepIInnerBound}) subsumes all current known inner bounds and is strictly larger for Ex.~\ref{Ex:3CQBCRoleOfCosetCds}. We remark that this combination is analogous to the early work of Ahlswede and Han \cite{198305TIT_AhlHan} in the context of distributed source coding.

\section{Preliminaries and Problem Statement}
\label{Sec:Preliminaries}

\subsection{Notation and Problem Statement}
\label{Subsec:NotationProblemStatement}
We supplement notation in \cite{BkWilde_2017} with the following. For $K\in \mathbb{N}$, $[K] \define \left\{1,\ldots,K \right\}$. For prime $\Prime \in \integers$, $\CalF_{\Prime}$ will denote the finite field of size $\Prime$ with $\oplus$ denoting field addition in $\CalF_{\Prime}$ (i.e., mod$-\Prime$). For a Hilbert space $\CalH$, $\mathcal{L}(\CalH),\CalP(\CalH)$ and $\CalD(\CalH)$ denote the collection of linear, positive and density operators acting on $\CalH$ respectively.
We let an \underline{underline} denote an appropriate aggregation of objects. For example, $\ulineCalV \define \CalV_{1}\times \CalV_{2} \times \CalV_{3}$, $\ulinev \define (v_{1},v_{2},v_{3}) \in \ulineCalV$ and in regards to Hilbert spaces $\CalH_{Y_{i}}: i \in [3]$, we let $\CalH_{\ulineY} \define \otimes_{i=1}^{3}\CalH_{Y_{i}}$. Let $*$ denote the binary convolution: $p * q \coloneqq p(1-q) + (1-p)q$. For $p \in [0,1]$, $h_b(p) \define -p\log_2 p -(1-p)\log_2(1-p)$ denote the binary entropy function. For $A \in \mathcal{L}(\mathcal{H})$, $\|A\|_{1}$ denotes the trace norm and $\|A\|_{\infty}$ denotes the operator norm. A positive operator valued measurement (POVM) is a collection of positive operators that sum to the identity operator. When we say a POVM $\theta_{\mathcal{A}} \define \{\theta_a : a \in \mathcal{A}\}$, we assume that $\theta_a$ is positive for $a \in \mathcal{A}$ and $\sum_{a \in \mathcal{A}} \theta_a=I$ where $\mathcal{A}$ is a finite set. We abbreviate probability mass function, conditional typical projector, and unconditional typical projector as PMF, C-Typ-Proj and U-Typ-Proj respectively.

\begin{note}
Our work lies in quantum information theory (QIT) and we have strived to employ notation that is consistent with related articles such as \cite{200401CommMathPhy_Win,201206TIT_FawHaySavSenWil,201512TIT_SavWil,BkWilde_2017}. As observed in these articles \cite{200401CommMathPhy_Win,201206TIT_FawHaySavSenWil,201512TIT_SavWil,BkWilde_2017}, the convention is to denote matrices such as density operators and POVMs via regular faced alphabets as against to bold-faced ones. In the interest of consistency, we have thus not adopted bold-faced characters for matrices as recommended \href{https://link.springer.com/journal/220/submission-guidelines?IFA#Instructions%20for%20Authors_Scientific%20style}{here}.
\end{note}

Consider a (generic) \textit{$3-$CQBC} $(\rho_{x} \in \mathcal{D}(\mathcal{H}_{\ulineY}): x \in \CalX,\kappa)$ specified through (i) a finite set $\mathcal{X}$, (ii) Hilbert spaces $\mathcal{H}_{Y_{j}}: j \in [3]$, (iii) a collection $( \rho_{x} \in \mathcal{D}(\mathcal{H}_{\ulineY} ): x \in \CalX )$ and (iv) a cost function $\kappa :\mathcal{X} \rightarrow [0,\infty)$. The cost function is assumed to be additive, i.e., the cost of preparing the state $\otimes_{t=1}^{n}\rho_{x_{t}}$ is $\olinekappa^{n}(x^{n}) \define \frac{1}{n}\sum_{t=1}^{n}\kappa(x_{t})$. Reliable communication on a $3-$CQBC entails identifying a code.
\begin{definition}
A \textit{$3-$CQBC code} $c=(n,\ulineCalM,e,\uline{\mu})$ consists of (i) three message index sets $\mathcal{M}_{j}: j \in [3]$, (ii) an encoder map $e: \mathcal{M}_{1} \times \mathcal{M}_{2} \times \mathcal{M}_{3} \rightarrow \mathcal{X}^{n}$ and (iii) POVMs $\mu_{j} \define \{ \mu_{j,m_{j}}\in\CalP(\mathcal{H}_{Y_{j}}^{\otimes n}) : m_{j} \in \mathcal{M}_{j} \} : j \in [3]$. The average probability of error of the $3-$CQBC code $(n,\ulineCalM,e,\uline{\mu})$ is
\begin{eqnarray}
 \label{Eqn:AvgErrorProb}
 \mathbf{P}(e,\uline{\mu}) \define 1-\frac{1}{|\mathcal{M}_{1}||\mathcal{M}_{2}||\mathcal{M}_{3}|}\sum_{\ulinem \in \ulineCalM}\tr\left\{\left( \mu_{1,m_1} \otimes \mu_{2,m_2} \otimes \mu_{3,m_3}\right)   \left(\otimes_{t=1}^{n}\rho_{x_{t}(\ulinem)}\right) \right\},
 \nonumber
\end{eqnarray}
where $\uline{\mu} \define \otimes _{j=1}^3 \mu_{j}$ and $x^n(\ulinem) \define (x_{t}(\ulinem):1 \leq t \leq n) \define  e(\ulinem)$. The average cost per symbol of transmitting message $\ulinem \in \ulineCalM$ is 
$\tau(e|\ulinem) \define \olinekappa^{n}(e(\ulinem))$, and the average cost per symbol of $3-$CQBC code is $\tau(e) \define \frac{1}{|\ulineCalM|}\sum_{\ulinem \in \ulineCalM}\tau(e|\ulinem)$.
\end{definition}
\begin{definition}
A rate-cost quadruple $(R_{1},R_{2},R_{3},\tau) \in [0,\infty)^{4}$ is \textit{achievable} if there exists a sequence of $3-$CQBC codes $(n,\ulineCalM^{(n)},e^{(n)},\uline{\mu}^{(n)})$ for which $\displaystyle\lim_{n \rightarrow \infty}\mathbf{P}(e^{(n)},\uline{\mu}^{(n)}) = 0$,
\begin{eqnarray}
 \label{Eqn:3CQBCAchievability}
 \lim_{n \rightarrow \infty} n^{-1}\log |\mathcal{M}_{j}^{(n)}| = R_{j} :j \in [3], \mbox{ and }\lim_{n \rightarrow \infty} \tau(e^{(n)}) \leq \tau .
 \nonumber
\end{eqnarray}
The capacity region $\mathscr{C}$ of a $3-$CQBC is the set of all achievable rate-cost vectors and $\mathscr{C}(\tau) \define \{ \ulineR : (\ulineR,\tau) \in \mathscr{C}\}$.
\end{definition}
Our focus in this article is the capacity region $\mathscr{C}(\tau)$ of a generic $3-$CQBC. Specifically, our goal is to provide a single-letter characterization of an inner bound to $\mathscr{C}(\tau)$. The main novelty of our work is the design of a coding scheme based on nested coset codes (NCCs) which we define below.
\begin{definition}
\label{Def:CosetCds}
    An $(n,k,g,b^n)$ coset code built over a finite field $\CalF_{\Prime}$ is specified via (i) a generator matrix $g \in \CalF_{\Prime}^{k \times n }$ and (ii) a bias $b^n \in \CalF_{\Prime}^n$. We let $u^n(a)=a g \oplus b^n$ for $a \in \CalF_{\Prime}^k$ denote the codewords of the coset code $(n,k,g,b^n)$, where as mentioned above $\oplus$ denotes field addition in $\mathcal{F}_{\Prime}$. The rate of this code is $\frac{k}{n} \log(\Prime)$. 
\end{definition}
\begin{definition}
\label{Defn:NestedCosetCode}
An $(n,k,l,g_{I},g_{O/I}, b^n)$ NCC over $\CalF_{\Prime}$  consists of (i) generator matrices $ g_{I} \in \CalF_{\Prime}^{k\times n}$ and $ g_{O/I} \in \CalF_{\Prime}^{l \times n}$, and
(ii) bias vector $ b^n \in \CalF_{\Prime}^n$. We let $u^n(a,m)= a g_{I}\oplus m g_{O/I} \oplus b^n$, for $(a,m) \in \CalF_{\Prime}^k \times \CalF_{\Prime}^l$ denote codewords of the coset code. The collection $\left( u^n(a,m) \in \mathcal{F}_{\Prime}^n : a \in \mathcal{F}_{\Prime}^k\right)$ is referred to as the coset corresponding to message $m \in \mathcal{F}_{\Prime}^l$.
\end{definition}

\subsection{Analysis of Marton's binning of channel code via a likelihood encoder}

\label{Subsec:2CQBC}
The technique of Marton's binning will play a central role throughout this article. As discussed in Sec.~\ref{Sec:Introduction} we need a robust/general approach to analyze binning in diverse quantum channel coding scenarios. In this section we describe a likelihood encoder coupled with corresponding analysis steps that can enable one to analyze binning in \textbf{\textit{any}} channel coding scenario, classical or quantum. We describe the same in the context of analyzing Marton's binning for the $2-$CQBC. As will be evident to an informed reader, the following steps decouple the encoding and decoding analysis in a way (see Rem.~\ref{Rem:BinningandChannelCodingAnalysis}) so as to permit its applicability in any channel coding scenario.  We begin with the corresponding necessary definitions of $2-$CQBC.

Consider a (generic) \textit{$2\mhyphen$CQBC} $(\rho_{x} \in \mathcal{D}(\mathcal{H}_{Y_1} \otimes \mathcal{H}_{Y_2}): x \in \CalX,\kappa)$ specified through (i) a finite set $\mathcal{X}$, (ii) Hilbert spaces $\mathcal{H}_{Y_{j}}: j \in [2]$, (iii) a collection $( \rho_{x} \in \mathcal{D}(\mathcal{H}_{Y_1} \otimes \mathcal{H}_{Y_2} ): x \in \CalX )$ and (iv) a cost function $\kappa :\mathcal{X} \rightarrow [0,\infty)$. The cost function is assumed to be additive, i.e., the cost of preparing the state $\otimes_{t=1}^{n}\rho_{x_{t}}$ is $\olinekappa^{n}(x^{n}) \define \frac{1}{n}\sum_{t=1}^{n}\kappa(x_{t})$. Reliable communication on a $2\mhyphen$CQBC entails identifying a code.
\begin{definition}
A \textit{$2\mhyphen$CQBC code} {$c=(n,\CalM_1,\CalM_2,e,\mu_1 \otimes \mu_2)$} consists of (i) two message index sets $\mathcal{M}_{j}: j \in [2]$, (ii) an encoder map $e: \mathcal{M}_{1} \times \mathcal{M}_{2} \rightarrow \mathcal{X}^{n}$ and (iii) POVMs $\mu_{j} \define \{ \mu_{j,m_{j}}\in\CalP(\mathcal{H}_{Y_{j}}^{\otimes n}) : m_{j} \in \mathcal{M}_{j} \} : j \in [2]$. The average probability of error of the $2\mhyphen$CQBC code $(n,\CalM_1,\CalM_2,e,\mu_1 \otimes \mu_2)$ is
\begin{eqnarray}
 \label{Eqn:AvgErrorProb}
 \mathbf{P}(e,\mu_1\otimes \mu_2) \define 1-\frac{1}{|\mathcal{M}_{1}||\mathcal{M}_{2}|}\sum_{m_1 \in \CalM_1} \sum_{m_2 \in \CalM_2} \tr\left\{\left(\mu_{1,m_1} \otimes \mu_{2,m_2} \right) \left(\otimes_{t=1}^n \rho_{x_t(m_1,m_2)}\right)\right\},
 \nonumber
\end{eqnarray}
where $x^n(m_1,m_2) \define (x_{t}(m_1,m_2) :1 \leq t \leq n) \define e(m_1,m_2)$. Average cost per symbol of transmitting message $(m_1,m_2) \in \CalM_1 \times \CalM_2$ is $\tau(e|m_1,m_2) \define \olinekappa^{n}(e(m_1,m_2))$ and the average cost per symbol of $2\mhyphen$CQBC code is $\tau(e) \define \frac{1}{|\CalM_1||\CalM_2|}\sum_{m_1 \in \CalM_1} \sum_{m_2 \in \CalM_2} \tau(e|m_1,m_2)$.
\end{definition}
\begin{definition}
A rate-cost triple $(R_{1},R_{2},\tau) \in [0,\infty)^{3}$ is \textit{achievable} if there exists a sequence of $2\mhyphen$CQBC codes $(n,\CalM_1^{(n)},\CalM_2^{(n)},e^{(n)}, \left( \mu_1 \otimes \mu_2 \right)^{(n)})$ for which $\displaystyle\lim_{n \rightarrow \infty}\mathbf{P}(e^{(n)},\left(\mu_1 \otimes \mu_2 \right)^{(n)}) = 0$,
\begin{eqnarray}
 \label{Eqn:3CQICAchievability}
 \lim_{n \rightarrow \infty} n^{-1}\log |\mathcal{M}_{j}^{(n)}| = R_{j} :j \in [2], \mbox{ and }\lim_{n \rightarrow \infty} \tau(e^{(n)}) \leq \tau .
 \nonumber
\end{eqnarray}
The capacity region $\mathscr{C}$ of a $2-$CQBC is the set of all achievable rate-cost vectors and $\mathscr{C}(\tau) \define \{ (R_1,R_2):(R_1,R_2,\tau) \in \mathscr{C}\}$.
\end{definition}
\begin{theorem}
\label{Thm:2CQBC}
A rate-cost triple $(R_1,R_2,\tau) \in [0, \infty)^3$ is achievable if there exists (i) two finite sets $\mathcal{V}_1$, $\mathcal{V}_2$, (ii) a PMF $p_{V_1 V_2}$ on $\mathcal{V}_1 \times \mathcal{V}_2$ and (iii) a mapping function $f:\mathcal{V}_1 \times \mathcal{V}_2 \rightarrow \mathcal{X} $ such that $\sum_{(v_1,v_2)} p_{V_1V_2}(v_1,v_2) \kappa(f(v_1,v_2)) \leq \tau$,
\begin{eqnarray}
R_1 + R_2 < \sum_{j=1}^2 I(V_j ;Y_j) -I(V_1;V_2), \mbox{ and } R_j < I(V_j;Y_j), \mbox{ for } j=1,2 \nonumber 
\end{eqnarray}
holds, where all the information quantities are computed with respect to the state 
\begin{eqnarray}
    &&\rho^{V_1 V_2 X Y_1 Y_2} \define \sum_{(v_1,v_2,x)}  p_{V_1 V_2 X}(v_1,v_2,x)  \ketbra{v_1,v_2,x} \otimes \rho_{x}^{Y_1Y_2}, \label{Eqn:2CQBCState}
\end{eqnarray}
with $p_{V_1V_2X}(v_1,v_2,x) = p_{V_1V_2}(v_1,v_2) \mathds{1}\{x=f(v_1,v_2)\},$ for $(v_1,v_2,x) \in \mathcal{V}_1 \times \mathcal{V}_2 \times \mathcal{X}$.
\end{theorem}
\begin{proof}
Consider a pair $\mathcal{V}_j : j =1,2$ of finite sets, a generic function $f: \mathcal{V}_1 \times \mathcal{V}_2 \rightarrow \mathcal{X}$ and a generic PMF $p_{V_{1}V_{2}}$ on $ \CalV_1 \times \CalV_2 $ that satisfy the constraints stated in the hypothesis, i.e. the average cost $\sum_{(v_1,v_2)} p_{V_1V_2}(v_1,v_2) \kappa(f(v_1,v_2)) \leq \tau$. Let $(R_1,R_2)$ be a rate pair that satisfy the corresponding bounds in Thm.~\ref{Thm:2CQBC}. We divide the proof into four parts entailing the code structure, encoding rule, decoding POVMs and error analysis.

\noindent \textbf{Code structure:}
For $j =1,2$, the code employed to communicate Rx $j$'s message comprises (i) a collection $c_j \define \{v_j^n(b_j,m_j) :  b_j \in [2^{nK_j}], m_j \in [2^{nR_j}]\}$ constructed over the alphabet $\mathcal{V}_j^n$ and (ii) a binning map {$b_j : [2^{nR_1}] \times [2^{nR_2}]  \rightarrow [2^{nK_j}]$.} For $j =1,2,$ let $c_j(m_j) \define (v_j^n(b_j,m_j) : b_j \in [2^{nK_j}])$ denote the bin corresponding to message $m_j$. Conventionally, the codewords in $c_j$ are picked IID with distribution $\prod_{t=1}^n p_{X_j}$. However, to demonstrate the generality of our approach, we pick these codewords, as will be formalized when we analyze the error probability, IID with an alternate distribution $\prod_{t=1}^n q_{V_j}$.
Secondly, the binning map, which identifies the chosen codewords, is conventionally based on typicality. In other words, $(b_1(m_1,m_2),b_2(m_1,m_2)) \in [2^{nK_1}] \times [2^{nK_2}]$ is chosen deterministically to ensure $v_j^n(b_j(m_1,m_2),m_j) : j=1,2$ is jointly typical with respect to $p_{V_1V_2}$. A corresponding randomization - uniformly  chosen amongst all joint typical pairs - is employed to analyze error probability. Instead of forcing typicality, as is done conventionally, we employ a soft probabilistic approach inspired by Cuff's likelihood encoder \cite{201311TIT_Cuf} for choosing the pair of codewords. This will be evident when we specify the distribution with respect to which the binning map $b_j: [2^{nR_1}] \times [2^{nR_2}] \rightarrow [2^{nK_j}]$ is chosen. 

\noindent \textbf{Encoding Rule:}
For each message pair $(m_1,m_2)$, the encoder chooses codeword pair $\left(v_j^n(b_j^n(m_1,m_2),m_j) : j=1,2\right)$.
To communicate the messages $(m_1,m_2)$, the encoder prepares the quantum state $\rho_{x^n(m_1,m_2)}^{Y_1Y_2}$ where 
\begin{eqnarray}
    x^n(m_1,m_2) \define f^n\left( v_1^n(b_1(m_1,m_2),m_1), v_2^n(b_2(m_1,m_2),m_2) \right). \nonumber
\end{eqnarray}
with $f$ evaluated letter-by-letter.

\noindent \textbf{Decoding POVMs:}
The decoding POVMs for Rx $1$ and Rx $2$ being identical, we present the same in terms of a generic index $j$. {Let $\pi_{v_j^n}$ and $\pi$ denote the C-TYP-Proj and U-TYP-Proj with respect to the state $\rho_{v_j^n}$ and $\rho$ respectively, where $\rho_{v_j^n} = \bigotimes_{t=1}^n \rho_{v_{j_t}}$ and for all $v_j \in \CalV_j$, $\rho_{v_j} \define \sum_{v_{\msout{j}}} p_{V_{\msout{j}}|V_j}(v_{\msout{j}}|v_{j}) \rho_{f(v_{1},v_{2})}^{Y_j}$, where $\msout{j}  \in \{1,2\} \backslash \{j\}$ denotes the complement index, i.e., $\{j, \msout{j}\} = \{1,2\}$, and $\rho \define \sum_{v_1,v_2} p_{V_1V_2}(v_1,v_2)  \rho_{f(v_1,v_2)}^{Y_j}$.}
We define the square-root measurement  \cite{BkWilde_2017,BkHolevo_2019} $\{\mu_{m_j} : m_j \in [2^{nR_j}]\}$ to decode $m_j$ as 
\begin{eqnarray}
    &&\hspace{-0.3in}\mu_{m_j} \define \left( \sum_{(\tilde{b}_j, \tilde{m}_j)} \gamma_{\tilde{b}_j, \tilde{m}_j} \right)^{-\frac{1}{2}} \left(\sum_{b_j} \gamma_{b_j,m_j} \right) \left( \sum_{(\tilde{b}_j, \tilde{m}_j)} \gamma_{\tilde{b}_j,\tilde{m}_j} \right)^{-\frac{1}{2}}\!\!, \mbox{ and } \mu_{-1}^{Y_j} \define I^{Y_j} - \sum_{m_j} \mu_{m_j},\mbox{ where } \gamma_{b_j,m_j} \define \pi \pi_{v_j^n(b_j,m_j)} \pi. \nonumber
\end{eqnarray}

\noindent \textbf{Error analysis:}
As is standard in information theory, we derive an upper bound on the error probability of a good code by averaging the error probability over an ensemble of codes. We begin by specifying the codebook distribution, i.e, the random code distribution, with respect to which this averaging is performed. Recall that our codes and the coding scheme are completely specified via the objects: $c_j \define \{v_j^n(b_j,m_j) :  b_j \in [2^{nK_j}], m_j \in [2^{nR_j}]\},$ and $b_j: [2^{nR_1}] \times [2^{nR_2}]  \rightarrow [2^{nK_j}]$ for $j=1,2$. It therefore suffices to specify a joint distribution of these objects. For any choice of the arguments in their respective range spaces, let 
{\begin{eqnarray}
    &&\hspace{0.5in}P\left(\begin{aligned}
    &\left(V_j^n(b_j,m_j)=v_j^n(b_j,m_j) : b_j \in[2^{nK_j}], m_j \in [2^{nR_j}]\right), \\&\left( B_j(m_1,m_2)=b_j(m_1,m_2) : (m_1,m_2) \in [2^{nR_1}] \times [2^{nR_2}]\right) : j =1,2
    \end{aligned}\right) \nonumber \\ &&\hspace{-0.3in}=\left[ \prod_{m_1=1}^{2^{nR_1}}\prod_{b_1=1}^{2^{nK_1}}
    q_{V_1}^n(v_1^n(b_1,m_1)) \right] \left[\prod_{m_2=1}^{2^{nR_2}} \prod_{b_2=1}^{2^{nK_2}} q_{V_2}^n(v_2^n(b_2,m_2))\right] \left[\prod_{m_1=1}^{2^{nR_1}}\prod_{m_2=1}^{2^{nR_2}}\frac{r_{V_1V_2}^n(v_1^n(b_1(m_1,m_2),m_1), v_2^n(b_2(m_1,m_2),m_2))}{\sum_{(l_1,l_2)} r_{V_1V_2}^n(v_1^n(l_1,m_1), v_2^n(l_2,m_2))}\right], \nonumber \\ &&\hspace{-0.3in} \mbox{ where } r_{V_1V_2}^n(v_1^n,v_2^n) \define \frac{p_{V_1V_2}^n(v_1^n,v_2^n)}{q_{V_1}^n(v_1^n) q_{V_2}^n(v_2^n)}, \mbox{ for } (v_1^n,v_2^n) \in \mathcal{V}_1 \times \mathcal{V}_2, \label{Eqn:Dist} 
\end{eqnarray}}
specify our joint codebook distribution.
{From the above distribution of the code, it is evident that each of the codewords in the two codes has been picked independently IID according to the respective distributions. Furthermore, the distribution of the chosen codewords conditional on the message and the coset is highlighted in the following remark.}
\begin{remark}
   As specified when we described the code structure, the distribution of our binning map conditioned on the code distinguishes our encoder from conventional joint typicality-based encoders. Specifically, note from \eqref{Eqn:Dist} that  
    \begin{eqnarray}
&& P\left(B_1(m_1,m_2)=b_1, B_2(m_1,m_2)=b_2 | C_j(m_j)=(v_j^n(b_j,m_j) : b_j \in [2^{nK_j}]) : j=1,2\right) \nonumber \\ 
&=&\frac{r_{V_1V_2}^n(v_1^n(b_1,m_1),v_2^n(b_2,m_2))}{\sum_{(l_1,l_2)}r_{V_1V_2}^n(v_1^n(l_1,m_1),v_2^n(l_2,m_2))}. \label{Eqn:Defr}
\end{eqnarray}
We highlight that the pair of $V_1,V_2-$codewords is chosen via a soft likelihood encoder whose distribution is proportional to $r_{V_1V_2}^n(.,.)$. The numerator and denominator of $r_{V_1V_2}(.)$ exemplify the acceptance of $p_{V_1V_2}$ and rejection of $q_{V_1}q_{V_2}$. This choice of $r_{V_1V_2}(.)$ is inspired from Cuff's likelihood encoder \cite{201311TIT_Cuf}.\end{remark}

We now derive an upper bound on the average error probability of a good code.
\begin{eqnarray}
 \mathbf{P}(e,\uline{\mu}) 
 &\define& \frac{1}{2^{n(R_1 + R_2)}} \sum_{m_1} \sum_{m_2} \tr\left\{ \left(I^{Y_1 Y_2} - \mu_{m_1} \otimes \mu_{m_2} \right)\rho_{e(m_1,m_2)}^{Y_1 Y_2}\right\} \nonumber \\
 &\leq& T_1 + T_2, \mbox{ where } \nonumber \\
 T_j &\define& \frac{1}{2^{n(R_1 + R_2)}}\sum_{m_1,m_2} \tr\left\{ \left(I^{Y_j} - \Gamma_{m_j}\right) \rho_{e(m_1,m_2)}^{Y_j}\right\},\mbox{and } 
 \nonumber \\
 \Gamma_{m_j} &\define& \left( \sum_{(\tilde{b}_j, \tilde{m}_j)}\gamma_{\tilde{b}_j,\tilde{m}_j} \right)^{-\frac{1}{2}} \gamma_{b_j(m_1,m_2),m_j} \left( \sum_{(\tilde{b}_j, \tilde{m}_j)} \gamma_{\tilde{b}_j,\tilde{m}_j} \right)^{-\frac{1}{2}}, \mbox{ for } j=1,2. \nonumber  
 \end{eqnarray}
 This can be verified via the standard quantum union bound \eqref{Eqn:IneqMeasurement} and the fact that $\mu_{m_j} \geq \Gamma_{m_j},$ for $j=1,2$.

\noindent For $j=1,2$, observe that,
\begin{eqnarray}
  T_j &=& \frac{1}{2^{n(R_1 + R_2)}}\sum_{m_1} \sum_{m_2} \sum_{b_1} \sum_{b_2} \mathds{1}\{b_1(m_1,m_2)=b_1, b_2(m_1,m_2)=b_2\} \tr\left\{ \left(I^{Y_j} - \Gamma_{m_j}\right) \rho_{f^n\left(v_1^n(b_1,m_1),v_2^n(b_2,m_2)\right)}^{Y_j}\right\} \nonumber \\
   &\overset{(*)}{=}& T_{j.1} + T_{j.2}, \mbox{ where} \nonumber\\ 
 T_{j.1} &\define&  \frac{1}{2^{n(R_1 + R_2)}}\sum_{m_1} \sum_{m_2} \sum_{b_1} \sum_{b_2} \left[ \mathds{1}\{b_1(m_1,m_2)=b_1, b_2(m_1,m_2)=b_2\}  - \frac{r_{V_1V_2}^n(v_1^n(b_1,m_1),v_2^n(b_2,m_2))}{2^{n(K_1 + K_2)}} \right] \nonumber \\
 &&\tr\left\{ \left(I^{Y_j} - \Gamma_{m_j}\right) \rho_{f^n\left(v_1^n(b_1,m_1),v_2^n(b_2,m_2)\right)}^{Y_j}\right\}, \mbox{ and }  \nonumber
\\  T_{j.2} &\define&  \frac{1}{2^{n(R_1 + R_2)}}\sum_{m_1} \sum_{m_2} \sum_{b_1} \sum_{b_2}
\frac{r_{V_1V_2}^n(v_1^n(b_1,m_1),v_2^n(b_2,m_2))}{2^{n(K_1 + K_2)}} \tr\left\{ \left(I^{Y_j} - \Gamma_{m_j}\right) \rho_{f^n\left(v_1^n(b_1,m_1),v_2^n(b_2,m_2)\right)}^{Y_j}\right\}. \nonumber 
\end{eqnarray}
The equality $(*)$ follows by adding and subtracting 
\begin{eqnarray}
\frac{r_{V_1V_2}^n\left(v_1^n(b_1,m_1),v_2^n(b_2,m_2)\right)}{2^{n(K_1 + K_2)}}, \label{Eqn:Ratio}    
\end{eqnarray}
thereby decomposing $T_j$ into two terms $T_{j.1}$ and $T_{j.2}$ described above. 
\begin{remark}
\label{Rem:BinningandChannelCodingAnalysis}
We highlight the above technique of adding and subtracting the term in \eqref{Eqn:Ratio} which enables us to decompose the error into two terms, $T_{j.1}$ is analyzed using the soft-covering lemma and $T_{j.2}$ using a conventional channel coding approach. This enables us to break the statistical dependence. Moreover, bounding $\frac{q_{V_j}(.)}{p_{V_j}(.)}$ appropriately via the divergence (see equation \eqref{Eqn:Divergenceterm}) enable us to obtain the right channel coding bound. This approach essentially decouples the binning analysis and the channel coding analysis, thus leaving the channel coding analysis oblivious to the presence of binning.    
\end{remark}

The term $T_{j.1}$ is simply bounded by a covering argument \cite{CuffPhDThesis} referred to theirin as `cloud mixing'. Observe that, 
{\begin{eqnarray}
    T_{j.1} &\leq& \frac{1}{2^{n(R_1 + R_2)}}\sum_{m_1} \sum_{m_2} \sum_{b_1} \sum_{b_2}\left|
      \mathds{1}\{b_1(m_1,m_2)=b_1, b_2(m_1,m_2)=b_2\} - \frac{r_{V_1V_2}^n(v_1^n(b_1,m_1),v_2^n(b_2,m_2))}{2^{n(K_1 + K_2)}}\right| \nonumber  \\  &&\left|\tr\left\{ \left(I^{Y_j} - \Gamma_{m_j}\right) \rho_{f^n\left(v_1^n(b_1,m_1),v_2^n(b_2,m_2)\right)}^{Y_j}\right\} \right| \nonumber \\ 
 &\leq& \frac{1}{2^{n(R_1 + R_2)}}\sum_{m_1} \sum_{m_2} \sum_{b_1} \sum_{b_2} \left| \mathds{1}\{b_1(m_1,m_2)=b_1, b_2(m_1,m_2)=b_2\}  - \frac{r_{V_1V_2}^n(v_1^n(b_1,m_1),v_2^n(b_2,m_2))}{2^{n(K_1 + K_2)}}\right| \nonumber .
\end{eqnarray}}
We evaluate the expectation over $C_j(m_j)=\left(V_j^n(b_j,m_j) : b_j \in [2^{nK_j}]\right) : j=1,2$. We obtain 
{\begin{eqnarray}
    \mathbb{E}[T_{1.1}] &\leq& \frac{1}{2^{n(R_1 + R_2)}}\sum_{m_1} \sum_{m_2}  \sum_{b_1} \sum_{b_2} \sum_{c_1(m_1)} \sum_{c_2(m_2)}  P\left(C_j(m_j)=\left(v_j^n(b_j,m_j) : b_j \in [2^{nK_j}]\right) : j=1,2\right) \nonumber \\
  &&\hspace{-0.8in}\left| P\left(B_1(m_1,m_2)=b_1, B_2(m_1,m_2)=b_2 \big| C_j(m_j)=(v_j^n(b_j,m_j) : b_j \in [2^{nK_j}]) : j=1,2 \right)   - \frac{r_{V_1V_2}^n(v_1^n(b_1,m_1),v_2^n(b_2,m_2))}{2^{n(K_1 + K_2)}}  \right| \nonumber \\ 
&=& \frac{1}{2^{n(R_1 + R_2)}}\sum_{m_1} \sum_{m_2}  \sum_{b_1} \sum_{b_2} \sum_{c_1(m_1)} \sum_{c_2(m_2)} P\left(C_j(m_j)=\left(v_j^n(b_j,m_j) : b_j \in [2^{nK_j}]\right) : j=1,2\right) \nonumber \\
  &&\left|
\frac{r_{V_1V_2}^n(v_1^n(b_1,m_1),v_2^n(b_2,m_2))}
{\sum_{(l_1,l_2)} r_{V_1V_2}^n(v_1^n(l_1,m_1),v_2^n(l_2,m_2))} - \frac{r_{V_1V_2}^n(v_1^n(b_1,m_1),v_2^n(b_2,m_2))}
{2^{n(K_1 + K_2)}}\right| \nonumber \\
&=& \frac{1}{2^{n(R_1 + R_2)}} \sum_{m_1} \sum_{m_2}  \sum_{b_1} \sum_{b_2} \sum_{c_1(m_1)} \sum_{c_2(m_2)} P\left(C_j(m_j)=\left(v_j^n(b_j,m_j) : b_j \in [2^{nK_j}]\right) : j=1,2\right) \nonumber \\ 
&& \frac{r_{V_1V_2}^n(v_1^n(b_1,m_1),v_2^n(b_2,m_2))}
{\sum_{(l_1,l_2)} r_{V_1V_2}^n(v_1^n(l_1,m_1),v_2^n(l_2,m_2))}\left|1 -  \frac{1}{2^{n(K_1 + K_2)}} \sum_{(l_1,l_2)}  r_{V_1V_2}^n(v_1^n(l_1,m_1),v_2^n(l_2,m_2))  \right| \nonumber \\
  &\overset{(b)}{=}&\frac{1}{2^{n(R_1 + R_2)}}\sum_{m_1} \sum_{m_2}  \sum_{c_1(m_1)} \sum_{c_2(m_2)} P\left(C_j(m_j)=\left(v_j^n(b_j,m_j) : b_j \in [2^{nK_j}]\right) : j=1,2\right) \nonumber \\ 
  &&\left| \frac{1}{2^{n(K_1 + K_2)}} \sum_{(l_1,l_2)}  r_{V_1V_2}^n(v_1^n(l_1,m_1),v_2^n(l_2,m_2)) -1 \right| \nonumber \\
&=&E\left\{
    \left|\frac{1}{2^{n(K_1 + K_2)}} \sum_{(l_1,l_2)} r_{V_1V_2}^n(V_1^n(l_1,M_1),V_2^n(l_2,M_2)) -1 \right|\right\}, \nonumber
\end{eqnarray}}
Now, from the `cloud mixing' lemma \cite{CuffPhDThesis}, we have $\mathbb{E}[T_{j.1}] \leq \epsilon$, if 
\begin{eqnarray}
  K_j > D(p_{V_j}\|q_{V_j}), \mbox{ for } j=1,2, \mbox{ and }  K_1+K_2 >  D(p_{V_1 V_2}\|q_{V_1} q_{V_2}). \nonumber
\end{eqnarray}
Next, to upper bound the term $T_{j.2}$, we apply the Hayashi-Nagaoka inequality \cite{200307TIT_HayNag}. 
We obtain 
\begin{eqnarray}
&&T_{j.2} \leq 2  \: T_{j.2.1} + 4 \: (T_{j.2.2} + T_{j.2.3}), \mbox{ where} \nonumber \\
T_{j.2.1} &\define& \frac{1}{2^{n(R_1 + R_2)}}\sum_{m_1} \sum_{m_2} \sum_{b_1} \sum_{b_2}
\frac{r_{V_1V_2}^n(v_1^n(b_1,m_1),v_2^n(b_2,m_2))}{2^{n(K_1 + K_2)}} \tr\left\{ (I^{Y_j} - \pi \: \pi_{v^n_j(b_j,m_j)} \: \pi) \rho_{f^n\left(v_1^n(b_1,m_1),v_2^n(b_2,m_2)\right)}^{Y_j}\right\}, \nonumber \\ 
T_{j.2.2} &\define& \frac{1}{2^{n(R_1 + R_2)}} \sum_{m_1} \sum_{m_2} \sum_{b_1} \sum_{b_2} \sum_{\tilde{m}_j \neq m_j} \sum_{\tilde{b}_j}
\frac{r_{V_1V_2}^n(v_1^n(b_1,m_1),v_2^n(b_2,m_2))}{2^{n(K_1 + K_2)}} \tr\left\{  \pi \: \pi_{v^n_j(\tilde{b}_j,\tilde{m}_j)} \: \pi \rho_{f^n\left(v_1^n(b_1,m_1),v_2^n(b_2,m_2)\right)}^{Y_j}\right\},  \nonumber \\
T_{j.2.3} &\define& \frac{1}{2^{n(R_1 + R_2)}} \sum_{m_1} \sum_{m_2} \sum_{b_1} \sum_{b_2} \sum_{\tilde{b}_j \neq b_j} \frac{r_{V_1V_2}^n(v_1^n(b_1,m_1),v_2^n(b_2,m_2))}{2^{n(K_1 + K_2)}}\tr\left\{ \pi \: \pi_{v^n_j(\tilde{b}_j,m_j)} \: \pi \rho_{f^n\left(v_1^n(b_1,m_1),v_2^n(b_2,m_2)\right)}^{Y_j}\right\}. \nonumber  
\end{eqnarray}
\begin{proposition}
    \label{Prop:2CQBC1}
    For all $\epsilon>0$, and $\delta> 0$ sufficiently small and $n$ sufficiently large, we have $\mathbb{E}[T_{j.2.1}] \leq \epsilon$.
\end{proposition}
\begin{proof}
    The proof is provided in Appendix \ref{App:2CQBC1}.
\end{proof}

\begin{proposition}
    \label{Prop:2CQBC2}
For any $\epsilon >0$, and for all $\delta>0$ sufficiently small and $n$ sufficiently large, we have $\mathbb{E}[T_{j.2.i}] \leq \epsilon$, for $j =1,2$ and $i=2,3$, if the following inequalities hold
\begin{eqnarray}
  R_j + K_j  < I(V_j;Y_j) + D(p_{V_j}||q_{V_j})  : j=1,2 \nonumber 
\end{eqnarray}
where all the mutual information quantities are computed with respect to the state {$\rho^{V_1V_2XY_1Y_2}$} defined in \eqref{Eqn:2CQBCState}.
\end{proposition}
\begin{proof}
    The proof is provided in Appendix \ref{App:2CQBC2}.
\end{proof}
  This completes the proof. The bounds stated in Theorem~\ref{Thm:2CQBC} can now be derived via Fourier–Motzkin elimination.
\end{proof}
\begin{remark}
As is evident from the above proof, the likelihood encoder and the proposed steps decouple the encoding and the decoding analysis. From this, it is evident that the likelihood encoder associated with the proposed steps enables us to analyze any channel coding scenario involving binning, classical or quantum. In particular, one can also achieve the general Marton's inner bound for the $2-$CQBC involving both superposition and binning. In the rest of this article, we will employ this likelihood encoder and the associated steps presented above to prove the achievability of all proposed inner bounds.
\end{remark}

\section{Role and Import of Coset Codes for Communication Over $3-$CQBC}
\label{Sec:IdeaSection}
\subsection{Role of Coset Codes in Quantum Broadcast Channels}
\label{SubSec:NeedForCosetCodes}
{\textit{How} and \textit{why} do coset codes satisfying algebraic closure yield higher rates over a $3-$CQBC? Communication on a BC entails fusing the codewords chosen for the different Rxs through a single input. From the perspective of any Rx, a specific aggregation of the codewords chosen for the other Rxs acts as interference. See Fig.~\ref{Fig:3BCInterference}. The Tx can precode for this interference via Marton's binning \cite{197905TIT_Mar}. In general, precoding entails a rate loss \cite[Problem 7.12]{BkNITElGamalKim_2011}. In other words, suppose $W$ is the interference seen by Rx $j$, then the rate that Rx $j$ can achieve by decoding $W$ and peeling it off can be strictly larger than what Rx $j$ can achieve if the Tx precodes for $W$. This motivates every Rx to decode as large a fraction of the interference that it can and precode only for the minimal residual uncertainty.
\begin{figure}
 \centering
\includegraphics[width=4in]{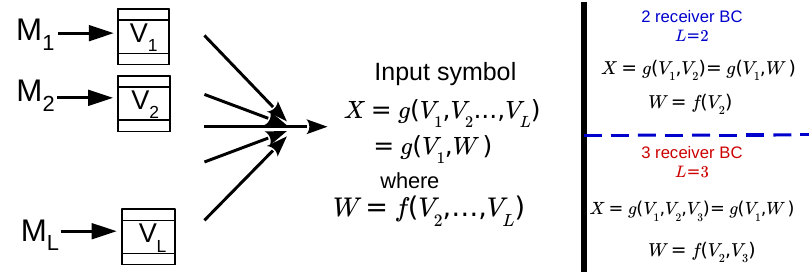}
    \caption{From Rx $j$’s perspective, the interference $W$ it experiences is a specific combination of the other user codewords. While on a $2-$user BC, $W$ is a univariate function of the lone interfering signal, on a $3-$user BC, $W$ is a bivariate function of $V_i, V_k$.}
    \label{Fig:3BCInterference}
\end{figure}
In contrast to a BC with $2$ Rxs, the interference encountered by Rx $j$ on a BC with $3$ Rxs is a \textit{bivariate} function of $V_{i},V_{k}$ - the signals of the other Rxs (Fig.~\ref{Fig:3BCInterference}). A joint design of the $V_{i},V_{k}-$codes endowing them with structure can enable Rx $j$ to decode the interference efficiently even while being unable to decode $V_{i}$ and $V_{k}$. This phenomenon is exemplified through Ex.~\ref{Ex:3CQBCRoleOfCosetCds}. Our discussion of Ex.~\ref{Ex:3CQBCRoleOfCosetCds} illustrates the role and import of coset codes for communication over a $3-$CQBC.}

\begin{Notation}
     For $x \in \{0,1\}$, $\delta \in (0,\frac{1}{2})$, $\sigma_{\delta}(x)\define(1-\delta)\ketbra{1-x}+\delta\ketbra{x}$, and for $k \in \{2,3\}$, $\gamma_{k}(x) = \mathds{1}_{\{x=0\}}\ketbra{0}+\mathds{1}_{\{x=1\}}\ketbra{v_{\varphi_{k}}}$, where $\ket{v_{\varphi_{k}}} = \left[\cos\varphi_{k}~\sin\varphi_{k} \right]^{t}$ and $0 < \varphi_{2}< \varphi_{3} < \frac{\pi}{2}$.
\end{Notation}
\begin{example}
 \label{Ex:3CQBCRoleOfCosetCds}
Let $\ulineCalX = \CalX_{1}\times \CalX_{2} \times \CalX_{3}$ be the input set with $\CalX_{j} = \{0,1\}$ for $j \in [3]$. For $\ulinex = (x_{1},x_{2},x_{3}) \in \ulineCalX$, let $\rho_{\ulinex} = \sigma_{\delta_1}(x_{1}\oplus x_{2} \oplus x_{3}) \otimes \beta_{2}(x_{2}) \otimes \beta_{3}(x_{3})$. The symbol $X_{1}$ is costed via a Hamming cost function i.e., the cost function $\kappa:\ulineCalX \rightarrow \{0,1\}$ is given by $\kappa(\ulinex) = x_{1}$. We investigate $\ScrC(\tau)$.
\end{example}
\noindent\textit{Commutative Case}: $\beta_{k}(x_{k}) = \sigma_{\delta}(x_{k})$ for $k =2,3$ with $\tau * \delta_1 \boldsymbol{\stackrel{(i)}{<}} \delta$ and $h_{b}(\delta) \boldsymbol{\stackrel{(ii)}{<}} \frac{1+h_{b}(\tau * \delta_1)}{2}$.

Since $\rho_{\ulinex} : \ulinex \in \ulineCalX$ are commuting, the $3-$CQBC for the commuting case can be identified via a $3-$user classical BC equipped with input set $\ulineCalX$ and output sets $\CalY_{1}=\CalY_{2}=\CalY_{3}=\{0,1\}$. Its input $X=(X_{1},X_{2},X_{3}) \in \ulineCalX$ and outputs $Y_{j}: j \in [3]$ are related via $Y_{1}=X_{1}\oplus X_{2} \oplus X_{3}\oplus N_{1}$, $Y_{j}=X_{j} \oplus N_{j}$ where $N_{1}, N_{2},N_{3}$ are mutually independent Ber$(\delta_1)$, Ber$(\delta)$, Ber$(\delta)$ random variables respectively. We study the question : \textit{What is the maximum achievable rate for Rx $1$ while Rxs $2$ and $3$ are fed at their respective capacities $C_{2}\define C_{3}\define 1-h_{b}(\delta)$?}

Choose any $k \in \{2,3\}$, any $j \in [3]\setminus\{k\}$ and any input PMF $p_{X}=p_{X_{1}X_{2}X_{3}}$. Observe $I(X_{j};Y_{k})=0$. \textcolor{black}{Rx $2$ and $3$ can be pumped information only through $X_{2}$ and $X_{3}$ respectively. }Thus, achieving capacities for Rxs $2$, $3$ forces $(X_{2},X_{3})\sim p_{X_{2}X_{3}}\!=\mbox{Ber}(\frac{1}{2})\otimes \mbox{Ber}(\frac{1}{2})$, i.e., independent uniform. This implies Rx $1$ suffers interference $X_{2}\oplus X_{3}\!\!\sim $Ber$(\frac{1}{2})$. With $X_{1}$ being Hamming weight constrained to $\tau<\frac{1}{2}$, Tx cannot cancel this interference. Can Rx $1$ achieve its interference-free cost constrained capacity $C_{1}\define h_{b}(\tau * \delta_1)-h_{b}(\delta_1)$?

The relationship $Y_{1}=X_{1}\oplus \left[ X_{2}\oplus X_{3}\right]\oplus N_{1}$ between input $X_{1}$ and output $Y_{1}$ with $X_{2}\oplus X_{3} \sim $Ber$(\frac{1}{2})$ known only to the Tx is a typical well studied case of \textit{rate loss} \cite{201804TIT_PadPra} in a binary point-to-point channel with Tx side information  \cite{BkNITElGamalKim_2011}. These imply that Rx $1$ can achieve a rate of $C_{1}$
\textit{only if} Rx $1$ recovers $X_{2}\oplus X_{3}$ \textit{perfectly}. Can it recover $X_{2}\oplus X_{3}$ \textit{perfectly}?

With \textbf{unstructured IID} codes, Rx $1$ can recover $X_{2}\oplus X_{3}$ perfectly \textit{only} by recovering $X_{2}$ \textit{and} $X_{3}$ \textit{perfectly} \cite{201804TIT_PadPra}. However, from inequality $\boldsymbol{(ii)}$, we have $1-h_{b}(\delta_1) < C_{1}+C_{2}+C_{3}$ implying the $X_{1}-Y_{1}$ or the $X-Y_{1}$ channel \textbf{cannot} support Rx $1$ decoding $X_{1},X_{2},X_{3}$. Let's consider linear/coset codes.

A simple linear coding scheme can achieve the rate triple $(C_{1},C_{2},C_{3})$. In contrast to building independent codebooks for Rxs $2$ and $3$, choose cosets $\lambda_{2},\lambda_{3}$ of a \textit{common linear code} $\lambda$ that achieves capacity $1-h_{b}(\delta)$ of both $X_{2}-Y_{2}$ and $X_{3}-Y_{3}$ binary symmetric channels. Since the sum of two cosets is another coset of the \textit{same linear} code, the interference patterns seen by Rx $1$ are constrained to another coset $\lambda_{2}\oplus \lambda_{3}$ of the same linear code of rate $1-h_{b}(\delta)$. Instead of attempting to decode the \textit{pair} of Rx $2,3$'s codewords, suppose Rx $1$ \textit{decodes only the sum of the Rx $2,3$ codewords} within $\lambda_{2}\oplus \lambda_{3}$. Specifically, suppose Rx ${1}$ attempts to jointly decode it's codeword and the sum of Rx $2$ and $3$'s codewords, the latter being present in $\lambda_{2}\oplus \lambda_{3}$, then it would be able to achieve a rate of $C_{1}$ for itself if $\mathscr{C}_{1}=1-h_{b}(\delta_1) > C_{1}+ \max\{C_{2},C_{3}\} = C_{1}+1-h_{b}(\delta)$. The latter inequality is guaranteed since $\tau * \delta_1 \boldsymbol{\stackrel{(i)}{<}} \delta$ holds. We conclude this case with the following fact proven in \cite{201804TIT_PadPra}.
\begin{fact}
 \label{Prop:UnstructuredCodingSchemeSub-Optimal}
For Ex.~\ref{Ex:3CQBCRoleOfCosetCds} with inequalities $\boldsymbol{(i),(ii)}$, $(C_{1},C_{2},C_{3})$ is \textbf{not} achievable via \textbf{any} known unstructured IID coding scheme. $(C_{1},C_{2},C_{3})$ is achievable via the above linear code strategy. 
\end{fact} 
\noindent\textit{Non-Commuting Case}: $\beta_{k}(x_{k}) = \gamma_{k}(x_{k})$ for $k=2,3$ with $h_{b}(\tau*\delta_1)+\tilde{h}_{b}(\cos\varphi_{2})+\tilde{h}_{b}(\cos\varphi_{3}) \boldsymbol{\stackrel{(a)}{>}} 1 \boldsymbol{\stackrel{(b)}{>}} h_{b}(\tau*\delta_1)+\tilde{h}_{b}(\cos\varphi_{3})$ where $\tilde{h}_{b}(x)\define h_{b}(\frac{1+x}{2})$ for $x \in [0,\frac{1}{2}]$.

We follow the same line of argument as above. For $k=2,3$, the capacity of user $k$ is $C_{k}\define \tilde{h}_{b}(\cos\varphi_{k})$ and is achieved \textit{only} by choosing $p_{X_{2}X_{3}}=p_{X_{2}}p_{X_{3}}$ with each marginal Ber$(\frac{1}{2})$\cite[Ex.5.6]{BkHolevo_2019}. This forces Rx $1$ to suffer Ber$(\frac{1}{2})$ interference $X_{2}\oplus X_{3}$. Identical arguments as above revolving around rate loss  imply that Rx $1$ can achieve a rate of $C_{1}\define h_{b}(\tau * \delta_1)-h_{b}(\delta_1)$ \textit{only if} Rx $1$ recovers $X_{2}\oplus X_{3}$ \textit{perfectly}. The only way unstructured IID codes can enable Rx $1$ recover $X_{2}\oplus X_{3}$ \textit{perfectly} is by recovering the \textit{pair} $X_{2},X_{3}$ \textit{perfectly}. This can be done \textit{only if} the unconstrained capacity $\CalC_{1}=1-h_{b}(\delta_1) \geq C_{1}+C_{2}+C_{3}$. However, from $\boldsymbol{(a)}$ we have $\CalC_{1}< C_{1}+C_{2}+C_{3}$. Unstructured IID codes cannot achieve rate triple $(C_{1},C_{2},C_{3})$.

We design capacity achieving linear codes $\lambda_{2},\lambda_{3}$ for Rxs $2,3$ respectively in such a way that $\lambda_{2}$ is a \textit{sub-coset} of $\lambda_{3}$. Then, the interference patterns are contained within $\lambda_{2}\oplus \lambda_{3}$ which is now a coset of $\lambda_{3}$. Rx $1$ can therefore decode into this collection, which is of rate at most $C_{3}=\tilde{h}_{b}(\cos\varphi_{3})$. Inequality $\boldsymbol{(b)}$ implies the unconstrained capacity $\CalC_{1}=1-h_{b}(\delta_1) \geq C_{1}+\max\{C_{2},C_{3}\} = C_{1}+C_{3}$ implying coset codes can achieve the rate triple $(C_{1},C_{2},C_{3})$. We conclude with a formal statement.
\begin{proposition}
 \label{Prop:UnstructuredCodingSchemeSub-Optimal}
For Ex.~\ref{Ex:3CQBCRoleOfCosetCds} with inequalities $\boldsymbol{(a),(b)}$, $(C_{1},C_{2},C_{3})$ is \textbf{not} achievable via \textbf{any} known unstructured IID coding scheme. $(C_{1},C_{2},C_{3})$ is achievable via the above linear code strategy. 
\end{proposition}
\subsection{Evolving Structure of a General Coding Strategy}
\label{SubSec:StructureOfAGeneralCodingStrategy}
Ex.~\ref{Ex:3CQBCRoleOfCosetCds} exemplifies two facts. Firstly, interference on a $3-$CQBC is in general a bivariate function of the interferer signals. Secondly, the presence of a \textit{rate loss} motivates each Rx to decode as large a component of the interference it suffers and minimize the residual component that is precoded for. Moreover, the coding strategy designed for Ex.~\ref{Ex:3CQBCRoleOfCosetCds} illustrates that jointly designed coset codes enables a Rx decode the bivariate interference component more efficiently, thereby minimizing the residual component precoded for. A general coding scheme must facilitate \textit{each Rx} to efficiently decode \textit{both} univariate and bivariate components of its two interfering signals. Unstructured IID codes and jointly designed coset codes are efficient at decoding the former and latter components, respectively.

A general coding strategy must therefore split each Tx's message into parts to enable each Rx to decode both the univariate and bivariate components of the interference it encounters. This will involve multiple codes, a Marton's precoding strategy to multiplex the different codewords and simultaneous decoding to be employed at the Rx to decode the interference and the corresponding messages. 
We develop this general coding strategy in three pedagogical steps. In the first step presented in Sec.~\ref{Sec:3to1CQBC}, we analyze a scenario where only one of the Rx decodes bivariate interference. Building on this, in Step II (Sec.~\ref{SubSec:StepICodingTheorem}) we analyze decoding of bivariate interferences at all Rxs using coset codes. Finally, we combine this with the unstructured IID codes strategy in Step III (Sec.~\ref{SubSec:Step2}) to derive the most general inner bound.

\section{Step I : A Coset Code Strategy to Manage Bivariate Interference at a Single Receiver}
\label{Sec:3to1CQBC}
\begin{figure}
 \centering
\includegraphics[width=3.5in]{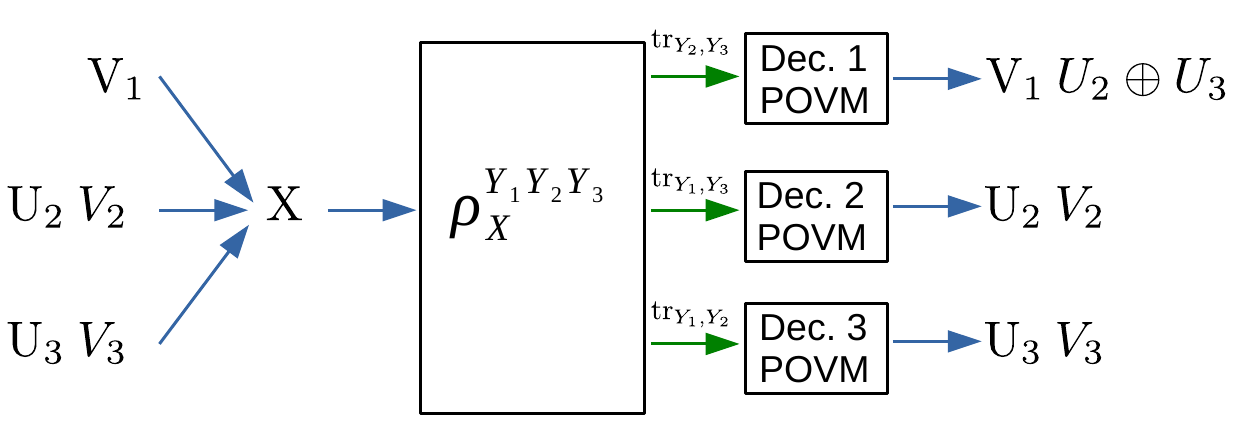}
    \caption{Depiction of random variables of the coding strategy}
    \label{Fig:3to1CQBC}
\end{figure}

In this section, we present the first step toward deriving an inner bound on the capacity region of the $3$-CQBC using coset codes. Specifically, we manage interference suffered by a single receiver - Rx $1$ - using coset codes. Tx encodes one part each of Rx $2$ and $3$'s messages using cosets of a common linear code (see Fig.~\ref{Fig:3to1CQBC}). By decoding the sum of these chosen codewords, Rx $1$ peels off bivariate interference. Thm.~\ref{Thm:3to1CQBC} characterizes the corresponding inner bound, and we provide a complete proof. The coding strategy we present here combine all the new elements $-$ likelihood encoder, simultaneous decoding of unstructured IID codes and coset codes using tilted decoding POVMs. Since this simplified setting involves fewer codebooks, it provides an ideal pedagogical step to present the combination of all new elements in a single proof.
\begin{theorem}
\label{Thm:3to1CQBC}
    A rate-cost quadruple $(R_1,R_2,R_3, \tau)$ is achievable if there exist (i) a finite field $\mathcal{U}_2=\mathcal{U}_3=\mathcal{F}_{\Prime}$, (ii) finite sets $\mathcal{V}_j : j\in[3]$, (iii) PMF $p_{U_2U_3\ulineV}$ on $\mathcal{U}_2 \times \mathcal{U}_3 \times \mathcal{\ulineV}$, (iv) non-negative numbers $K_1, K_2, K_3,  L_2, S_2, T_2, L_3, S_3$ and $T_3$ and (v) a mapping function $f: \mathcal{U}_2 \times \mathcal{U}_3 \times \mathcal{\ulineV} \rightarrow \mathcal{X}$ such that $R_j= T_j + L_j$ for $j=2,3$, $\sum_{u_2,u_3,\ulinev} p_{U_2U_3\ulineV}(u_2,u_3,\ulinev) \kappa(f(u_2,u_3,\ulinev))\leq \tau$ and 
\begin{eqnarray}
    S_{\mathcal{A}} + K_{\mathcal{B}} &>& |\mathcal{A}|\log\left(\Prime\right) + \sum_{\beta \in \mathcal{B}} H(V_{\beta}) - H(U_{\mathcal{A}},V_{\mathcal{B}}), \nonumber \\
    \max\{S_2 + T_2, S_3 + T_3\} &>& \log(\Prime) - \min_{\theta \in \mathcal{F}_{\Prime} \setminus \{0\}} H(U_2 \oplus \theta U_3), \nonumber\\
    K_{\mathcal{B}} + \max\{S_2 + T_2, S_3 + T_3\} &>& \log(\Prime) + \sum_{\beta \in \mathcal{B}} H(V_{\beta}) - \min_{\theta \in \mathcal{F}_{\Prime} \setminus \{0\}} H(U_2 \oplus \theta U_3, V_{\mathcal{B}}), \nonumber\\
    R_1 + K_1 &<& H(V_1) - H(V_1| Y_1 , U),\nonumber \\
    \max\{S_2 + T_2,S_3 + T_3\} &<&  \log (\Prime)    - H(U | Y_1,V_1), \nonumber\\
   R_1 + K_1 + \max\{S_2 + T_2, S_3 + T_3\} &<& \log(\Prime) + H(V_1) - H(V_1,U| Y_1) , \nonumber \\
    L_j + K_j  &<& H(V_j) - H(V_j|Y_j,U_j), \nonumber \\
        S_j + T_j &<&   \log(\Prime) - H(U_j|Y_j,V_j), \nonumber \\
        L_j + K_j + S_j + T_j &<&  \log(\Prime)  + H(V_j) -H(U_j;V_j|Y_j) \nonumber
   \end{eqnarray}
   holds for $j=2,3$, where $\mathcal{A} \subseteq \{2,3\}$, $ \mathcal{B} \subseteq \{1,2,3\}$, $S_{\mathcal{A}} \define \sum_{\alpha \in \mathcal{A}} S_{\alpha}$,  $U_{\mathcal{A}} \define (U_{\alpha}: \alpha \in \mathcal{A})$, $K_{\mathcal{B}} \define \sum_{\beta \in \mathcal{B}} K_{\beta}$, $V_{\mathcal{B}} \define (V_{\beta} : \beta \in \mathcal{B})$, and all the information quantities are computed with respect to the state 
    \begin{eqnarray}
        &&\rho^{UU_2U_3\ulineV  \ulineY} \define \sum_{u,u_2,u_3,\ulinev} p_{UU_2U_3\ulineV}(u,u_2,u_3, \ulinev) \ketbra{u,u_2,u_3,\ulinev} \otimes \rho^{Y_1Y_2Y_3}_{f(u_2,u_3,\ulinev)}.  
        \label{Eqn:Step1ProofSinlgeLetterState}
    \end{eqnarray}
with $p_{UU_2U_3\ulineV}(u,u_2,u_3, \ulinev) = p_{U_2U_3\ulineV}(u_2,u_3, \ulinev) \mathds{1}\{u=u_2\oplus u_3\}$.    

\end{theorem}

\begin{proof}
As stated in the hypothesis, choose a generic finite field and let $\Prime$ denote its cardinality. Set $\CalU_{2}=\CalU_{3}=\mathcal{F}_{\Prime}$ be this finite field of cardinality $\upsilon$. Choose finite sets $\mathcal{V}_j : j \in [3]$, a generic function $f: \mathcal{U}_2 \times \mathcal{U}_3 \times \mathcal{\ulineV} \rightarrow \mathcal{X}$ and a generic PMF $p_{U_{2}U_{3}\ulineV}$ on $ \CalU_2 \times \CalU_3 \times \mathcal{\ulineV}$, that satisfy the cost constraint $\sum_{u_2,u_3,\ulinev} p_{U_2U_3\ulineV}(u_2,u_3,\ulinev) \kappa(f(u_2,u_3,\ulinev))\leq \tau$. Note that $p_{U U_2 U_3 \ulineV} = p_{U_2 U_3 \ulineV} \mathds{1}\{U=U_2 \oplus U_3\}$.   
Having chosen these objects, fix the same throughout the rest of the discussion and let the joint quantum state be defined as in \eqref{Eqn:Step1ProofSinlgeLetterState} with respect to this PMF $p_{U_2U_3\ulineV}$. Let $(R_{1},R_{2},R_{3}) \in [0,\infty)^{3}$ be a rate triple for which there exists non negative numbers $K_1, K_2, K_3,  L_2, S_2, T_2, L_3, S_3$ and $T_3$ such that $R_j=T_j + L_j$ for $j=2,3$, and all the bounds stated in the hypothesis of Thm.~\ref{Thm:3to1CQBC} statement holds with respect to the joint single-letter quantum state in \eqref{Eqn:Step1ProofSinlgeLetterState}. Having identified the parameters that define our coding strategy, we now describe the code structure.
\begin{figure}
 \centering
\includegraphics[width=5in]{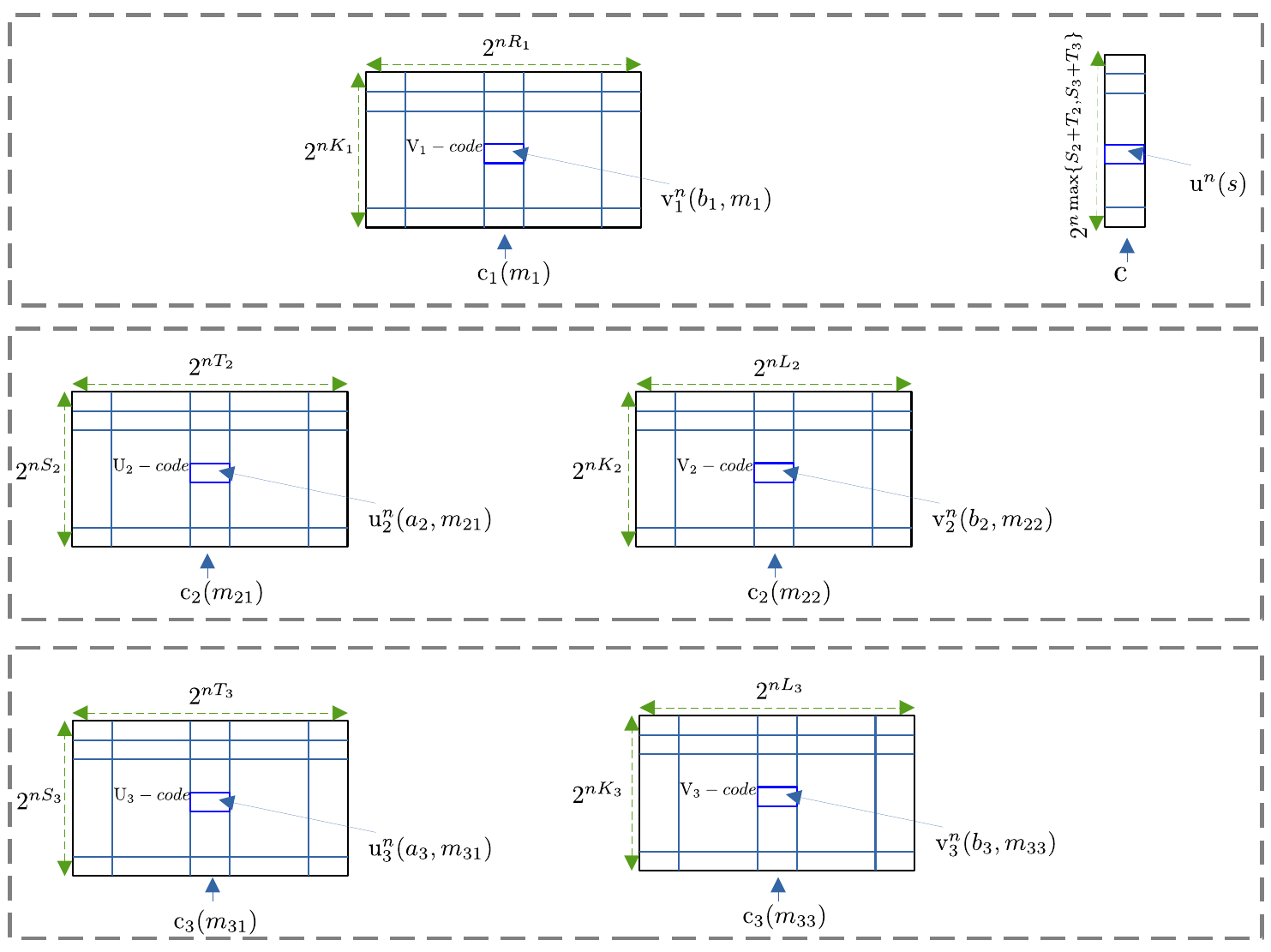}
    \caption{The coding strategy involves five codes - two of them being NCCs built over the finite field $\mathcal{F}_{\Prime}$ and three are unstructured IID codes built on $\CalV_j : j\in[3]$. The $5$ codebooks on the left are used by Tx.  The rightmost coset is obtained by adding $U_2$ and $U_3$. For $j\in [3]$, Rx $j$ decodes into the two codes within the gray dotted box.}
    \label{Fig:Codestructure3to1CQBC}
    \vspace{-0.15in}
\end{figure}

\noindent \textbf{Code structure:}
The coding strategy involves five codebooks - one to encode Rx $1$'s message and two each for encoding Rx $2$ and Rx $3$'s messages respectively. See Fig.~\ref{Fig:Codestructure3to1CQBC}.
Rx $1$’s message is communicated using (i) an IID code 
\begin{eqnarray}
c_1 \define \bigl\{ v_1^n(b_1,m_1) : b_1 \in [2^{nK_1}], \:  m_1 \in [2^{nR_1}] \bigr\}, \nonumber 
\end{eqnarray}
constructed over the alphabet $\mathcal{V}_1$ and (ii) a binning map $b_1 : [2^{nR_1}] \times [2^{nR_2}] \times [2^{nR_3}] \rightarrow [2^{nK_1}]$. Let $v_1^n(b_1(\ulinem),m_1)$ denote the chosen codeword from the bin $c_1(m_1) \define \{v_1^n(b_1,m_1) : b_1 \in [2^{nK_1}]\}$ corresponding to message $m_1$.
For $j=2,3$, Rx $j$’s message is split into two parts
$m_j = (m_{j1}, m_{jj})$.
The message $m_{j1}$ is communicated using a NCC over the finite field $\mathcal{F}_{\Prime}$ and the message $m_{jj}$ is communicated using an IID code built over $\CalV_j$.
Note that the NCCs are chosen such that the smaller of the two is a subcoset of the larger. Specifically, for $j=2,3$ 
\begin{eqnarray}
    \mbox{set } s_j \define \left \lceil \frac{n S_j}{\log(\Prime)} \right \rceil \mbox{ and } t_j \define \left \lfloor \frac{n T_j}{\log(\Prime)} \right \rfloor, \mbox{ and define } \sfb, \sfs \mbox{ such that } \{\sfb,\sfs\}=\{2,3\} \mbox{ and } s_{\sfb} + t_{\sfb} \geq s_{\sfs} + t_{\sfs}.  \label{Eqn:Step1ProofBiggerOfkPlusl} 
\end{eqnarray}
For $j=2,3$, the message $m_{j1}$ is communicated through a (i) NCC $(n,s_j,t_j,g_I^j,g_{O|I}^j,\bias_j^n)$ with generator matrices
$g_I^j \in \mathcal{F}_{\Prime}^{s_j \times n}$,
$g_{O|I}^j \in \mathcal{F}_{\Prime}^{t_j \times n}$, a bias vector $\bias_j^n \in \mathcal{F}_{\Prime}^n$, $s_j \define \left \lceil \frac{n S_j}{\log(\Prime)} \right \rceil$ and $t_j \define \left \lfloor \frac{n T_j}{\log(\Prime)} \right \rfloor$ and (ii) a binning map $a_j : [2^{nR_1}] \times [2^{nR_2}] \times [2^{nR_3}] \rightarrow [2^{nS_j}]$.
Let $g_{j}\define \left[\!\! \begin{array}{c}g_{I}^{j}\\g_{O/I}^{j}\end{array}\!\!\right]$ for $j=2,3$. These generator matrices are identified such that  $g_{\sfb}\define \left[\!\! \begin{array}{c}g_{\sfs}\\g_{\Delta}\end{array}\!\!\right]$. This choice of the generator matrices ensures that the smaller of {these} NCCs is a subcoset of the larger NCC. This will also be imposed by the distribution of the corresponding random codebooks with respect to which we analyze the average error probability. For $j=2,3$, let 
$u^n_j(a_j(\ulinem), m_{j1})$ denote the chosen codeword from the coset $c_{j1}(m_{j1}) \define \{u^n_j(a_j,m_{j1}) : a_j \in \mathcal{F}_{\Prime}^{s_j}\}$ corresponding to message $m_{j1}$.
The message $m_{jj}$ is communicated using (i) the codebook
\begin{eqnarray}
c_{jj} \define \bigl\{ v_j^n(b_j,m_{jj}) : b_j \in [2^{nK_j}], \: m_{jj} \in [2^{nL_j}] \bigr\}, \nonumber
\end{eqnarray}
constructed over the alphabet $\mathcal{V}_j$ and (ii) a binning map $b_j : [2^{nR_1}] \times [2^{nR_2}] \times [2^{nR_3}] \rightarrow [2^{nK_j}]$. For $j=2,3$, let $v_j^n(b_j(\ulinem), m_{jj})$ denote the chosen codeword from the bin $c_{jj}(m_{jj}) \define \{v_j^n(b_j,m_{jj}) : b_j \in [2^{nK_j}]\}$ corresponding to message $m_{jj}$.
\begin{remark}
 \label{Rem:InterAndIntraStructure}
 We highlight the intra- and inter-structural properties of the codes. Firstly, $c_{21}$ and $c_{31}$ are coset codes. This endows them with individual algebraic closure properties that are absent in unstructured IID codes. Secondly, the rows of $g_{\sfs}$ being contained in the rows of $g_{\sfb}$ implies that by adding all codewords of Rx $2$'s NCC to all codewords of Rx $3$'s NCC, we obtain a collection of vectors of cardinality at most $\Prime^{s_{\sfb}+t_{\sfb}}$ and not $\Prime^{s_{1}+t_{1}+s_{2}+t_{2}}$. We crucially exploit this sum containment in the decoding specified below.
\end{remark}

\noindent \textbf{Encoding Rule:}
To communicate the message $\ulinem=(m_1,m_2,m_3)$, the encoder prepares the quantum state $\rho^{Y_1 Y_2 Y_3}_{x^n(\ulinem)}$ where
\begin{eqnarray}
    x^n(\ulinem) \define f^n \left( v_1^n(b_1(\ulinem),m_1),u_2^n(a_2(\ulinem),m_{21}), v_2^n(b_2(\ulinem),m_{22}), u_3^n(a_3(\ulinem),m_{31}), v_3^n(b_3(\ulinem),m_{33}) \right), \nonumber 
\end{eqnarray}
with $f$ evaluated letter-by-letter.

\noindent \textbf{Decoding POVMs:}
We begin by describing the decoding POVM for Rx $1$. In addition to decoding $m_1$, Rx $1$ aims to decode the sum of the chosen $U_2-, U_3-$codewords simultaneously. Naturally, $m_{1}$ is deciphered by decoding into $c_{1}$. In regards to the sum of the chosen $U_{2}-,U_{3}-$codewords, the decoder identifies the `sum coset code'  $c$ with parameters $(n,s_{\sfb}+t_{\sfb},g_{\sfb},\bias^n)$, where $\bias^n=\bias_2^n \oplus \bias_3^n$.
For $ a \in \mathcal{F}_{\Prime}^{s_{\sfb}+t_{\sfb}}$, we define $u^n(a) \define a g_{\sfb} \oplus \bias^n$ to represent a  generic codeword in this `sum coset code'.\footnote{Strictly speaking, one must refer to codewords of this sum coset code as vectors since none of the Txs choose any codeword in this sum coset code explicitly.} 

\noindent We now employ Sen’s TSA technique \cite{202102SAD_Sen} to enable simultaneous decoding. The role of TSA is to handle unions and intersections of typical subspaces, which may overlap in a nontrivial manner. This is achieved by enlarging the Hilbert space, which allows the typical subspaces to be tilted in different orthogonal directions, thereby increasing their separation. In our setting, Rx $1$ simultaneously decodes two codewords $-$ $v_1^n(b_1(\ulinem),m_1)$ and {$u^n(a)$}. In this case, there are three possible error events, each corresponding to a distinct subspace. Therefore, we introduce two tilting maps to rotate two of these subspaces in different orthogonal directions, while the third subspace, corresponding to the event in which both codewords are decoded incorrectly, remains untilted. Moreover, we augment the space to ensure the smoothing property. In what follows, we define the auxiliary spaces and the tilting maps.
Consider two auxiliary finite sets $\CalD_1$ and $\CalD_2$ along with corresponding Hilbert space ${\CalD_{1}}$ and ${\CalD_{2}}$ of dimension $|\CalD_{1}|$ and $|\CalD_{2}|$, respectively. We define the extended space $\CalH_{Y_1}^e$ as follows
\[\left(\CalH_{Y_1}^e\right)^{\otimes n} \define \left(\CalH_{Y_1}^G\right)^{\otimes n} \oplus \left( \left(\CalH_{Y_1}^G\right)^{\otimes n} \otimes \CalD_1^{\otimes n} \right) \oplus \left(\left(\CalH_{Y_1}^G\right)^{\otimes n} \otimes \CalD_2^{\otimes n}\right),\]
where $\CalH_{Y_1}^G=\CalH_{Y_1} \otimes \mathbb{C}^2$.
In the sequel, we let $\boldsymbol{\CalH_{Y_1}}= \CalH_{Y_1}^{\otimes n}$, $\boldsymbol{\CalH_{Y_1}^G}=\left(\CalH_{Y_1}^G\right)^{\otimes n}$,
$\boldsymbol{\CalH_{Y_1}^{e}}=\left(\CalH_{Y_1}^{e}\right)^{\otimes n}$ and $\boldsymbol{\CalD_{j}}= \CalD_{j}^{\otimes n} : j \in [2]$.
For $j \in [2]$, we define the tilting map $\CalT_{ d_{j}^{n}, \eta} : \boldsymbol{\CalH_{Y_1}^G}\rightarrow
\boldsymbol{\CalH_{Y_1}^G} \oplus \left(\boldsymbol{\CalH_{Y_1}^G}  \otimes   \boldsymbol{\CalD_{j}}\right)
$ as
\begin{eqnarray}
\CalT_{ d_{j}^{n}, \eta}(\ket{h}) = \frac{1}{\sqrt{1+\eta^2} } \left(\ket{h} +\eta \ket{h} \otimes  \ket{d_{j}^{n}}\right).
\nonumber
\end{eqnarray}
Consider the classical-quantum state 
\begin{eqnarray}
\rho^{V_1UY_1} &=& \sum_{v_1} \sum_{u} p_{V_1U}(v_1,u) \ketbra{v_1,u} \otimes \rho_{v_1,u}^{Y_1}, \mbox{ where} \label{Eqn:3to1CQBCstateDec1} \\
\rho^{Y_1}_{v_1,u} &\define& \sum_{u_2,u_3} \sum_{v_2,v_3} p_{U_2U_3V_2V_3|V_1U}(u_2,u_3,v_2,v_3|v_1,u) \rho^{Y_1}_{f(u_2,u_3,\ulinev)}. \nonumber
\end{eqnarray}
We define the following associated density operators
\begin{eqnarray}
&& \rho^{Y_1}_{v_1} = \sum_{u} p_{U|V_1}(u|v_1) \rho^{Y_1}_{v_1,u}, \:
\rho^{Y_1}_{u} = \sum_{v_1} p_{V_1|U}(v_1|u) \rho^{Y_1}_{v_1,u},  \mbox{ and }  \rho^{Y_1} = \sum_{v_1} \sum_{u} p_{V_1U}(v_1,u) \rho^{Y_1}_{v_1,u}. 
\label{Eqn:3to1CQBCtheassociatedstateforuser1}
\end{eqnarray}
Let $\pi_{v_1^n,u^n}^{Y_1}, \pi_{v_1^n}^{Y_1}, \pi_{u^n}^{Y_1}$ and $\pi^{Y_1}$ be the C-Typ-Proj with respect to the states $ \rho^{Y_1}_{v_1^n,u^n} \define \bigotimes_{t=1}^n  \rho^{Y_1}_{v_{1_t},u_t}$, $\rho^{Y_1}_{v_1^n} \define \bigotimes_{t=1}^n \rho^{Y_1}_{v_{1_t}}$, $\rho^{Y_1}_{u^n} \define \bigotimes_{t=1}^n \rho^{Y_1}_{u_t}$ and $\bigotimes_{t=1}^n \rho^{Y_1}$ respectively.
Consider the following POVM defined in the Hilbert space $\boldsymbol{\CalH_{Y_1}}$ 
\begin{eqnarray}
    G^1_{v_1^n,u^n} \define \pi_{u^n}^{Y_1}  \pi^{Y_1}_{v_1^n,u^n}  \pi_{u^n}^{Y_1}, \: G^2_{v_1^n,u^n} \define \pi_{v_1^n}^{Y_1}  \pi_{v_1^n,u^n}^{Y_1}  \pi_{v_1^n}^{Y_1}, \mbox{ and }  G^0_{v_1^n,u^n} \define \pi^{Y_1}  \pi_{v_1^n,u^n}^{Y_1}  \pi^{Y_1}.\label{Eqn:3to1CQBCtheoriginalpovmelements}
\end{eqnarray}

\noindent Using Gelfand–Naimark’s Theorem \cite[Thm.~3.7]{BkHolevo_2019}, we construct a projector $\olineG^{\CalJ}_{v_1^n,u^n}$ in $\boldsymbol{\CalH_{Y_1}^G}$ that yields identical measurement statistics on states in $\boldsymbol{\CalH_{Y_1}^G}$ as $G^{\CalJ}_{v_1^n,u^n}$ gives on states in $\boldsymbol{\CalH_{Y_1}}$ for $\CalJ \in \{0,1,2\}$.
We define the complement orthogonal projector as 
\begin{eqnarray}
\olineB_{v_1^n,u^n}^{\CalJ} \define I_{\boldsymbol{\CalH_{Y_1}^G}}-\olineG^{\CalJ}_{v_1^n,u^n}  \mbox{ for } \CalJ \in \{0,1,2\}. \label{Eqn:3to1CQBCcomplementprojector}
\end{eqnarray}
Now, we define the tilted projectors by  $\beta_{v_1^n,d_1^n,u^n,d_2^n}^{\CalJ}=\CalT_{d^n_{\CalJ^c},\eta}\left(\olineB^{\CalJ}_{v_1^n,u^n}\right)$ for $\CalJ \in\{1,2\}$ and the untilted projector as $\beta^0_{v_1^n,d_1^n,u^n,d_2^n}=\olineB^0_{v_1^n,u^n}$.
Next, we define $\beta^{*}_{v_1^n,d_1^n,u^n,d_2^n}$ as the projector in $\boldsymbol{\CalH_{Y_1}^{e}}$ whose support is the union of the supports of $\beta^{\CalJ}_{v_1^n,d_1^n,u^n,d_2^n}$ for all $\CalJ \in \{0,1,2\}$.
Let $\pi_{\boldsymbol{\CalH_{Y_1}^G}}$ be the orthogonal projector in $\boldsymbol{\CalH_{Y_1}^{e}}$ onto $\boldsymbol{\CalH_{Y_1}^G}$ . Finally, we define the square-root measurement \cite{BkWilde_2017,BkHolevo_2019} $\{\mu_{m_1,a}^{Y_1}: (m_1,a) \in [2^{nR_1}] \times \mathcal{F}_{\upsilon}^{s_{\sfb} + t_{\sfb}} \}$ as
\begin{eqnarray}
&&\mu_{m_1,a}^{Y_1} \define \left(\sum_{\widehat{m}_1,\widehat{b}_1,\widehat{a}} \gamma^{*}_{(v_1^n,d_1^n)(\widehat{b}_1,\widehat{m}_1),(u^n,d_2^n)(\widehat{a})}\right)^{-\frac{1}{2}}
\left(\sum_{b_1} \gamma^{*}_{(v_1^n,d_1^n)(b_1,m_1),(u^n,d_2^n)(a)} \right) \left(\sum_{\widehat{m}_1,\widehat{b}_1,\widehat{a}} \gamma^{*}_{(v_1^n,d_1^n)(\widehat{b}_1, \widehat{m}_1),(u^n,d_2^n)(\widehat{a})}\right)^{-\frac{1}{2}} \nonumber, 
\\ &&\mbox{and } \mu^{Y_1}_{-1} \define I^{Y_1} - \sum_{m_1} \sum_{a} \mu^{Y_1}_{m_1,a}, \mbox{ where } \nonumber \\
&&\gamma^{*}_{(v_1^n,d_1^n)(b_1,m_1),(u^n,d_2^n)(a)} \define \left(I_{\boldsymbol{\CalH_{Y_1}^{e}}}-  \beta^{*}_{(v_1^n,d_1^n)(b_1,m_1),(u^n,d_2^n)(a)}\right) \pi_{\boldsymbol{\CalH_{Y_1}^G} }\left(I_{\boldsymbol{\CalH_{Y_1}^{e}}}-  \beta^{*}_{(v_1^n,d_1^n)(b_1,m_1),(u^n,d_2^n)(a)}\right). \label{Eqn:3to1CQBCPovmelementfordecoder1}  
\end{eqnarray}

\med\textit{Decoding POVM for Rxs $2$ and $3$}:
Having described the decoding POVM for Rx $1$, we now turn to the decoding POVMs for Rx $2$ and Rx $3$. Since their constructions are essentially identical, we present only the decoding POVM for Rx $2$.
Rx $2$ aims to decode the message $m_2=(m_{21},m_{22})$ corresponding to the codewords $u_2^n(a_2(\ulinem),m_{21})$ and $v_2^n(b_2(\ulinem),m_{22})$. We adopt the joint decoding strategy proposed in \cite{201206TIT_FawHaySavSenWil}. To this end, we first define the associated classical-quantum state
\begin{eqnarray}
    \rho^{U_2V_2Y_2} &\define& \sum_{u_2} \sum_{v_2} p_{U_2V_2}(u_2,v_2) \ketbra{u_2,v_2} \otimes \rho^{Y_2}_{u_2,v_2}, \mbox{ where}  \label{Eqn:3to1CQBCstateDec2} \\
     \rho^{Y_2}_{u_2,v_2} &\define& \sum_{v_1, u_3, v_3} p_{V_1U_3V_3|U_2V_2}(v_1,u_3,v_3|u_2,v_2) \rho^{Y_2}_{f(u_2,u_3,\ulinev)}. \nonumber 
\end{eqnarray}
We define the marginal density operators as
\begin{eqnarray}
    \rho^{Y_2}_{u_2} = \sum_{v_2} p_{V_2|U_2}(v_2|u_2) \rho^{Y_2}_{u_2,v_2}, \: \rho^{Y_2}_{v_2} = \sum_{u_2} p_{U_2|V_2}(u_2|v_2) \rho^{Y_2}_{u_2,v_2}, \mbox{ and }  \rho^{Y_2} = \sum_{u_2} \sum_{v_2} p_{U_2V_2}(u_2,v_2) \rho^{Y_2}_{u_2,v_2}.  \label{Eqn:3to1CQBCtheassociatedstateforuser2}
\end{eqnarray}
Let $\pi^{Y_2}_{u_2^n, v_2^n}$, $\pi^{Y_2}_{u_2^n}$, $\pi^{Y_2}_{v_2^n}$ and $\pi^{Y_2}$ be the C-Typ-Proj
with respect to the states $\rho^{Y_2}_{u_2^n,v_2^n}= \bigotimes_{t=1}^n \rho^{Y_2}_{u_{2_t},v_{2_t}}$, $\rho^{Y_2}_{u_2^n}= \bigotimes_{t=1}^n \rho^{Y_2}_{u_{2_t}}$, $\rho^{Y_2}_{v_2^n}= \bigotimes_{t=1}^n \rho^{Y_2}_{v_{2_t}}$, and $\bigotimes_{t=1}^n \rho^{Y_2}$ respectively. Define the operator 
\begin{eqnarray}
\Upsilon_{u_2^n(a_2,m_{21}),v_2^n(b_2,m_{22})} \define \pi^{Y_2} \: \pi^{Y_2}_{u_2^n(a_2,m_{21})} \: \pi^{Y_2}_{u_2^n(a_2,m_{21}),v_2^n(b_2,m_{22})} \: \pi^{Y_2}_{u_2^n(a_2,m_{21})} \pi^{Y_2}. \label{Eqn:3to1CQBCPovmelementfordecoder2}  
\end{eqnarray}

\noindent Finally, we define the square-root measurement \cite{BkWilde_2017,BkHolevo_2019} $\{\mu^{Y_2}_{m_2} : m_2 \in [2^{nR_2}]\}$ as  
\begin{eqnarray}
&&\hspace{-0.25in}\mu^{Y_2}_{m_2} \define \left( \sum_{\tilde{m}_{22},\tilde{b_2}, \tilde{m}_{21}, \tilde{a}_2} \Upsilon_{u_2^n(\tilde{a}_2,\tilde{m}_{21}),v_2^n(\tilde{b_2},\tilde{m}_{22})}\right)^{-\frac{1}{2}} \left( \sum_{b_2,a_2} \Upsilon_{u_2^n(a_2,m_{21}),v_2^n(b_2,m_{22})}\right)
\left( \sum_{\tilde{m}_{22},\tilde{b_2}, \tilde{m}_{21}, \tilde{a}_2} \Upsilon_{u_2^n(\tilde{a}_2,\tilde{m}_{21}),v_2^n(\tilde{b_2},\tilde{m}_{22})}\right)^{-\frac{1}{2}},
\nonumber \\
&&\hspace{-0.25in}\mbox{and } \mu_{-1}^{Y_2} \define I^{Y_2} - \sum_{m_2} \mu_{m_2}^{Y_2}. \nonumber
\end{eqnarray}
\noindent \textbf{Error analysis:}
As is standard, we derive an upper bound on the error probability of a good code by averaging the error probability over an ensemble of codes. We begin by specifying the codebook distribution, i.e, the random code distribution, with respect to which this averaging is performed. Towards that end, recall that $g_{j} = \left[\!\! \begin{array}{c}g_{I}^{j}\\g_{O/I}^{j}\end{array}\!\!\right]$ for $j=2,3$ and these generator matrices satisfy $g_{\sfb}\define \left[\!\! \begin{array}{c}g_{\sfs}\\g_{\Delta}\end{array}\!\!\right]$, where $\sfb$ is defined as in \eqref{Eqn:Step1ProofBiggerOfkPlusl}. From the code structure, encoding and decoding it is evident that our coding strategy is completely specified via the following objects : $c_{1}=(v_{1}^{n}(b_1,m_{1}): b_1 \in [2^{nK_1}], m_{1} \in [2^{nR_{1}}] ), c_{jj} =\{v_j^n(b_j,m_{jj}) : b_j \in [2^{nK_j}], m_{jj} \in [2^{nL_j}]\}$ for $j \in \{2,3\}$, $g_{\sfb}$, $\bias_{j}^{n}$ for $j \in \{2,3\}$, $(d_{1}^{n}(b_1,m_{1}): b_1 \in [2^{nK_1}], m_{1} \in [2^{nR_{1}}])), (d_{2}^{n}(a): a \in \CalF_{\Prime}^{s_{\sfb}+t_{\sfb}}))$, $b_j : [2^{nR_1}] \times [2^{nR_2}] \times [2^{nR_3}] \rightarrow [2^{nK_j}]$, for $j \in [3]$ and finally maps $a_{j}: [2^{nR_1}] \times [2^{nR_2}] \times [2^{nR_3}] \rightarrow [2^{nS_j}]$ for $j \in \{2,3\}$. It suffices therefore to specify the joint distribution of these objects. For any choice of the arguments in their respective range spaces, let
{\begin{eqnarray}
&&\hspace{-0.3in}P\left(
\begin{array}{c}
\left(V_1^n(b_1,m_1)=v_1^n(b_1,m_1) : b_1 \in [2^{nK_1}], m_1 \in [2^{nR_1}]\right),   \left(\left(V_j^n(b_j, m_{jj}) :b_j \in [2^{nK_j}], m_{jj} \in [2^{nL_j}]\right): j=2,3\right), \\
G_b=g_b, \left(\Bias_j^n=\bias_j^n : j=2,3\right), \left(D_1^n(b_1,m_1)=d_1^n : b_1 \in [2^{nK_1}], m_1 \in [2^{nR_1}]\right), 
\left(D_2^n(a)=d_2^n : a \in \mathcal{F}_{\Prime}^{s_{\sfb} + t_{\sfb}}\right), \\
\left(\left(B_j(\ulinem) = b_j(\ulinem) : \ulinem \in [2^{nR_1}] \times [2^{nR_2}] \times[2^{nR_3}] \right) : j \in [3] \right),\\ \left(\left(A_j(\ulinem)=a_j(\ulinem) : \ulinem \in [2^{nR_1}] \times [2^{nR_2}] \times [2^{nR_3}]\right) :  j=2,3 \right)
\end{array}\right) \nonumber  \\
&&\hspace{-0.3in}= \left[\prod_{m_1=1}^{2^{nR_1}} \prod_{b_1=1}^{2^{nK_1}} p_{V_1}^n(v_1^n(b_1,m_1)) \right] \left[\prod_{j=2}^3 \prod_{m_{jj}=1}^{2^{nL_j}} \prod_{b_j=1}^{2^{nK_j}} p_{V_j}^n(v_j^n(b_j,m_{jj})) \right]\frac{1}{\Prime^{s_{\sfb}t_{\sfb}}} \frac{1}{\Prime^{2n}} 
\left[\prod_{m_1=1}^{2^{nR_1}} \prod_{b_1=1}^{2^{nK_1}} \frac{1}{|\mathcal{D}_1|^n} \right] \left[\prod_{a=1}^{\Prime^{s_{\sfb}+t_{\sfb}}}\frac{1}{|\CalD_{2}|^{n}} \right]\nonumber \\
&&\hspace{-0.3in}\left[\prod_{m_1=1}^{2^{nR_1}} \prod_{j=2}^3 \prod_{m_{j1}=1}^{2^{nT_j}} \prod_{m_{jj}=1}^{2^{nL_j}}
\frac{r_{U_2U_3\ulineV}^n\left(u_2^n(a_2(\ulinem),m_{21}), u_3^n(a_3(\ulinem),m_{31}), v_1^n(b_1(\ulinem),m_1), v_2^n(b_2(\ulinem), m_{22}), v_3^n(b_3(\ulinem),m_{33})\right)}{\sum_{\tilde{a}_2, \tilde{a}_3} \sum_{\tilde{b}_1,\tilde{b}_2,\tilde{b}_3} r_{U_2U_3\ulineV}^n\left(u_2^n(\tilde{a}_2,m_{21}), u_3^n(\tilde{a}_3,m_{31}), v_1^n(\tilde{b}_1,m_1), v_2^n(\tilde{b}_2, m_{22}), v_3^n(\tilde{b}_3,m_{33})\right)} \right]\label{Eqn:DistStepI}
\end{eqnarray}
}
specify our joint codebook distribution, where 
\begin{eqnarray}
&&r_{U_2U_3\ulineV}^n\left(u_2^n(a_2,m_{21}), u_3^n(a_3,m_{31}), v_1^n(b_1,m_1), v_2^n(b_2, m_{22}), v_3^n(b_3,m_{33})\right)  \nonumber \\
&\define&\frac{p_{U_2U_3\ulineV}^n\left(u_2^n(a_2,m_{21}), u_3^n(a_3,m_{31}), v_1^n(b_1,m_1), v_2^n(b_2, m_{22}), v_3^n(b_3,m_{33})\right)}{q^n_{U_2}(u_2^n(a_2,m_{21})) q^n_{U_3}(u_3^n(a_3,m_{31})) p_{V_1}^n(v_1^n(b_1,m_1)) p_{V_2}^n(v_2^n(b_2, m_{22})) p_{V_3}^n(v_3^n(b_3,m_{33}))} \nonumber 
\end{eqnarray}
and for $j=2,3$, $q_{U_j}^n(u_j^n) \define \frac{1}{\Prime^n}$ is the uniform distribution. 
We now derive an upper bound on the average error probability of a good code, by averaging the error probability of every code with respect to the above specified codebook distribution. Note that, upon receiving the quantum state $\rho^{Y_1}_{x^n(\ulinem)},$
Rx~1 prepares the auxiliary state $\ketbra{0}$, concatenates the same with the received state and measures this concatenated state via the square root measurement \cite{BkWilde_2017,BkHolevo_2019} $\{\mu^{Y_1}_{m_1,a}: (m_1,a) \in [2^{nR_1}] \times \mathcal{F}_{\Prime}^{s_{\sfb}+ t_{\sfb}}\}$ on the combined state $\left(\rho^{Y_1}_{x^n(\ulinem)} \otimes \ketbra{0}\right)$. 
Hence, the average error probability of the code is given by
\begin{eqnarray}
    &&\hspace{-0.3in}\mathbf{P}(e,\uline{\mu})=\frac{1}{|\mathcal{\ulineM}|} \frac{1}{\upsilon^{s_{\sfb} + t_{\sfb}}}  \sum_{\ulinem} \sum_{a}  \tr \left\{ 
\left( I^{Y_1Y_2Y_3} - \mu^{Y_1}_{m_1,a}
 \otimes \mu_{m_2}^{Y_2}
 \otimes \mu_{m_3}^{Y_3} \right)
\left(
\rho_{x^n(\ulinem)}^{Y_1Y_2Y_3}\otimes \ketbra{0} 
\right)\right\}. \nonumber 
\end{eqnarray}
Let $E$ and $F$ be operators on the quantum systems $A$ and $B$, respectively, satisfying $0 \leq E \leq I^A$ and $0 \leq F \leq I ^B$. Then, observe that 
    \begin{eqnarray}
       I^{AB}  - E \otimes F &=& (I^A-E) \otimes F + I^A \otimes (I^B -F) \nonumber \\
       &\leq& (I^A - E) \otimes I^B + I^{AB} - (I^A \otimes F). \label{Eqn:IneqMeasurement}
    \end{eqnarray}
From \eqref{Eqn:IneqMeasurement}, we have
\begin{eqnarray}
    I^{Y_1Y_2Y_3} - \mu^{Y_1}_{m_1,a} \otimes \mu_{m_2}^{Y_2} \otimes \mu_{m_3}^{Y_3} &\leq&  \left(I^{Y_1Y_2Y_3} - \mu^{Y_1}_{m_1,a} \otimes I^{Y_2} \otimes I^{Y_3} \right) + \left(I^{Y_1Y_2Y_3} - I^{Y_1} \otimes  \mu_{m_2}^{Y_2} \otimes I^{Y_3} \right) \nonumber  \\&&+ \left(I^{Y_1Y_2Y_3} -I^{Y_1} \otimes I^{Y_2} \otimes  \mu_{m_3}^{Y_3} \right).\nonumber 
\end{eqnarray}
Using the above inequality, we obtain 
\begin{eqnarray}
  &&\mathbf{P}(e,\uline{\mu})\leq  T_1 + T_2 + T_3, \mbox{ where } 
  \nonumber \\
  T_1 &\define&   \frac{1}{|\mathcal{\ulineM}|} \frac{1}{\upsilon^{s_{\sfb} + t_{\sfb}}} \sum_{\ulinem} \sum_{a} \tr \left\{ 
\left( I - \mu^{Y_1}_{m_1,a} \right)
\left(
\rho_{x^n(\ulinem)}^{Y_1}\otimes \ketbra{0} 
\right) 
\right\},\nonumber \\
T_2 &\define& \frac{1}{|\mathcal{\ulineM}|} \sum_{\ulinem} \tr \left\{ \left( I - \mu_{m_2}^{Y_2} \right)  \rho_{x^n(\ulinem)}^{Y_2} \right\},  \nonumber \\ T_3 &\define& \frac{1}{|\mathcal{\ulineM}|} \sum_{\ulinem} \tr \left\{ \left( I - \mu_{m_3}^{Y_3} \right)  \rho_{x^n(\ulinem)}^{Y_3} \right\}.  \nonumber 
    \end{eqnarray}

\med\textit{\underline{Rx $1$'s Error Analysis}}: We begin by analyzing the first term $T_1$, which corresponds to Rx $1$'s error analysis. Observe that \begin{eqnarray} 
&& \hspace{2.3in}\mu_{m_1,a}^{Y_1} \geq \Gamma^{Y_1}_{m_1,a}, \mbox{ where} \nonumber \\
&&\Gamma^{Y_1}_{m_1,a} \define \left(\sum_{\widehat{m}_1,\widehat{b}_1,\widehat{a}} \gamma^{*}_{(v_1^n,d_1^n)(\widehat{b}_1,\widehat{m}_1),(u^n,d_2^n)(\widehat{a})}\right)^{-\frac{1}{2}}
\gamma^{*}_{(v_1^n,d_1^n)(b_1(\ulinem),m_1),(u^n,d_2^n)(a)}\left(\sum_{\widehat{m}_1,\widehat{b}_1,\widehat{a}} \gamma^{*}_{(v_1^n,d_1^n)(\widehat{b}_1,\widehat{m}_1),(u^n,d_2^n)(\widehat{a})}\right)^{-\frac{1}{2}} , \nonumber
\end{eqnarray}
Therefore, we have 
\begin{eqnarray}
T_1 &\leq& \frac{1}{|\mathcal{\ulineM}|} \frac{1}{\upsilon^{s_{\sfb} + t_{\sfb}}} \sum_{\ulinem} \sum_{a}
\tr \left\{ \left( I - \Gamma^{Y_1}_{m_1,a} \right)\left(
\rho_{x^n(\ulinem)}^{Y_1}\otimes \ketbra{0} 
\right) \right\} \nonumber \\
&=& \frac{1}{|\mathcal{\ulineM}|} \frac{1}{\upsilon^{s_{\sfb} + t_{\sfb}}} \sum_{\ulinem}  \sum_{a} \sum_{a_2,a_3} \sum_{b_1,b_2,b_3} \mathds{1}\left\{{a_i(\ulinem)=a_i : i=2,3, b_j(\ulinem)=b_j : j \in [3]}\right\} \tr \left\{  \left( I - \Gamma^{Y_1}_{m_1,a} \right)  \left( \rho_{x^n(\ulinem)}^{Y_1} \otimes \ketbra{0}\right)  \!\right\} \nonumber \\ 
&\overset{(*)}{=}& T_{1.1} + T_{1.2},  \mbox{ where} \nonumber \\
T_{1.1} &\define& \frac{1}{|\mathcal{\ulineM}|} \frac{1}{\upsilon^{s_{\sfb} + t_{\sfb}}}\sum_{\ulinem}  \sum_{a}\sum_{a_2,a_3}  \sum_{b_1,b_2,b_3} \tr \left\{  \left( I - \Gamma^{Y_1}_{m_1,a} \right)  \left( \rho_{x^n(\ulinem)}^{Y_1} \otimes \ketbra{0}\right)  \right\} \nonumber \\
&&\hspace{-0.65in}\left[ \begin{aligned}
\mathds{1}\left\{a_i(\ulinem)=a_i : i=2,3, b_j(\ulinem)=b_j : j \in [3]\right\} -\frac{r_{U_2U_3\ulineV}^n\left(u_2^n(a_2,m_{21}),u_3^n(a_3,m_{31}),v_1^n(b_1,m_1),v_2^n(b_2,m_{22}),v_3^n(b_3,m_{33})\right)}{2^{n(S_2 + S_3 + K_1 + K_2 + K_3)}} 
\end{aligned} \right], \nonumber \\  
T_{1.2} &\define& \frac{1}{|\mathcal{\ulineM}|} \frac{1}{\upsilon^{s_{\sfb} + t_{\sfb}}}\frac{1}{2^{n(S_2 + S_3 + K_1 + K_2 + K_3 )}} \sum_{\ulinem}  \sum_{a} \sum_{a_2,a_3} \sum_{b_1,b_2,b_3} \tr \left\{  \left( I - \Gamma^{Y_1}_{m_1,a} \right)  \left( \rho_{x^n(\ulinem)}^{Y_1} \otimes \ketbra{0}\right)  \right\} \nonumber \\ 
&&r_{U_2U_3\ulineV}^n\left(u_2^n(a_2,m_{21}),u_3^n(a_3,m_{31}),v_1^n(b_1,m_1),v_2^n(b_2,m_{22}),v_3^n(b_3,m_{33})\right). \nonumber 
    \end{eqnarray}
The equality $(*)$ follows by adding and subtracting 
\begin{eqnarray}
\frac{r_{U_2U_3\ulineV}^n\left(u_2^n(a_2,m_{21}),u_3^n(a_3,m_{31}),v_1^n(b_1,m_1),v_2^n(b_2,m_{22}),v_3^n(b_3,m_{33})\right)}{2^{n(S_2 + S_3 + K_1 + K_2 + K_3)}}, \nonumber    
\end{eqnarray}
thereby decomposing the expression into two terms $T_{1.1}$ and $T_{1.2}$ described above.
The term $T_{1.1}$ is simply bounded by `cloud mixing' \cite{CuffPhDThesis}. We provide the following proposition.
\begin{proposition}
\label{Prop:Dec1Cuffterm}
For any $\epsilon \in (0,1)$, and for all $\delta$ sufficiently small and $n$ sufficiently large, we have $\mathbb{E}[T_{1.1}] \leq \epsilon$ if
\begin{eqnarray}
  S_{\mathcal{A}} + K_{\mathcal{B}} &>& |\mathcal{A}|\log\left(\Prime\right) + \sum_{\beta \in \mathcal{B}} H(V_{\beta}) - H(U_{\mathcal{A}},V_{\mathcal{B}}), \nonumber \\
    \max\{S_2 + T_2, S_3 + T_3\} &>& \log(\Prime) - \min_{\theta \in \mathcal{F}_{\Prime} \setminus \{0\}} H(U_2 \oplus \theta U_3), \mbox{ and }\nonumber\\
     K_{\mathcal{B}} + \max\{S_2 + T_2, S_3 + T_3\} &>& \log(\Prime) + \sum_{\beta \in \mathcal{B}} H(V_{\beta}) - \min_{\theta \in \mathcal{F}_{\Prime} \setminus \{0\}} H(U_2 \oplus \theta U_3, V_{\mathcal{B}}), \nonumber
\end{eqnarray}
where $\mathcal{A} \subseteq \{2,3\}$, $ \mathcal{B} \subseteq \{1,2,3\}$, $S_{\mathcal{A}} \define \sum_{\alpha \in \mathcal{A}} S_{\alpha}$,  $U_{\mathcal{A}} \define (U_{\alpha}: \alpha \in \mathcal{A})$, $K_{\mathcal{B}} \define \sum_{\beta \in \mathcal{B}} K_{\beta}$ and $V_{\mathcal{B}} \define (V_{\beta} : \beta \in \mathcal{B})$.
\end{proposition}
\begin{proof}
    See Appendix \ref{App:Dec1Cuffterm} for a proof.
\end{proof}

{Next, to bound the term $T_{1.2}$, we first employ an alternate `proxy' state. Specifically, we substitute the original received state ($\rho_{v_1^n,u^n}^{Y_1} \otimes \ketbra{0}$) by a specific `tilted state'. The effect of this substitution on the error probability can be suppressed by ensuring that the tilted state is close to the original received state in the $\mathbb{L}_{1}-$norm . Towards identifying this tilted state, we define a new titling map
$\CalT_{d_{1}^{n},d_{2}^{n}, \eta} : \boldsymbol{\CalH_{Y_1}^G}\rightarrow
\boldsymbol{\CalH_{Y_1}^e}$ as
\begin{eqnarray}
\CalT_{d_{1}^{n},d_{2}^{n}, \eta}(\ket{h}) = \frac{1}{\sqrt{1+2\eta^2} } \left(\ket{h} +\eta \ket{h} \otimes  \ket{d_{1}^{n}}+\eta \ket{h} \otimes  \ket{d_{2}^{n}}\right),
\nonumber
\end{eqnarray}
to tilt the state in all directions. As discussed earlier, this tilting map is chosen carefully so as to ensure that the tilted state remains close to the original state in $\mathbb{L}_{1}$ norm. Proposition \ref{Prop:3to1CQBCclosnessofstates} establishes this thereby guaranteeing that the two states induce approximately the same measurement outcome statistics. Let
\begin{eqnarray}
    \theta_{v_1^n,d_1^n,u^n,d_2^n} \define \mathcal{T}_{d_1^n,d_2^n,\eta}\left( \rho^{Y_1}_{v_1^n,u^n} \otimes \ketbra{0}\right) , \label{Eqn:3to1CQBCtiltedstate}
\end{eqnarray}
be the tilted state along all directions, where  $\CalT_{d_1^n,d_2^n,\eta}$ acts on each pure state in the mixture individually.
\begin{proposition}
    \label{Prop:3to1CQBCclosnessofstates}
    For n sufficiently large, we have 
    \begin{eqnarray}
        \norm{\theta_{v_1^n,d_1^n,u^n,d_2^n} - \left( \rho_{v_1^n,u^n}^{Y_1} \otimes \ketbra{0}\right) }_1 \leq 4 \eta. \nonumber 
    \end{eqnarray}
\end{proposition}
\begin{proof}
    The proof is provided in Appendix \ref{App:3to1CQBCClosenessOfStates}.
\end{proof}

Now, we evaluate the expectation with respect to the random choice of the codebook. We obtain 
\begin{eqnarray}
\mathbb{E}[T_{1.2}] &=& \frac{1}{|\mathcal{\ulineM}|} \frac{1}{\upsilon^{s_{\sfb} + t_{\sfb}}} \frac{1}{2^{n(S_2 + S_3 + K_1 + K_2 + K_3  )}} \sum_{\ulinem} \sum_{a} \sum_{a_2,a_3} \sum_{b_1,b_2,b_3}  \sum_{u_2^n} \sum_{u_3^n} \sum_{\ulinev^n} \sum_{u^n} q_{U_2}^n(u_2^n) q_{U_3}^n(u_3^n) p_{V_1}^n(v_1^n) p_{V_1}^n(v_2^n) p_{V_3}^n(v_3^n) \nonumber \\
&& r^n_{U_2 U_3 \ulineV}(u_2^n,u_3^n, \ulinev^n)
\mathds{1}\{u^n=u_2^n \oplus u_3^n\} \tr \left\{  \left( I - \Gamma^{Y_1}_{m_1,a} \right)  \left( \rho_{f^n\left(u_2^n,u_3^n,\ulinev^n\right)}^{Y_1} \otimes \ketbra{0}\right)  \right\} \nonumber \\
&\overset{(a)}{=}& \frac{1}{|\mathcal{\ulineM}|} \frac{1}{\upsilon^{s_{\sfb} + t_{\sfb}}} \frac{1}{2^{n(S_2 + S_3 + K_1 + K_2 + K_3)}} \sum_{\ulinem} \sum_{a} \sum_{a_2,a_3}  \sum_{b_1,b_2,b_3} \sum_{u_2^n} \sum_{u_3^n} \sum_{\ulinev^n} \sum_{u^n}  p_{U_2U_3 \ulineV}^n(u_2^n,u_3^n,\ulinev^n) \mathds{1}\{u^n=u_2^n \oplus u_3^n\} \nonumber \\ && 
   \tr \left\{  \left( I - \Gamma^{Y_1}_{m_1,a} \right)  \left( \rho_{f^n\left(u_2^n,u_3^n,\ulinev^n\right)}^{Y_1} \otimes \ketbra{0}\right)  \right\} \nonumber \\
 &\overset{(b)}{=}& \frac{1}{|\mathcal{\ulineM}|} \frac{1}{\upsilon^{s_{\sfb} + t_{\sfb}}} \frac{1}{2^{n(S_2 + S_3 + K_1 + K_2 + K_3)}} \sum_{\ulinem} \sum_{a} \sum_{a_2,a_3} \sum_{b_1,b_2,b_3}  \sum_{u_2^n} \sum_{u_3^n} \sum_{\ulinev^n} \sum_{u^n} p_{U_2,U_3 \ulineV U}^n(u_2^n,u_3^n,\ulinev^n,u^n) \nonumber \\ && 
   \tr \left\{  \left( I - \Gamma^{Y_1}_{m_1,a} \right)  \left( \rho_{f^n\left(u_2^n,u_3^n,\ulinev^n\right)}^{Y_1} \otimes \ketbra{0}\right)  \right\} \nonumber \\
&\overset{(c)}{=}& \frac{1}{|\mathcal{M}_1|} \frac{1}{\upsilon^{s_{\sfb} + t_{\sfb}}} \frac{1}{2^{nK_1}} \sum_{m_1} \sum_{a}  \sum_{b_1}  \sum_{v_1^n} \sum_{u^n} p_{V_1U}^n(v_1^n,u^n)  \tr \left\{  \left( I - \Gamma^{Y_1}_{m_1,a} \right)  \left( \rho_{v_1^n,u^n}^{Y_1} \otimes \ketbra{0}\right)  \right\} \nonumber \\
&\overset{(d)}{=}& \frac{1}{|\mathcal{M}_1|} \frac{1}{\upsilon^{s_{\sfb} + t_{\sfb}}} \frac{1}{2^{nK_1}} \sum_{m_1} \sum_{a}  \sum_{b_1}  \sum_{v_1^n} \sum_{u^n} p_{V_1U}^n(v_1^n,u^n)  \norm{ \theta_{v_1^n,D_1^n(b_1,m_1),u^n,D_2^n(a)} - \left( \rho_{v_1^n,u^n}^{Y_1} \otimes \ketbra{0}\right)}_1\nonumber \\
&+&\frac{1}{|\mathcal{M}_1|} \frac{1}{\upsilon^{s_{\sfb} + t_{\sfb}}} \frac{1}{2^{nK_1}} \sum_{m_1} \sum_{a}  \sum_{b_1}  \sum_{v_1^n} \sum_{u^n} p_{V_1U}^n(v_1^n,u^n)
   \tr \left\{  \left( I - \Gamma^{Y_1}_{m_1,a} \right) \theta_{v_1^n,D_1^n(b_1,m_1),u^n,D_2^n(a)} \right\} \label{Eqn:3to1CQBCDec1secondterm}
   \end{eqnarray}}
where (a) follows from $q_{U_2}^n(u_2^n) q_{U_3}^n(u_3^n) p_{V_1}^n(v_1^n) p_{V_1}^n(v_2^n) p_{V_3}^n(v_3^n) r^n_{U_2^n U_3^n \ulineV^n}(u_2^n,u_3^n,\ulinev^n)= p_{U_2U_3\ulineV}^n(u_2^n, u_3^n,\ulinev^n)$, (b) follows from $p^n_{U_2,U_3 \ulineV U}(u_2^n,u_3^n,\ulinev^n,u^n) = p^n_{U_2,U_3 \ulineV}(u_2^n,u_3^n,\ulinev^n) \mathds{1}\{u^n=u_2^n \oplus u_3^n\}$, (c) follows by using the definition $\rho_{v_1^n,u^n}^{Y_1} = \sum_{u_2^n,u_3^n} \sum_{v_2^n,v_3^n} p^n_{U_2U_3V_2V_3|V_1U}(u_2^n,u_3^n,v_2^n,v_3^n|v_1^n,u^n) \rho^{Y_1}_{f^n(u_2^n,u_3^n,\ulinev^n)}$, and (d) follows from the trace inequality $\tr(\Delta \rho) \leq \tr(\Delta \sigma) + \frac{1}{2} \norm{\rho - \sigma}_1$, where $0 \leq \Delta,\rho,\sigma \leq I$.

\noindent By evaluating the expectation over $(D_1^n(b_1,m_1), D_2^n(a))$ and using Proposition \ref{Prop:3to1CQBCclosnessofstates}, we bound the first term in \eqref{Eqn:3to1CQBCDec1secondterm} by $4 \eta$. Next, to bound the second term in \eqref{Eqn:3to1CQBCDec1secondterm}, we apply Hayashi-Nagaoka inequality \cite{200307TIT_HayNag}. We obtain
\begin{eqnarray}
    &&\mathbb{E}[T_{1.2}] \leq 4 \eta + 2 \: T_{1.2.1} + 4 \: \left(\sum_{i=2}^6 T_{1.2.i}\right), \mbox{ where}\nonumber \\
     T_{1.2.1} &\define& \frac{1}{|\mathcal{M}_1|} \frac{1}{\upsilon^{s_{\sfb} + t_{\sfb}}} \frac{1}{2^{nK_1}} \sum_{m_1} \sum_{a}  \sum_{b_1} \sum_{v_1^n} \sum_{u^n} p_{V_1U}^n(v_1^n,u^n)
    \tr \left\{  \left( I - \gamma^{*}_{v_1^n,D_1^n(b_1,m_1),u^n,D_2^n(a)} \right) \theta_{v_1^n,D_1^n(b_1,m_1),u^n,D_2^n(a)} \right\},\nonumber \\
    T_{1.2.2} &\define& \frac{1}{|\mathcal{M}_1|} \frac{1}{\upsilon^{s_{\sfb} + t_{\sfb}}} \frac{1}{2^{nK_1}} \sum_{m_1} \sum_{a}  \sum_{b_1} \sum_{\tilde{m}_1 \neq m_1}  \sum_{\tilde{b}_1} \sum_{v_1^n} \sum_{u^n} p_{V_1U}^n(v_1^n,u^n) \nonumber \\
    &&\tr \left\{ \gamma^{*}_{(V_1^n,D_1^n)(\tilde{b}_1,\tilde{m}_1),u^n,D_2^n(a)} \: \theta_{v_1^n,D_1^n(b_1,m_1),u^n,D_2^n(a)} \right\}\!\!,\nonumber \\
    T_{1.2.3} &\define& \frac{1}{|\mathcal{M}_1|} \frac{1}{\upsilon^{s_{\sfb} + t_{\sfb}}} \frac{1}{2^{nK_1}} \sum_{m_1} \sum_{a} \sum_{b_1} \sum_{\Tilde{a} \neq a} \sum_{v_1^n} \sum_{u^n} p_{V_1U}^n(v_1^n,u^n) \tr \left\{ \gamma^{*}_{v_1^n,D_1^n(b_1,m_1),(U^n,D_2^n)(\tilde{a})} \: \theta_{v_1^n,D_1^n(b_1,m_1),u^n,D_2^n(a)} \right\},\nonumber \\
    T_{1.2.4} &\define& \frac{1}{|\mathcal{M}_1|} \frac{1}{\upsilon^{s_{\sfb} + t_{\sfb}}} \frac{1}{2^{nK_1}} \sum_{m_1} \sum_{a} \sum_{b_1} \sum_{\tilde{m}_1 \neq m_1} \sum_{\tilde{b}_1} \sum_{\Tilde{a} \neq a}  \sum_{v_1^n} \sum_{u^n} p_{V_1U}^n(v_1^n,u^n) \nonumber \\
    &&\tr \left\{ \gamma^{*}_{(V_1^n,D_1^n)(\tilde{b}_1,\tilde{m}_1),(U^n,D_2^n)(\tilde{a})} \: \theta_{v_1^n,D_1^n(b_1,m_1),u^n,D_2^n(a)} \right\},\nonumber \\
    T_{1.2.5} &\define& \frac{1}{|\mathcal{M}_1|} \frac{1}{\upsilon^{s_{\sfb} + t_{\sfb}}} \frac{1}{2^{nK_1}} \sum_{m_1} \sum_{a}  \sum_{b_1} \sum_{\tilde{b}_1 \neq b_1} \sum_{v_1^n} \sum_{u^n} p_{V_1U}^n(v_1^n,u^n) \tr \left\{ \gamma^{*}_{(V_1^n,D_1^n)(\tilde{b}_1,m_1),u^n,D_2^n(a)} \: \theta_{v_1^n,D_1^n(b_1,m_1),u^n,D_2^n(a)} \right\},\nonumber \\
    T_{1.2.6} &\define& \frac{1}{|\mathcal{M}_1|} \frac{1}{\upsilon^{s_{\sfb} + t_{\sfb}}} \frac{1}{2^{nK_1}} \sum_{m_1} \sum_{a} \sum_{b_1} \sum_{\tilde{b}_1 \neq b_1} \sum_{\Tilde{a} \neq a}  \sum_{v_1^n} \sum_{u^n} p_{V_1U}^n(v_1^n,u^n) \nonumber \\
    &&\tr \left\{ \gamma^{*}_{(V_1^n,D_1^n)(\tilde{b}_1,m_1),(U^n,D_2^n)(\tilde{a})} \: \theta_{v_1^n,D_1^n(b_1,m_1),u^n,D_2^n(a)} \right\}.\nonumber
    \end{eqnarray}
In the following propositions, we provide the rate constraints required to bound these error terms.
\begin{proposition}
    \label{Prop:Dec1firsttermHay}
    For all $\epsilon>0$, and $\delta, \eta> 0$ sufficiently small and $n$ sufficiently large, we have $\mathbb{E}[T_{1.2.1}] \leq \epsilon$.
\end{proposition}
\begin{proof}
    The proof is provided in Appendix \ref{App:Dec1FirsttermHay}.
\end{proof}
\begin{proposition}
    \label{Prop:Dec1SecondtermHay}
For any $\epsilon >0$, and for all $\delta,\eta >0$ sufficiently small and $n$ sufficiently large, we have $\sum_{i=2}^{6}\mathbb{E}[T_{1.2.i}] \leq \epsilon$ if the following inequalities hold.
\begin{eqnarray}
    R_1 + K_1 &<& H(V_1) - H(V_1|Y_1,U_2 \oplus U_3), \nonumber \\
    \max\{S_2 + T_2,S_3 + T_3\} &<&  \log (\Prime)  - H(U_2 \oplus U_3 |Y_1,V_1), \mbox{ and} \nonumber \\
R_1 + K_1 + \max\{S_2 + T_2, S_3 + T_3\} &<& \log(\Prime) + H(V_1) - H(V_1,U_2 \oplus U_3 | Y_1) ,
 \nonumber 
\end{eqnarray}
where all the mutual information quantities are computed with respect to the state $\rho^{V_1U_2 \oplus U_3Y_1}$ defined in \eqref{Eqn:3to1CQBCstateDec1}.
\end{proposition}
\begin{proof}
    The proof is provided in Appendix \ref{App:Dec1SecondtermHay}.
\end{proof}
This completes analysis of error at Rx $1$. We now proceed to analyze $T_2$ and $T_3$.
\med\textit{\underline{Rx $2$ and $3$'s Error Analysis}}: Since Rx $2$ and $3$'s decoding POVMs and error probability terms are identical, we provide the analysis for Rx $2$.  Consider the second term $T_2$, which corresponds to Rx $2$'s error analysis. Observe that 
\begin{eqnarray}
    T_2 &=& \frac{1}{|\mathcal{\ulineM}|} \sum_{\ulinem} \tr \left\{ \left( I - \mu_{m_2}^{Y_2} \right)  \rho_{x^n(\ulinem)}^{Y_2} \right\} \nonumber \\
     &\leq&\frac{1}{|\mathcal{\ulineM}|} \sum_{\ulinem} \tr \left\{ \left( I - \Gamma_{m_2}^{Y_2} \right)  \rho_{x^n(\ulinem)}^{Y_2} \right\} \nonumber \\
     &=&\frac{1}{|\mathcal{\ulineM}|} \sum_{\ulinem} \sum_{a_2, a_3} \sum_{b_1,b_2,b_3} \mathds{1}\{a_2(\ulinem)=a_2, a_3(\ulinem)=a_3,b_1(\ulinem)=b_1,b_2(\ulinem)=b_2, b_3(\ulinem)=b_3\} \tr \left\{ \left( I - \Gamma_{m_2}^{Y_2} \right)  \rho_{x^n(\ulinem)}^{Y_2} \right\} \nonumber \\
&=&T_{2.1} + T_{2.2}, \mbox{ where} \nonumber \\
 T_{2.1} &\define& \frac{1}{|\mathcal{\ulineM}|} \sum_{\ulinem} \sum_{a_2,a_3}  \sum_{b_1,b_2,b_3}
   \tr \left\{  \left( I - \Gamma^{Y_2}_{m_2}\right) \rho_{x^n(\ulinem)}^{Y_2}  \right\} \nonumber \\ 
&&\hspace{-0.6in}\left[\begin{aligned}
\mathds{1}\left\{a_i(\ulinem)=a_i : i=2,3, b_j(\ulinem)=b_j : j \in [3]\right\} - \frac{r_{U_2U_3\ulineV}^n\left(u_2^n(a_2,m_{21}),u_3^n(a_3,m_{31}), v_1^n(b_1,m_1),v^n_2(b_2,m_{22}),v^n_3(b_3,m_{33})\right)}{2^{n(S_2 + S_3 + K_1 + K_2 + K_3)}} 
\end{aligned}\right]\nonumber \\ 
T_{2.2} &\define& \frac{1}{|\mathcal{\ulineM}|} \frac{1}{2^{n(S_2 + S_3 + K_1 +K_2 + K_3)}} \sum_{\ulinem} \sum_{a_2,a_3}  \sum_{b_1,b_2,b_3} \tr \left\{  \left( I - \Gamma^{Y_2}_{m_2}\right) \rho_{x^n(\ulinem)}^{Y_2}\right\} \nonumber \\ 
&&r_{U_2U_3\ulineV}^n\left(u_2^n(a_2,m_{21}),u_3^n(a_3,m_{31}), v_1^n(b_1,m_1),v^n_2(b_2,m_{22}),v^n_3(b_3,m_{33})\right). \nonumber 
\end{eqnarray}

\noindent The first inequality follows from $\mu_{m_2}^{Y_2} \geq \Gamma_{m_2}^{Y_2}$, where 
\begin{eqnarray}
    \Gamma_{m_2}^{Y_2} \define \left( \sum_{\tilde{m}_{21}, \tilde{a}_2,\tilde{m_{22}}, \tilde{b}_2} \Upsilon_{u_2^n(\tilde{a}_2,\tilde{m}_{21}),v_2^n(\tilde{b}_2,\tilde{m}_{22})}\right)^{-\frac{1}{2}}  \Upsilon_{u_2^n(a_2(\ulinem),m_{21}),v_2^n(b_2(\ulinem),m_{22})}
\left( \sum_{\tilde{m}_{21}, \tilde{a}_2,\tilde{m_{22}}, \tilde{b}_2} \Upsilon_{u_2^n(\tilde{a}_2,\tilde{m}_{21}),v_2^n(\tilde{b}_2,\tilde{m}_{22})}\right)^{-\frac{1}{2}}.
\nonumber 
\end{eqnarray}
The last equality follows by adding and subtracting 
\begin{eqnarray}
\frac{r_{U_2U_3\ulineV}^n\left(u_2^n(a_2,m_{21}),u_3^n(a_3,m_{31}), v_1^n(b_1,m_1),v^n_2(b_2,m_{22}),v^n_3(b_3,m_{33})\right)}{2^{n(S_2 + S_3 + K_1 + K_2 + K_3)}}, \nonumber     
\end{eqnarray}
thereby decomposing the expression into two terms. The term $T_{2.1}$ follows the same analysis as $T_{1.1}$ using a cloud mixing argument. We state the following proposition.

\begin{proposition}
\label{Prop:Dec2Cuffterm}
For any $\epsilon \in (0,1)$, and for all $\delta$ sufficiently small and $n$ sufficiently large, we have $\mathbb{E}[T_{2.1}] \leq \epsilon$ if

\begin{eqnarray}
   S_{\mathcal{A}} + K_{\mathcal{B}} &>& |\mathcal{A}|\log\left(\Prime\right) + \sum_{\beta \in \mathcal{B}} H(V_{\beta}) - H(U_{\mathcal{A}},V_{\mathcal{B}}), \nonumber \\
        \max\{S_2 + T_2, S_3 + T_3\} &>& \log(\Prime) - \min_{\theta \in \mathcal{F}_{\Prime} \setminus \{0\}} H(U_2 \oplus \theta U_3), \mbox{ and}\nonumber\\   K_{\mathcal{B}} + \max\{S_2 + T_2, S_3 + T_3\} &>& \log(\Prime) + \sum_{\beta \in \mathcal{B}} H(V_{\beta}) - \min_{\theta \in \mathcal{F}_{\Prime} \setminus \{0\}} H(U_2 \oplus \theta U_3, V_{\mathcal{B}}), \nonumber
\end{eqnarray}
where $\mathcal{A} \subseteq \{2,3\}$, $ \mathcal{B} \subseteq \{1,2,3\}$, $S_{\mathcal{A}} \define \sum_{\alpha \in \mathcal{A}} S_{\alpha}$,  $U_{\mathcal{A}} \define (U_{\alpha}: \alpha \in \mathcal{A})$, $K_{\mathcal{B}} \define \sum_{\beta \in \mathcal{B}} K_{\beta}$ and $V_{\mathcal{B}} \define (V_{\beta} : \beta \in \mathcal{B})$.
\end{proposition}
\begin{proof}
    The proof follows the same arguments as in Appendix \ref{App:Dec1Cuffterm}.
\end{proof}

We now evaluate the expectation of $T_{2.2}$ with respect to the random codebook. We obtain 
\begin{eqnarray}
    \mathbb{E}[T_{2.2}] &=& \frac{1}{|\mathcal{\ulineM}|} \frac{1}{2^{n(S_2 + S_3 + K_1 + K_2 + K_3)}} \sum_{\ulinem} \sum_{a_2,a_3} \sum_{b_1,b_2,b_3} \sum_{u_2^n} \sum_{u_3^n} \sum_{\ulinev^n} q_{U_2}^n(u_2^n) q_{U_3}^n(u_3^n) p_{V_1}^n(v_1^n) p_{V_2}^n(v_2^n) p_{V_3}^n(v_3^n)  \nonumber \\
    &&r_{U_2U_3\ulineV}^n\left(u_2^n, u_3^n, \ulinev^n\right)  
    \tr \left\{  \left( I - \Gamma^{Y_2}_{m_2}\right) \rho_{f^n\left(u_2^n,u_3^n,\ulinev^n\right)}^{Y_2}   \right\} \nonumber  \\
    &\overset{(a)}{=}& \frac{1}{|\mathcal{\ulineM}|} \frac{1}{2^{n(S_2 + S_3+ K_1 + K_2 + K_3)}} \sum_{\ulinem} \sum_{a_2,a_3} \sum_{b_1,b_2,b_3} \sum_{u_2^n} \sum_{u_3^n} \sum_{\ulinev^n} p_{U_2U_3\ulineV}^n(u_2^n,u_3^n,\ulinev^n)
    \tr \left\{  \left( I - \Gamma^{Y_2}_{m_2}\right) \rho_{f^n\left(u_2^n,u_3^n,\ulinev^n\right)}^{Y_2}   \right\} \nonumber  \\
    &\overset{(b)}{=}& \frac{1}{2^{n(R_2 + S_2 + K_2)}} \sum_{m_2} \sum_{a_2} \sum_{b_2} \sum_{u_2^n} \sum_{v_2^n}  p_{U_2V_2}^n(u_2^n,v_2^n)
    \tr \left\{  \left( I - \Gamma^{Y_2}_{m_2}\right) \rho_{u_2^n, v_2^n}^{Y_2}   \right\} \nonumber  \\
&\overset{(c)}{\leq}& \frac{1}{2^{n(R_2 + S_2 + K_2)}} \sum_{m_2} \sum_{a_2} \sum_{b_2} \sum_{u_2^n} \sum_{v_2^n}  p_{U_2V_2}^n(u_2^n,v_2^n) \nonumber \\
&&\left[
    \tr \left\{  \left( I - \Gamma^{Y_2}_{m_2}\right) \: \pi^{Y_2}_{v_2^n} \: \rho_{u_2^n, v_2^n}^{Y_2} \: \pi^{Y_2}_{v_2^n}  \right\} + \norm{ \: \pi^{Y_2}_{v_2^n} \: \rho_{u_2^n, v_2^n}^{Y_2} \: \pi^{Y_2}_{v_2^n} - \rho_{u_2^n, v_2^n}^{Y_2} }_1 \right] \nonumber \\
 &\overset{(d)}{\leq}& \frac{1}{2^{n(R_2 + S_2 + K_2)}} \sum_{m_2} \sum_{a_2} \sum_{b_2} \sum_{u_2^n} \sum_{v_2^n}  p_{U_2V_2}^n(u_2^n,v_2^n)
    \tr \left\{  \left( I - \Gamma^{Y_2}_{m_2}\right) \: \pi^{Y_2}_{v_2^n} \: \rho_{u_2^n, v_2^n}^{Y_2} \: \pi^{Y_2}_{v_2^n}  \right\} + 2 \sqrt{\epsilon}, \nonumber 
\end{eqnarray}
where (a) follows from $q_{U_2}^n(u_2^n) q_{U_3}^n(u_3^n) p_{V_1}^n(v_1^n) p_{V_2}^n(v_2^n) p_{V_3}^n(v_3^n)  r_{U_2U_3\ulineV}^n\left(u_2^n, u_3^n, \ulinev^n\right) = p_{U_2U_3\ulineV}^n(u_2^n,u_3^n,\ulinev^n)$, (b) follows from 
$\rho^{Y_2}_{u_2^n, v_2^n} \define \sum_{v_1^n,u_3^n,v_3^n} p_{V_1U_3V_3|U_2V_2}^n(v_1^n,u_3^n,v_3^n | u_2^n,v_2^n) \rho^{Y_2}_{f^n(u_2^n,u_3^n,\ulinev^n)}$, (c) follows from the inequality $\tr(\Delta \rho) \leq \tr(\Delta \sigma) + \frac{1}{2} \norm{\rho - \sigma}_1$, for $0 \leq \Delta, \rho, \sigma \leq I$, and (d) follows from $\tr(\pi^{Y_2}_{v_2^n} \rho_{u_2^n,v_2^n}^{Y_2}) \geq 1 - \epsilon$,
and from the Gentle Operator Lemma \cite{BkWilde_2017}, for  sufficiently large $n$. 
Applying Hayashi-Nagaoka inequality \cite{200307TIT_HayNag}, we obtain 
\begin{eqnarray}
    &&\mathbb{E}[T_{2.2}] \leq 2 \: T_{2.2.1} + 4 \left( \sum_{i=2}^9 T_{2.2.i}\right) + 2 \sqrt{\epsilon}, \mbox{ where} \nonumber \\
      T_{2.2.1} &\define& \frac{1}{2^{n(R_2 + S_2 + K_2)}} \sum_{m_2} \sum_{a_2} \sum_{b_2} \sum_{u_2^n} \sum_{v_2^n}  p_{U_2V_2}^n(u_2^n,v_2^n)
    \tr \left\{  \left( I - \Upsilon_{u_2^n,v_2^n}\right) \: \pi^{Y_2}_{v_2^n} \: \rho_{u_2^n, v_2^n}^{Y_2} \: \pi^{Y_2}_{v_2^n}  \right\}, \nonumber \\
    T_{2.2.2} &\define& \frac{1}{2^{n(R_2 + S_2 + K_2)}} \sum_{m_2} \sum_{a_2} \sum_{b_2} \sum_{\tilde{m}_{22} \neq m_{22}} \sum_{\tilde{b}_2} \sum_{u_2^n} \sum_{v_2^n}  p_{U_2V_2}^n(u_2^n,v_2^n)
    \tr \left\{ \Upsilon_{u_2^n,V_2^n(\tilde{b}_2,\tilde{m}_{22})} \: \pi^{Y_2}_{v_2^n} \: \rho_{u_2^n, v_2^n}^{Y_2} \: \pi^{Y_2}_{v_2^n}  \right\}, \nonumber \\
    T_{2.2.3} &\define& \frac{1}{2^{n(R_2 + S_2 + K_2)}} \sum_{m_2} \sum_{a_2} \sum_{b_2} \sum_{\tilde{m}_{21} \neq m_{21}} \sum_{\tilde{a}_2} \sum_{u_2^n} \sum_{v_2^n}  p_{U_2V_2}^n(u_2^n,v_2^n)
    \tr \left\{ \Upsilon_{U_2^n(\tilde{a}_2, \tilde{m}_{21}),v_2^n} \: \pi^{Y_2}_{v_2^n} \: \rho_{u_2^n, v_2^n}^{Y_2} \: \pi^{Y_2}_{v_2^n}  \right\}, \nonumber \\
    T_{2.2.4} &\define& \frac{1}{2^{n(R_2 + S_2 + K_2)}} \sum_{m_2} \sum_{a_2} \sum_{b_2} \!\sum_{\tilde{m}_{22} \neq m_{22}} \! \sum_{\tilde{b}_2} \! \sum_{\tilde{m}_{21} \neq m_{21}} \! \sum_{\tilde{a}_2} \sum_{u_2^n} \sum_{v_2^n}  p_{U_2V_2}^n(u_2^n,v_2^n) \nonumber \\
    &&\tr \left\{  \Upsilon_{U_2^n(\tilde{a}_2, \tilde{m}_{21}),V_2^n(\tilde{b}_2, \tilde{m}_{22})} \: \pi^{Y_2}_{v_2^n} \: \rho_{u_2^n, v_2^n}^{Y_2} \: \pi^{Y_2}_{v_2^n}  \right\}. \nonumber
\end{eqnarray}
Here, we presented only four terms of $\left( \sum_{i=2}^9 T_{2.2.i}\right)$. The remaining terms can be bounded similarly. In the following propositions, we provide the rate constraints required to bound these terms.
\begin{proposition}
    \label{Prop:Dec2firsttermHay}
    For all $\epsilon>0$, and $\delta, \eta> 0$ sufficiently small and $n$ sufficiently large, we have $\mathbb{E}[T_{2.2.1}] \leq \epsilon$.
\end{proposition}
\begin{proof}
    The proof is provided in Appendix \ref{App:Dec2FirsttermHay}.
\end{proof}
\begin{proposition}
    \label{Prop:Dec2SecondtermHay}
    For any $\epsilon >0$, and for all $\delta,\eta >0$ sufficiently small and $n$ sufficiently large, we have $\sum_{i=2}^{9}\mathbb{E}[T_{2.2.i}] \leq \epsilon$ if the following inequalities holds.
    \begin{eqnarray}
    L_2 + K_2  &<& I(V_2;Y_2,U_2), \nonumber \\ 
        S_2 + T_2 &<&   \log(\Prime) - H(U_2|Y_2,V_2), \mbox{ and} \nonumber \\
        L_2 + K_2 + S_2 + T_2 &<& \log(\Prime) + H(V_2) -H(U_2;V_2|Y_2) , \nonumber
    \end{eqnarray}
where all the mutual information quantities are computed with respect to the state $\rho^{U_2V_2Y_2}$ defined in \eqref{Eqn:3to1CQBCstateDec2}.
\end{proposition}
\begin{proof}
    The proof is provided in Appendix \ref{App:Dec2SecondtermHay}.
\end{proof}
\noindent As stated earlier, the error analysis of Rx $3$ is identical to the error analysis of Rx $2$. Therefore, the analysis of $T_3$ is omitted. This completes the proof.
\end{proof}

\begin{lemma}
\label{Lem:CosetCdsInnBndStrictlyLargerForEx1}
Consider Ex.~\ref{Ex:3CQBCRoleOfCosetCds}. Under the condition $\beta_{k}(x_{k}) = \sigma_{\delta}(x_{k})$ for $k =2,3$ with $\tau * \delta_1 \boldsymbol{\stackrel{(i)}{<}} \delta$ and $h_{b}(\delta) \boldsymbol{\stackrel{(ii)}{<}} \frac{1+h_{b}(\tau * \delta_1)}{2}$, the rate triple $(h_{b}(\tau * \delta_1)-h_{b}(\delta_1), 1-h_{b}(\delta),1-h_{b}(\delta),\tau) $ is contained within the inner bound characterized in Thm.~\ref{Thm:3to1CQBC} but not within the inner bound achievable via unstructured IID random codes. Under the condition $\beta_{k}(x_{k}) = \gamma_{k}(x_{k})$ for $k=2,3$ with $h_{b}(\tau*\delta_1)+\tilde{h}_{b}(\cos\varphi_{2})+\tilde{h}_{b}(\cos\varphi_{3}) \boldsymbol{\stackrel{(a)}{>}} 1 \boldsymbol{\stackrel{(b)}{>}} h_{b}(\tau*\delta_1)+\tilde{h}_{b}(\cos\varphi_{3})$ where $\tilde{h}_{b}(x)\define h_{b}(\frac{1+x}{2})$  for $x \in [0,\frac{1}{2}]$, the rate triple $( h_{b}(\tau * \delta_1)-h_{b}(\delta_1),\tilde{h}_{b}(\cos\varphi_{2}),\tilde{h}_{b}(\cos\varphi_{3}))$ is contained within the inner bound characterized in Thm.~\ref{Thm:3to1CQBC} but not within the inner bound achievable via unstructured IID random codes.
\end{lemma}

\begin{proof}
    A proof follows from the discussion provided in Sec.~\ref{SubSec:NeedForCosetCodes} in the context of Ex.~\ref{Ex:3CQBCRoleOfCosetCds}.
\end{proof}

Having provided a proof of Thm.~\ref{Thm:3to1CQBC} that includes all the new elements in a simplified setting and demonstrated its import via Lemma \ref{Lem:CosetCdsInnBndStrictlyLargerForEx1}, we now proceed to obtain a generalized inner bound. 
{\section{New Inner Bound to the Capacity Region of the $3-$CQBC using Coset Codes}}
\label{Sec:SimultDecOfCosetCds}
We begin by stating some common notation that we employ throughout this section, followed by a discussion of our goals.

\begin{Notation}
\label{Not:3CQBCStep2}
In any context, when used in conjunction, $i,j,k$ will denote distinct indices in $[3]$, hence $\{i,j,k\}=[3]$ and we let $\dbrackthree \define \{(1,2),(1,3),(2,1),(2,3),(3,1),(3,2)\}$.
\end{Notation}

\begin{figure}
 \centering
\includegraphics[width=5in]{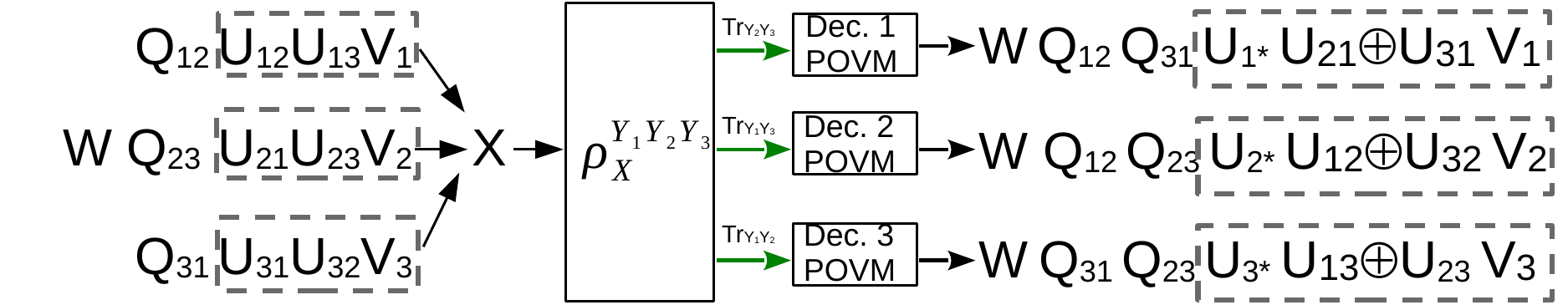}
    \caption{Depiction of all random variables in the full blown coding strategy, where, $U_{j*} = (U_{ji},U_{jk}) $ for $j \in [3]$, . In Sec.~\ref{SubSec:StepICodingTheorem} (Step II) only random variables in the gray dashed box are non-trivial, with the rest trivial.}
    \label{Fig:3CQBCMapOfRVs}
\end{figure}

Our goal in this section is to design and analyze a general coding strategy that enables \textit{each} Rx decode \textit{efficiently} \textbf{both} bivariate and univariate interference components that affects its reception. As we discussed in Sec.~\ref{SubSec:StructureOfAGeneralCodingStrategy}, unstructured IID codes and jointly designed coset codes enable efficient decoding of univariate and bivariate interference components respectively. Analogous to how Marton \cite{197905TIT_Mar} combined Cover's superposition coding \cite{197201TIT_Cov} with her precoding via binning \cite{1980MMPCIT_GelPin,197905TIT_Mar} technique through a two layer code, our general coding strategy will also involve two layers - one built using jointly designed coset codes to enable efficient decoding of bivariate interference components and the second involving unstructured IID codes. Refer to Fig.~\ref{Fig:3CQBCMapOfRVs} for a depiction of the random variables involved in our general coding strategy. $W,Q_{12},Q_{23},Q_{31},V_{1},V_{2},V_{3}$ constitute the unstructured IID codes employed in the natural generalization of Marton's coding strategy to the $3-$CQBC. See \cite[Thm.~2 in Sec.~II.D]{201804TIT_PadPra}. To this, we append the layer of coset codes represented via random variables $U_{ij} : (i,j) \in \dbrackthree$. As we will elaborate in Sec.~\ref{SubSec:StepICodingTheorem}, $U_{ij}$ and $U_{kj}$ are supported on the same finite field $\CalF_{\upsilon_{j}}$ permitting the decoding of $U_{ij}\oplus U_{kj}$. The codebooks decoded into by each Rx is indicated by the corresponding random variables depicted on the right.

We present this general coding strategy and the corresponding inner bound in two pedagogical steps. Step II, presented in Sec.~\ref{SubSec:StepICodingTheorem} will present, design and analysis of only the layer involving coset codes. Viewed
through the lens of Fig.~\ref{Fig:3CQBCMapOfRVs}, the Step II coding strategy corresponds to random variables $W=Q_{12}=Q_{23}=Q_{31}=\phi$ being trivial. In Step III, presented in Sec.~\ref{SubSec:Step2}, we combine this Step II design with a base layer involving unstructured IID codes, as depicted via random variables $W,Q_{12},Q_{23},Q_{31}$ in Fig.~\ref{Fig:3CQBCMapOfRVs}. Lemma \ref{Lem:CosetCdsInnBndStrictlyLargerForEx1} proves that the inner bound characterized in Thm.~\ref{Thm:3to1CQBC} can be strictly larger for certain examples. Our analysis of the coding strategy presented in Step III yields an inner bound that, in addition to be strictly larger for certain examples, subsumes all previous known inner bounds. See Rem.~\ref{Rem:afterstep3}.

\subsection{{Step II : A Coset Code Strategy to manage Bivariate Interference at all Rxs over a $3-$CQBC}}
\label{SubSec:StepICodingTheorem}

In this step, we design and analyze a coding strategy that enables \textit{each} Rx decode efficiently a bivariate interference component that hurts them. In contrast to Sec.~\ref{Sec:3to1CQBC}, where only Rx $2$ and $3$ employed jointly designed coset code to assist Rx $1$ decode bivariate interference, in this step, each Rx employs coset codes that are carefully designed jointly with inter- and intra- structural properties to enable each Rx efficienntly decode bivariate interference.

Refer to Fig.~\ref{Fig:3CQBCMapOfRVs}. In this second step, we activate only the `bivariate' $\ulineU \define (U_{1*},U_{2*},U_{3*})$ and private parts $\ulineV \define (V_{1},V_{2},V_{3})$ choosing them to be non-trivial, with the rest $W = Q_{12}= Q_{23}=Q_{31}=\phi$ trivial. The coding strategy and its analysis presented in the proof of Thm.~\ref{Thm:3CQBCStepIInnerBound} involves simultaneous decoding by the Rxs of a combination of unstructured IID codes and jointly designed coset codes, including decoding of the field addition and hence addresses all the new elements.
\begin{theorem}
\label{Thm:3CQBCStepIInnerBound}
Let $\hat{\alpha}_{S} \in [0,\infty)^{4}$ be the set of all rate-cost quadruples $(R_{1},R_{2},R_{3},\tau) \in [0,\infty)^{4}$ for which there exists (i) finite sets $\CalV_{j}: j \in [3]$, (ii) finite fields $\SemiPrivateRVSet_{ij}=\CalF_{\upsilon_{j}}$ for each $ij \in \dbrackthree$, (iii) a PMF $p_{\ulineU~\!\!\ulineV X}=p_{U_{1*}U_{2*}U_{3*}V_{1}V_{2}V_{3} X}$ on $\ulineCalU \times \ulineCalV\times\CalX$, (iv) nonnegative numbers $S_{ij},T_{ij}:ij \in \dbrackthree, K_{j},L_{j}:j\in [3]$, such that
$R_{1}=T_{12}+T_{13}+L_{1},
R_{2}=T_{21}+T_{23}+L_{2},
R_{3}=T_{31}+T_{32}+L_{3}$, $\Expectation\left\{ \kappa(X) \right\} \leq \tau$,
\begin{eqnarray}
\label{Eqn:ManyToManySourceCodingBounds}
&& S_{A}+M_{B}+K_{C} >\Theta(A,B,C),\mbox{ where } 
\end{eqnarray}
\begin{eqnarray}
\Theta(A,B,C)\define \max_{\substack{(\theta_{j}:j \in B) \in \underset{j \in B}{\prod} \fieldpij}}  \left\{
\begin{array}{c}\sum_{a \in A}\log |\mathcal{U}_{a}| + \sum_{j \in B}\log \upsilon_{j} +\sum_{c \in C} H(V_{c}) - H(U_{A}, U_{ij}\oplus \theta_{j}U_{kj}:j \in B,V_{C})\end{array}
 \right\}
\nonumber
\end{eqnarray}
for all $A \subseteq \left\{12,13,21,23,31,32\right\}, B \subseteq \left\{ 1,2,3 \right\}, C \subseteq \left\{ 1,2,3 \right\}$, that satisfy $A \cap A(B) = \phi$, where $A({B}) = \cup_{j \in B}\{ ij,kj\}$, $U_{A} = (U_{jk}:jk \in A)$, $V_{C}=(V_{c}:c \in C)$, $S_{A} = \sum _{jk \in A}S_{jk}, M_{B}\define \sum_{j \in B} \max\{ S_{ij}+T_{ij},S_{kj}+T_{kj}\}, K_{C} = \sum_{c \in C}K_{c}$, and
\begin{eqnarray}
\label{Eqn:CQBCChannelCodingStep1Bounds2}
S_{\mathcal{A}_{j}}+T_{\mathcal{A}_{j}} &<& \sum_{a \in \mathcal{A}_{j}}\!\!\log |\mathcal{U}_{a}| - H(U_{\mathcal{A}_{j}}|U_{\mathcal{A}_{j}^{c}},U_{ij}\oplus U_{kj},V_{j},Y_{j})
\\\label{Eqn:CQBCChannelCodingStep1Bounds3}
S_{\mathcal{A}_{j}}+T_{\mathcal{A}_{j}}+S_{ij}+T_{ij} &<& \sum_{a \in \mathcal{A}_{j}}\log |\mathcal{U}_{a}| + \log \upsilon_{j} - H(U_{\mathcal{A}_{j}},U_{ij}\oplus
U_{kj}|U_{\mathcal{A}_{j}^{c}},V_{j},Y_{j}) \\
\label{Eqn:CQBCChannelCodingStep1Bounds4}
S_{\mathcal{A}_{j}}+T_{\mathcal{A}_{j}}+S_{kj}+T_{kj} &<& \sum_{a \in \mathcal{A}_{j}}\log |\mathcal{U}_{a}| + \log \upsilon_{j}
- H(U_{\mathcal{A}_{j}},U_{ij}\oplus
U_{kj}|U_{\mathcal{A}_{j}^{c}},V_{j},Y_{j}) \\
\label{Eqn:CQBCChannelCodingStep1Bounds5}
S_{\mathcal{A}_{j}}+T_{\mathcal{A}_{j}}+K_{j}+L_{j} &<& \sum_{a \in \mathcal{A}_{j}}\log |\mathcal{U}_{a}|+H(V_{j})-H(U_{\mathcal{A}_{j}},V_{j}|U_{\mathcal{A}_{j}^{c}},U_{ij}\oplus
U_{kj},Y_{j}) \\
\label{Eqn:CQBCChannelCodingStep1Bounds6}
S_{\mathcal{A}_{j}}+T_{\mathcal{A}_{j}}+K_{j}+L_{j}+S_{ij}+T_{ij} &<& \sum_{a \in \mathcal{A}_{j}}\log |\mathcal{U}_{a}| + \log \upsilon_{j} +H(V_{j}) -
H(U_{\mathcal{A}_{j}},V_{j},U_{ij}\oplus U_{kj}|U_{\mathcal{A}_{j}^{c}},Y_{j}) \\
\label{Eqn:CQBCChannelCodingStep1Bounds7}
S_{\mathcal{A}_{j}}+T_{\mathcal{A}_{j}}+K_{j}+L_{j}+S_{kj}+T_{kj} &<& \sum_{a \in \mathcal{A}_{j}}\!\!\log |\mathcal{U}_{a}| + \log  \upsilon_{j} +H(V_{j}) - H(U_{\mathcal{A}_{j}},V_{j},U_{ij}\oplus
U_{kj}|U_{\mathcal{A}_{j}^{c}},Y_{j}), 
\end{eqnarray}
for every $\mathcal{A}_{j} \subseteq \left\{ ji,jk\right\}$ with
distinct indices $i,j,k$ in $\left\{ 1,2,3 \right\}$, where
$S_{\mathcal{A}_{j}} \define \sum_{a \in \mathcal{A}_{j}}S_{a},
T_{\mathcal{A}_{j}} \define \sum_{a \in \mathcal{A}_{j}}T_{a},
U_{\mathcal{A}_{j}} = (U_{a}:a \in \mathcal{A}_{j})$ and all the information quantities are evaluated with respect to the state 
\begin{eqnarray}
 \label{Eqn:StageITestChnl}
 \Psi^{\ulineU\!\!~\ulineU^{\oplus}\!\!~\ulineV X\!\!~\ulineY} &\define&
 \sum_{\substack{\ulineu, \ulinev,x\\u_{1}^{\oplus},u_{2}^{\oplus},u_{3}^{\oplus} }}p_{\ulineU \: \ulineV X}(u_{1*},u_{2*},u_{3*},\ulinev,x)\mathds{1}{\left\{u_{ij}\oplus u_{kj}=u_{j}^{\oplus}:j\in[3]\right\}} \nonumber \\ 
 &&\ketbra{u_{1*}~\! u_{2*}~\! u_{3*}~\! u_{1}^{\oplus}~\! u_{2}^{\oplus}~\! u_{3}^{\oplus}~\! \ulinev~\! x} \!\otimes\! \rho_{x}. \nonumber
\end{eqnarray}
Let $\alpha_{S}$ denote the convex closure of $\hat{\alpha}_{S}$. Then $\alpha_{S} \subseteq \ScrC(\tau)$ is an achievable rate region.

\end{theorem}
\begin{remark}
 \label{Rem:Step1CodingThmRemarks}
 Inner bound $\alpha_{S}$ (i) subsumes \cite[Thm.~1]{202202arXiv_Pad3CQBC, 202206ISIT_Pad3CQBC}, (ii) is the CQ analogue of \cite[Thm.~7]{201603TIT_PadSahPra}, (iii) does \textit{not} include a time-sharing random variable and \textit{includes} the `don't care' inequalities \cite{200702ITA_KobHan,200807TIT_ChoMotGarElg} - \eqref{Eqn:CQBCChannelCodingStep1Bounds3}, \eqref{Eqn:CQBCChannelCodingStep1Bounds4} with $\mathcal{A}_{j}=\phi$. Thus, $\alpha_{S}$ can be enlarged.
\end{remark}

\noindent \textbf{Discussion of the bounds}:
{We \textit{explain} how (i) each bound arises and (ii) their role in driving the error probability down. All the source coding bounds are captured through lower bound \eqref{Eqn:ManyToManySourceCodingBounds} by choosing different $A,B,C$. To understand these bounds, let's begin with a simple example. Suppose we have $K$ codebooks $C_{k}=(G_{k}^{n}(m_{k}):m_{k} \in [2^{nR_{k}}]):k \in [K]$ with all of them mutually independent and picked IID in the standard fashion with distribution $G_{k}^{n}(m_{k})\sim q_{G_{k}}^{n}$ for $k \in [K]$. What must the bounds satisfy so that we can find \textit{one} among the $2^{n(R_{1}+\cdots +R_{K})}$ $K-$tuples of codewords to be jointly typical with respect to a joint PMF $r_{\ulineG}=r_{G_{1}G_{2}\cdots G_{K}}$. Denoting $q_{\ulineG}= \prod_{k=1}^{K}q_{G_{k}}$, we know from standard information theory that if $\sum_{l \in S}R_{l} > D(r_{G_{S}}||\otimes_{l\in S}r_{G_{l}})+D(\otimes_{l\in S}r_{G_{l}}||\otimes_{l\in S}q_{G_{l}})$ for all $S\subseteq [K]$, then\footnote{As is standard, \!here \!$r_{G_{S}}\!$ is the marginal of $r_{\ulineG}$ over the RVs indexed in S.} via the standard second moment method \cite{198101TIT_GamMeu} we can find the desired $K-$tuple. The bounds in \eqref{Eqn:ManyToManySourceCodingBounds} are just these. We summarize here the detailed explanation provided in \cite{202202arXiv_Pad3CQBC}. We have $9$ codebooks (Fig.~\ref{Fig:FigCodeStructureCQBC(StepI)}), however we should also consider the sum of $U_{ij}$ and $U_{kj}$ codebooks, therefore 12 codebooks, hence $K=12$. $r_{G_{1}\cdots G_{6}}=q_{\ulineU}$ and $q_{U_{ji}}=\frac{1}{|\CalU_{ji}|}$ is uniform. Indeed, owing to pairwise independence and uniform distribution of codewords in the random coset code \cite{BkNIT_PraPadShi}. This explains the first term within the $\max$ defining $\Theta(\cdots)$. $r_{G_{7}G_{8}G_{9}}=p_{\ulineV}$ and the $V-$codewords have been picked independently $\prod p_{V_{j}}$. This explains the third term in the definition of $\Theta(\cdots)$. Owing to the coset structure, it turns out that \cite[Bnds.~(45), (46)]{201804TIT_PadPra} the $U_{ij}$-codeword plus $\theta_{j}$ times $U_{kj}$-codeword must be found in the sum of the $U_{ij}, U_{kj}$ cosets (Fig.~\ref{Fig:FigCodeStructureCQBC(StepI)} rightmost codes). These vectors being uniformly distributed $\sim \frac{1}{\upsilon_{j}}$, we have the second term and the presence of $U_{ji}\oplus \theta_{j}U_{jk}$ in the entropy term in the defn.~of $\Theta(\cdots)$.\footnote{The curious reader may look at the two bounds in \cite[Bnds.~(72)]{201804TIT_PadPra} with the $\max_{\theta \neq 0}$ which in fact yields \cite[Bnds.~(45), (46)]{201804TIT_PadPra}.} }

{Now to channel coding bounds \eqref{Eqn:CQBCChannelCodingStep1Bounds2}-\eqref{Eqn:CQBCChannelCodingStep1Bounds7}. The code of Rx $j$ is illustrated in line $j$ of Fig.~\ref{Fig:FigCodeStructureCQBC(StepI)}. Throughout the rest of this discussion, we rename the Rvs $Z_1=U_{ji}, Z_2=U_{jk}, Z_3=V_j, Z_4 = U_j^{\oplus} = U_{ij} \oplus U_{kj}, Z_5=Y_j,\olineZ=Z_{1}Z_{2}Z_{3}Z_{4}Z_{5},$ PMFs $r_{\ulineZ} = r_{Z_1Z_2Z_3} = p_{U_{ji} U_{jk} V_j}, q_{\ulineZ}=q_{Z_1Z_2Z_3}= \frac{p_{V_j}}{|\CalU_{ji}| |\CalU_{jk}|}$. Our decoding analysis is a simple CQMAC decoding at each user. We shall therefore relate \eqref{Eqn:CQBCChannelCodingStep1Bounds2}-\eqref{Eqn:CQBCChannelCodingStep1Bounds7} to the bounds one obtains on a CQMAC with $4$ Txs.
\begin{eqnarray}
  \label{Eqn3CQIC:Step1ExplanationChnlBnds}
  \ScrR^{\olineZ}\define \Psi^{U_{ji}U_{jk}V_{j}U_{j}^{\oplus}Y_{j}} =\tr_{U_{i*}U_{k*}V_{i}V_{k}U_{i}^{\oplus}U_{k}^{\oplus}Y_{i}Y_{k}}\{\Psi^{\ulineU\ulineU^{\oplus}\ulineV\ulineY}\},
  \nonumber\\
    \ScrQ^{Z_{1}Z_{2}Z_{3}Z_{4}}\define \sum_{\substack{(z_{1},z_{2},z_{3},z_{4})\\\in \CalF_{\Prime_{i}}\times \CalF_{\Prime_{k}}\times \CalV_{j}\times \CalF_{\Prime_{j}}}}\frac{p_{V_{j}}(z_{3})\ketbra{z_{1}~z_{2}~z_{3}~z_{4}}}{|\CalU_{ji}|\cdot |\CalU_{jk}| \cdot| \CalU_{ij}|}
    \nonumber
 \end{eqnarray}
$C_{1}=S_{ji}+T_{ji}, C_{2}=S_{jk}+T_{jk},C_{3}=K_{j}+L_{j}, C_{4}=\max\{S_{ij}+T_{ij},S_{kj}+T_{kj}\}$. The $7$ bounds in \eqref{Eqn:CQBCChannelCodingStep1Bounds2}, \eqref{Eqn:CQBCChannelCodingStep1Bounds5} are in fact the $7$ bounds $C_{A} < D(\ScrR^{\olineZ}||\ScrR^{\olineZ_{A}} \otimes \ScrR^{\olineZ_{A^{C}}} )+D(\ScrR^{\olineZ_{A}}||\ScrQ^{\olineZ_{A}})$ obtained by different choices of $A \subseteq\{1,2,3\}$, and the 16 bounds in \eqref{Eqn:CQBCChannelCodingStep1Bounds3}, \eqref{Eqn:CQBCChannelCodingStep1Bounds4}, \eqref{Eqn:CQBCChannelCodingStep1Bounds6}, \eqref{Eqn:CQBCChannelCodingStep1Bounds7} are in fact the $8$ bounds $C_{A\cup\{4\}} < D(\ScrR^{\olineZ}||\ScrR^{\olineZ_{A\cup \{4\}}} \otimes \ScrR^{\olineZ_{A^{C}\setminus\{4\}}} )+D(\ScrR^{\olineZ_{A\cup\{4\}}}||\ScrQ^{\olineZ_{A\cup \{4\}}})$ obtained by different choices of $A \subseteq\{1,2,3\}$ but now replicated twice owing to the $\max$ in the definition of $C_{4}$. Having identified the bounds, let us interpret the upper-bound $D(\ScrR^{\olineZ}||\ScrR^{\olineZ_{A}} \otimes \ScrR^{\olineZ_{A^{C}}} )+D(\ScrR^{\olineZ_{A}}||\ScrQ^{\olineZ_{A}})$. Observe that, if $\ScrQ^{Z_{1}Z_{2}Z_{3}Z_{4}}=\ScrR^{Z_{1}Z_{2}Z_{3}Z_{4}}$, then the upper-bound will just have one term which is essentially the bound obtained on a CQMAC. The addition of the $D(\ScrR^{\olineZ_{A}}||\ScrQ^{\olineZ_{A}})$ is due to the fact that codewords of random code are \textit{not} distributed with PMF $p_{U_{ji}U_{jk}V_{j}U_{j}^{\oplus}}$, i.e., $\Psi^{U_{ji}U_{jk}V_{j}}$, but with respect to $\ScrD^{Z_{1}Z_{2}Z_{3}Z_{4}}$. This explains \eqref{Eqn:CQBCChannelCodingStep1Bounds2}-\eqref{Eqn:CQBCChannelCodingStep1Bounds7}. Imposing the same guarantees correct decoding at each Rx.}

\begin{proof}
Consider (i) $\SemiPrivateRVSet_{ji} = \CalF_{\Prime_{i}}$ for $ji \in \llbracket 3 \rrbracket$, (ii) PMF $p_{\ulineU\: \ulineV X} = p_{U_{1*}U_{2*}U_{3*}V_1V_2V_3X}$ on $\uline{\mathcal{U}} \times \uline{\mathcal{V}} \times\mathcal{X}$ as described in the hypothesis. Let $(R_1,R_2,R_3)$ be a rate triple for which there exist non-negative numbers $S_{ji},T_{ji}$, for $ji \in \llbracket 3 \rrbracket$, and $K_{j},L_{j}$, for $j \in [3]$ such that $\mathbb{E}[\kappa(X)]\leq \tau$, $R_{j}=T_{ji} +T_{jk} +L_{j}$ for $ji \in \llbracket 3 \rrbracket$, and the bounds in (\ref{Eqn:ManyToManySourceCodingBounds})-(\ref{Eqn:CQBCChannelCodingStep1Bounds7}) hold.
We divide the proof into four parts - code structure, encoding, decoding POVMs, and error analysis. Since the error analysis at each Rx is essentially the error analysis of an effective $4-$user classical-quantum multiple access channel ($4-$CQMAC), we present the error analysis only for this $4-$CQMAC.

\noindent \textbf{Code structure:}
For $j \in [3]$, Rx $j$'s message is split into three parts $m_j=(m_{ji},m_{jk},m_{jj})$. See Fig.~\ref{Fig:FigCodeStructureCQBC(StepI)}. The semi private parts $m_{ji}$ and $m_{jk}$ are communicated using coset codes built over $\mathcal{U}_{ji}$ and $\mathcal{U}_{jk}$ respectively. The private part $m_{jj}$ is communicated using a conventional IID random code built over $\mathcal{V}_j$. $U_{j*} \define (U_{ji},U_{jk})$ represent information about the semi private parts $(m_{ji},m_{jk})$ and $V_{j}$ represents information about the private part $m_{jj}$. Note that $(U_{ji},U_{jk},V_j) \in \mathcal{F}_{\Prime_i} \times \mathcal{F}_{\Prime_k}
\times \mathcal{V}_j$.
Specifically, the message $m_{jj}$ is communicated using (i) $c_{jj}  \define \{v_{j}^{n}(b_j,m_{jj}) : b_j \in [2^{n K_{j}}],  m_{jj}\in [2^{nL_{j}}] \}$ constructed over $\CalV_{j}$, and (ii) a binning map {$b_j : [2^{nR_1}] \times [2^{nR_2}] \times [2^{nR_3}]  \rightarrow [2^{nK_j}]$.} Let $v^n_j(b_j(\ulinem),m_{jj})$ denote the chosen codeword from the bin $c_{jj}(m_{jj}) \define \{v_j^n(b_j,m_{jj}) : b_j \in [2^{nK_j}]\}$ corresponding to message $m_{jj}$.
For $ji \in \llbracket 3 \rrbracket$, the message $m_{ji}$ is communicated using (i) a NCC over the finite field $\CalF_{\Prime_i}$ denoted by $(n,s_{ji},t_{ji},g^{ji}_{I},g^{ji}_{O/I}, b_{ji}^n)$ where $g^{ji}_{I} \in \mathcal{F}_{\Prime_i}^{s_{ji} \times n}$, $g^{ji}_{O/I} \in \mathcal{F}_{\Prime_i}^{t_{ji} \times n}$, $b_{ji}^n \in \mathcal{F}_{\Prime_i}^n$, $s_{ji} \define \left \lceil \frac{nS_{ji}}{\log(\Prime_i)} \right \rceil$ and $t_{ji} \define \left \lfloor \frac{nT_{ji}}{\log(\Prime_i)} \right \rfloor$, and (ii) a binning map $a_{ji}: [2^{nR_1}] \times [2^{nR_2}] \times [2^{nR_3}] \rightarrow [2^{nS_{ji}}]$. We emphasize that the coset codes built over $\CalU_{ij}$ and $\CalU_{kj}$ intersect. In other words, the smaller of these two codes is a sub-coset of the larger. This is accomplished by ensuring that the rows of the generator matrix of the larger code contain all the rows of the generator matrix of the smaller code i.e for $j \in [3]$ the generator matrices $g_{ij} = \left[\!\! \begin{array}{c}g_{I}^{ij}\\g_{O/I}^{ij}\end{array}\!\!\right]$ satisfy $g_{\sfb}\define \left[\!\! \begin{array}{c}g_{\sfs}\\g_{\Delta}\end{array}\!\!\right]$ where $\sfb, \sfs$ are defined such that 
$\{\sfb, \sfs\}=\{ij, kj\} \mbox{ and } s_{\sfb} + t_{\sfb} \geq s_{\sfs} + t_{\sfs}$.
\begin{figure}
\centering
\includegraphics[width=4in]{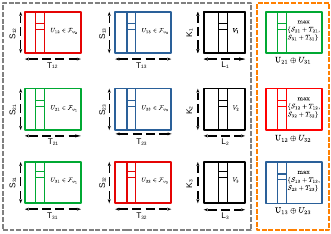}
    \caption{$9$ codes employed in the coding strategy are depicted in the grey box above. The $U_{ji} : ji \in \dbrackthree$ are coset codes built over finite fields. Codes with the same color are built over the same finite field, and the smaller of the two is a sub-coset of the larger. The black codes are built over auxiliary finite sets $\mathcal{V}_j$ using the conventional IID random code structure. Row $j$ depicts the codes of Rx $j$. Rx $j$, in addition to decoding into codes depicted in row $j$ also decodes $U_{ij} \oplus U_{kj}$. The three codes in the dotted orange box are employed only by the decoders $1$, $2$, and $3$ respectively. The three collections of codewords in the orange dotted box on the right are employed only by the corresponding Rx. Specifically, the first collection of the codewords at the top of the orange dotted box is obtained by adding all pairs of codewords in the $U_{21}$ and $U_{31}$ codes.}
    \label{Fig:FigCodeStructureCQBC(StepI)}
    \vspace{-0.15in}
\end{figure}
For $ji \in \dbrackthree$, let $u_{ji}^n(a_{ji}(\ulinem), m_{ji})$ denote the chosen codeword from the coset $c_{ji}(m_{ji}) \define \{u_{ji}^n(a_{ji}, m_{ji}) : a_{ji} \in \mathcal{F}_{\Prime_i}^{s_{ji} }\}$, where $u_{ji}^n(a_{ji},m_{ji}) \define a_{ji} g_{I}^{ji} \oplus m_{ji} g_{O|I}^{ji} \oplus b_{ji}^n$. {Finally, let $c \define \{x^n(\ulinem) : \ulinem \in [2^{nR_1}] \times [2^{nR_2}] \times [2^{nR_3}]\}$ be a codebook constructed over $\CalX$.}

\noindent \textbf{Encoding Rule:}
To communicate the message triple $\ulinem=(m_1,m_2,m_3)$, the encoder prepares the quantum state $\rho^{Y_1Y_2Y_3}_{x^n(\ulinem)}$. 
\noindent \textbf{Decoding POVM:} For $j \in [3],$ Rx $j$ decodes a bivariate component \( u^n_{ij}(a_{ij}(\ulinem),m_{ij}) \oplus u^n_{kj}(a_{kj}(\ulinem),m_{kj}) \) in addition to the codewords \( u^n_{ji}(a_{ji}(\ulinem),m_{ji}), u^n_{jk}(a_{jk}(\ulinem),m_{jk}), v^n_j(b_{j}(\ulinem),m_{jj}) \) corresponding to its own message. Specifically, Rx $j$ attempt to decode the quartet
\begin{eqnarray}
\left(u^n_{ji}(a_{ji}(\ulinem),m_{ji}), u^n_{jk}(a_{jk}(\ulinem),m_{jk}), v^n_j(b_j(\ulinem),m_{jj}), u^n_{ij}(a_{ij}(\ulinem),m_{ij}) \oplus u^n_{kj}(a_{kj}(\ulinem),m_{kj})\right)
 \nonumber 
\end{eqnarray}
using simultaneous decoding. In particular, the construction of our simultaneous decoding POVM adopts the \textit{tilting, smoothing, and augmentation} technique of Sen \cite{202103SAD_Sen}. A careful consideration of this setting reveals that if one considers this setting carefully from a receiver's channel coding error event, it will be evident that communication is occurring over an induced  $4-$CQMAC. We now proceed towards identifying this induced $4-$CQMAC. 
\begin{eqnarray}
\label{Eqn3CQBC:EffectCQMACOrigNotation}
\xi^{U_{j*}V_{j}U_{j}^{\oplus}Y_{j}} &=& \sum_{u_{j*},v_{j},u_{j}^{\oplus}} p_{U_{j*}V_{j}U_{j}^{\oplus}}(u_{j*},v_{j}, u_{j}^{\oplus}) \ketbra{u_{j*}~v_{j}~u_{j}^{\oplus}} \otimes \xi_{u_{j*},v_{j},u_{j}^{\oplus}}, \mbox{ where }
\label{decoding3CQBC} 
\\
 \xi_{u_{j*}v_{j}u_{j}^{\oplus}}&=& \sum_{u_{i*}} \sum_{u_{k*}}  \sum_{v_i} \sum_{v_k} p_{U_{i*}U_{k*}V_iV_k|U_{j*}V_jU_j^{\oplus}}(u_{i*},u_{k*},v_i,v_k|u_{j*},v_j,u_j^{\oplus}) \rho^{Y_j}_{f(\ulineu,\ulinev)}.  \nonumber 
\end{eqnarray}
Rx $j$'s error analysis is essentially the error analysis of the above MAC, where Rx $j$ attempts to decode \( U_{ji}, U_{jk}, V_{j} \), and \( U^{\oplus}_j = U_{ij} \oplus U_{kj} \). We may, therefore, restrict our attention to the effective 4$-$CQMAC (see Fig.~\ref{Fig4CQMAC}) specified via \eqref{Eqn3CQBC:EffectCQMACOrigNotation}. To reduce clutter and simplify notation, we henceforth use underlined symbols to denote quadruples instead of triples. In particular, for any four objects $a_1,a_2,a_3,a_4$, we write $\ulinea \define (a_1,a_2,a_3,a_4)$. Accordingly, we we rename $Z_{1}\define U_{ji}$, $Z_{2} \define U_{jk}$, $Z_{3} \define V_{j}$, $Z_{4} \define U_{ij}\oplus U_{kj}$, $\ulineZ \define (Z_{1}, Z_{2}, Z_{3} , Z_{4})$, a PMF $p_{\ulineZ}(\ulinez)= \sum_{u_{i*}} \sum_{u_{k*}} \sum_{v_i} \sum_{v_k} p_{\ulineU \: \ulineV \: U_j^{\oplus}} (z_1,z_2,u_{i*},u_{k*},z_3,v_i,v_k,z_4)
$, $\tilde{R}_1 \define S_{ji} +T_{ji}$, $\tilde{R}_2 \define S_{jk} + T_{jk}$, $\tilde{R}_3 \define K_j + L_j$, $\tilde{R}_4 \define \max\{S_{ij} + T_{ij}, S_{kj} + T_{kj}\}$ and $Y \define Y_j$ (see Fig.~\ref{TabNotataion4CQMAC}). Throughout the rest of the proof, we restrict our attention to the analysis of the error probability of the above 4$-$CQMAC on which the decoder attempts to decode $z_1, z_2, z_3$ and $z_4$.
Towards that end, we relabel the state in (\ref{decoding3CQBC}) as

\begin{eqnarray}
 \xi^{\ulineZ Y} \define \sum_{z_{1},z_{2},z_{3},z_{4}} p_{\ulineZ}(\ulinez)\ketbra{z_{1}~z_{2}~z_{3}~z_{4}}\otimes \xi_{z_{1}z_{2}z_{3}z_{4}}. \nonumber
\end{eqnarray}
\begin{figure}
\centering
\begin{minipage}{0.45\textwidth}
    \centering
    \includegraphics[width=\textwidth]{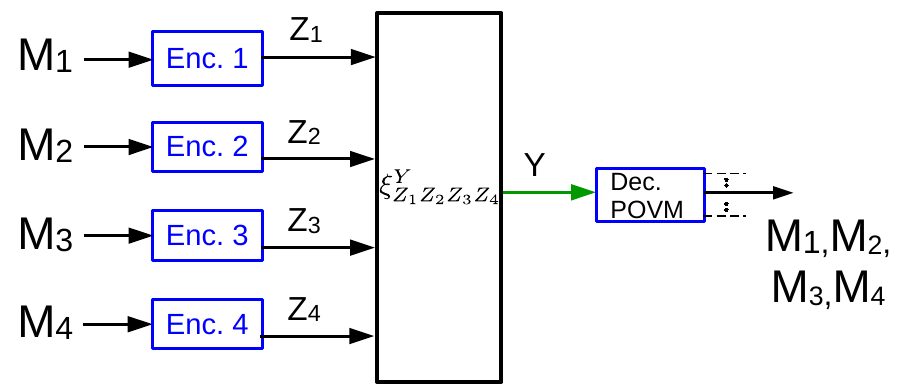}
    \caption{Communication over the effective four user CQMAC}
    \label{Fig4CQMAC}
\end{minipage}
\hfill\begin{minipage}{0.5\textwidth}
\vspace{0.2in}
    \centering
    \resizebox{\textwidth}{!}{
        \begin{tabular}{|c|c||c|c|}
            \hline
            New Notation & Original Notation & New Notation & Original Notation  \\
            \hline
            $m_{1}$ & $m_{ji}$ & $\tilde{R}_1$ & $S_{ji} + T_{ji}$ \\
            \hline
            $m_{2}$ & $m_{jk}$ & $\tilde{R}_2$ & $S_{jk} + T_{jk}$ \\
            \hline
            $m_3$ & $m_{jj}$ & $\tilde{R}_3$ & $K_j + L_j$ \\
            \hline
            $m_4$ & $m_{ij} \oplus m_{kj}$ & $\tilde{R}_4$ & $\max\{S_{ij} + T_{ij}, S_{kj} +T_{kj}\}$ \\
            \hline
            $Z_1$ & $U_{ji}$ & $p_{Z_1Z_2Z_3Z_4}$ & $p_{U_{j*}V_jU_j^{\oplus}}$  \\
            \hline
            $Z_{2}$ & $U_{jk}$ &$Y$ & $Y_j$ \\
            \hline
            $Z_3$ & $V_j$ & $\xi_{\ulinez}$ & $\xi_{u_{j*},v_j,u_j^{\oplus}}$   \\
            \hline 
            $Z_4$ & $U_j^{\oplus} = U_{ij} \oplus U_{kj}$ & $\ulineZ=(Z_1,Z_2,Z_3,Z_4)$ & $\left( U_{ji},U_{jk},V_j,U_j^{\oplus} \right)$  \\
            \hline
        \end{tabular}
    }
    \caption{Mapping between the new notation and the original notation}
    \label{TabNotataion4CQMAC}
\end{minipage}
\end{figure}

\noindent To analyze the 4$-$CQMAC, we adopt Sen's \cite{202103SAD_Sen} technique of \textit{tilting, smoothing, and augmentation}. In this setting, there are $15=2^4-1$ possible error events, each corresponding to a different subspace, and these subspaces can overlap in a nontrivial way. Sen's \cite{202103SAD_Sen} idea is to enlarge the space so that we can tilt each error subspace in an orthogonal direction and increase their separation. We also augment the space to ensure the smoothness property of the states obtained by averaging. 
Following this approach, we define $14=(2^4-1)-1$ tilting maps, one for each error subspace except the last, where all four messages $\ulinem \define (m_1,m_2,m_3,m_4)$ are wrong and remains untilted. We also define an additional tilting map for the quantum state itself, which tilts the state slightly in all directions. This tilting map should be chosen carefully so that the tilted state remains close to the original state in $\mathbb{L}_1$ norm and satisfies the required smoothing properties, that is, the difference between the average of the original state and the average of the tilted state\footnote{When we say `average of $\cdots$' we mean the expectation of the tilted state over the random codebook.} is small.
In what follows, we present this approach in detail. First, we define the new enlarged and augmented space. Next, we define the tilting maps. Then, we define the tilted state and state how close it remains to the original state. Finally, we describe the tilted subspace and construct the final POVM, which can simultaneously decode all messages.
{\begin{Notation}
In the rest of this proof, $S$ \footnote{Throughout the remaining part of the proof, $S \subseteq [4]$ will denote a generic non-empty subset of $\{1,2,3,4\}$. We alert the reader to distinguish $S$ from the $S_{ij}$ that appears in the rates (see Fig.~\ref{Fig:FigCodeStructureCQBC(StepI)}). The latter quantity always appears with a subscript, while the former never appears with a subscript.} will always refer to a \textbf{nonempty} subset of $[4]$. The following objects are defined for nonempty subset of $S$.
\end{Notation}
}

\noindent Consider four auxiliary finite sets  $\CalD_{i}$ for $i \in [4]$ along with corresponding four auxiliary Hilbert spaces ${\CalD_{i}}$ of dimension $|\CalD_{i}|$ for each $i \in [4]$.\footnote{$\CalD_{i}$ denotes both the finite set and the corresponding auxiliary Hilbert space. The specific reference will be clear from context.} Define the extended space $\CalH_{Y}^{e}$ as follows
\begin{eqnarray}
    \nonumber 
 \left(\CalH_{Y}^{e}\right)^{\otimes n} \define \CalH_{Y_{G}}^{\otimes n}
\oplus \bigoplus_{S \subsetneq [4]} \left(\CalH_{Y_{G}}^{\otimes n}
 \otimes \CalD_{S}^{\otimes n} \right)\mbox{ where }\CalH_{Y_{G}}^{\otimes n} =\CalH_{Y}^{\otimes n} \otimes \complex^{2}, \CalD_{S}^{\otimes n} = \bigotimes_{s \in S} \CalD_{s}^{\otimes n}.
\end{eqnarray}
Note that $\dim(\CalD_{S}) = \prod_{s \in S}\dim(\CalD_{s})$. For $ S \subseteq [4]$, let  $\ket{d_{S}}=\otimes_{s \in S} \ket{d_s}$ be a computational basis vector of  ${\CalD_{S}}$, $Z_{S}:=\left(Z_{s}\right)_{s \in S}$, and let $\eta>0$ be chosen appropriately in the sequel. In the sequel, we let $\boldsymbol{\CalH_{Y}}= \CalH_{Y}^{\otimes n}$,         $\boldsymbol{\CalH_{Y_G}}=\CalH_{Y_G}^{\otimes n}$,
$\boldsymbol{\CalH_{Y}^{e}}=(\CalH_{Y}^{e})^{\otimes n}$ and $\boldsymbol{\CalD_{S}}= \CalD_{S}^{\otimes n}$. For $S \subsetneq [4]$ we define the tilting map $\CalT^{S}_{ d_{S}^{n}, \eta} : \boldsymbol{\CalH_{Y_G}}\rightarrow
\boldsymbol{\CalH_{Y_G}} \oplus \left(\boldsymbol{\CalH_{Y_G}}  \otimes   \boldsymbol{\CalD_{S}}\right)
$ as
\begin{eqnarray}
\CalT^{S}_{ d_{S}^{n}, \eta}(\ket{h}) \define \frac{1}{\sqrt{\Omega(S,\eta)} } \left(\ket{h} +\eta^{ |S| } \ket{h} \otimes  \ket{d_{S}^{n}}\right), \label{Eqn:3CQICTiltingMaps}
\end{eqnarray}
where $\Omega(S,\eta) \define 1+ \eta^{2 |S|}$. $\CalT^{S}_{ d_{S}^{n}, \eta}$ is an isometric embedding of $\boldsymbol{\CalH_{Y_G}}$  into $
\boldsymbol{\CalH_{Y_G}} \oplus \left(\boldsymbol{\CalH_{Y_G}}  \otimes   \boldsymbol{\CalD_{S}}\right)$.
Next, we construct the decoding POVM. For $S \subseteq [4]$, we consider 
\begin{eqnarray}
   G^{S}_{\ulinez^{n}}\define \pi_{z_{S^{c}}^{n}} \pi_{\ulinez^{n}}\pi_{z_{S^{c}}^{n}}, \label{Eqn:3CQBCOriginalPOVM}
\end{eqnarray}
where $\pi_{\ulinez^{n}}$ is the C-Typ-Proj with respect to the state $\otimes_{t=1}^{n}\xi_{z_{1t}z_{2t}z_{3t}z_{4t}}$
and $\pi_{z_{S^c}^{n} }$ is C-Typ-Proj with respect to the state $\otimes_{t=1}^{n}\left(\xi_{z_{st}:s \in S^c}= \otimes_{t=1}^{n}\sum_{z^n_S} p_{Z_S | Z_{S^c}}^n(z^n_{S}|z^n_{S^c}) \xi_{z_{1t}z_{2t}z_{3t}z_{4t}}\right)$.

By Gelfand–Naimark’s Thm.~\cite[Thm.~3.7]{BkHolevo_2019}, there exists orthogonal projector $\olineG^{S}_{\ulinez^{n}}\in \CalP(\boldsymbol{\CalH_{Y_G}})$ that yields identical measurement statistic on states in $\boldsymbol{\CalH_{Y_G}}$ that $G^{S}_{\ulinez^{n}}$ gives on states in $\boldsymbol{\CalH_{Y}}$, for $S \subseteq [4]$. Let 
\begin{eqnarray}
\olineB_{\ulinez^{n}}^{S} \define I_{\boldsymbol{\CalH_{Y_G}}}-\olineG^{S}_{\ulinez^{n}},\mbox{ for } S \subseteq [4] \label{Eqn:3CQBCComplementprojector} 
\end{eqnarray}
be the complement projector. Now, consider
\begin{eqnarray}
 \beta^{S}_{\ulinez^{n},\ulined^{n}} \define \CalT_{d^{n}_{S^c},\eta}^{S^c} ( \olineB_{\ulinez^{n}}^{S}), \mbox{ for } S \subsetneq [4], \mbox{ and } \beta^{[4]}_{\ulinez^{n},\ulined^{n}} \define  \olineB_{\ulinez^{n}}^{[4]}, \label{Eqn3CQBC:Tiltingofprojectors}
\end{eqnarray}
where $\beta^{S}_{\ulinez^{n},\ulined^{n}}$ denotes the tilted projector along direction $d_{S^c}^{n}$ for $S \subsetneq [4]$ and $\beta^{[4]}_{\ulinez^{n},\ulined^{n}}$ denotes the untilted projector.
Next, we define $\beta^{*}_{\ulinez^{n},\ulined^{n}}$ as the projector in $\boldsymbol{\CalH_{Y}^{e}}$ whose support is the union of the supports of $\beta^{S}_{\ulinez^{n},d_{S}^{n}}$ for all $S \subseteq [4]$.
Let $\pi_{\boldsymbol{\CalH_{Y_G}}}$ be the orthogonal projector in $\boldsymbol{\CalH_{Y}^{e}}$ onto $\boldsymbol{\CalH_{Y_G}}$. 
We define the square root measurement  \cite{BkWilde_2017,BkHolevo_2019}, to decode $\{\mu_{\ulinem} : \ulinem\}$, as 
\begin{eqnarray}
\mu_{\ulinem} \define \left(\sum_{\uline{\widehat{m}}} \gamma^{*}_{(\ulinez^{n},\ulined^{n})(\widehat{\ulinem})}\right)^{-\frac{1}{2}}
\gamma^{*}_{(\ulinez^{n},\ulined^{n})(\ulinem)} \left(\sum_{\uline{\widehat{m}}} \gamma^{*}_{(\ulinez^{n},\ulined^{n})(\widehat{\ulinem})}\right)^{-\frac{1}{2}}, \mbox{ and } \mu_{-1} \define I - \sum_{\ulinem} \mu_{\ulinem}, \nonumber  \\ \mbox{where } 
\gamma^{*}_{(\ulinez^{n},\ulined^{n})(\ulinem)} \define \left(I_{\boldsymbol{\CalH_{Y}^{e}}}-  \beta^{*}_{(\ulinez^{n},\ulined^{n})(\ulinem)}\right) \pi_{\boldsymbol{\CalH_{Y_G}} }\left(I_{\boldsymbol{\CalH_{Y}^{e}}}-  \beta^{*}_{(\ulinez^{n},\ulined^{n})(\ulinem)}\right). \label{Eqn:3CQBCgammadef}
\end{eqnarray}

\noindent \textbf{Distribution of the random code:} As is standard in information theory, to derive an upper bound on the average error probability, we average over the ensemble of all codes.
The distribution of the random code is analogous to \eqref{Eqn:DistStepI}, that is employed in step I analysis involving $5$ codes. This is naturally extended to the scenarios involving $9$ codes, so in essence the unstructured codebooks are picked independently, and the coset codes are picked such that the larger of the codes in the same field contains the smaller one. For more elaboration, please look at the distribution \eqref{Eqn:DistStepI} in step I.

\noindent \textbf{Error Analysis:}
Based on the exposition provided in Remark \ref{Rem:BinningandChannelCodingAnalysis} ,we recognize that our approach enables us to separate the binning analysis and the channel coding analysis, thereby breaking the independence. We therefore assume no binning in the following analysis. 
As is standard in information theory, we derive an upper-bound on the probability of error of a good code by averaging the error probability over an ensemble of codes. Towards that end , we first introduce an alternate `proxy' state. Specifically, we substitute the state $\xi_{\ulinez^n}^{\otimes n} \otimes \ketbra{0}$ by a specific `tilted state' defined as 
\begin{eqnarray}
    &&\theta_{\ulinez^n, \ulined^n}^{\otimes n} \define \mathcal{T}_{\ulined^n, \eta} \{\xi_{\ulinez^n}^{\otimes n} \otimes \ketbra{0}\},  \nonumber 
\end{eqnarray} 
where the tilting map $\mathcal{T}_{\ulined^n, \eta}$ acts on each pure state in the mixture individually and is defined as 
\begin{eqnarray}
    && \CalT_{ \ulined^{n}, \eta}(\ket{h}) \define \frac{1}{\sqrt{\Omega(\eta)} } \left(\ket{h} + \sum_{S \subsetneq [4]} \eta^{ |S| } \ket{h} \otimes \ket{d_{S}^{n}}\right) \nonumber 
\end{eqnarray}
with $\Omega(\eta) \define 1+16 \eta^{2}+ 36 \eta^{4}+ 16 \eta^{6}$.
The effect of this substitution on the error probability can be suppressed by ensuring that the tilted state $\theta_{\ulinez^n, \ulined^n}^{\otimes n}$ is close to the state $\xi_{\ulinez^n}^{\otimes n} \otimes \ketbra{0}$ in the $\mathbb{L}_1-$norm, as established in the following Proposition.

\begin{proposition}
\label{Prop:3CQBCClosnessOFstates}
    For $n$ sufficiently large, we have 
\begin{eqnarray}
     \norm{\theta_{\ulinez^{n}, \ulined^{n}}^{\otimes n} - \left(\xi_{\ulinez^{n}}^{\otimes n} \otimes \ket{0}\bra{0}\right) }_{1} \leq 12 \eta. 
\end{eqnarray}
\end{proposition}
\begin{proof}
    The proof is analogous to the proof in Appendix \ref{App:3to1CQBCClosenessOfStates}.
\end{proof}

We decompose the average probability of error of the 4-user CQMAC (Fig.~\ref{Fig4CQMAC}) as 
\begin{eqnarray}
    &&\mathbf{P}(\ulinee,\mu)= \frac{1}{|\mathcal{\ulineM}|} \sum_{\ulinem}\tr \left\{ \left(I-\mu_{\ulinem}\right)  \left(\xi_{\ulinez}^n(\ulinem) \otimes \ketbra{0}\right)  \right) \leq T_1 + T_2, \nonumber 
\end{eqnarray}
where the inequality follows from using the inequality $\tr\left(\Delta \rho \right) \leq \tr\left(\Delta \sigma\right) + \frac{1}{2} \norm{\rho-\sigma}_1$, with $0 \leq \Lambda, \rho, \sigma \leq I$, which gives us 
\begin{eqnarray}
    && T_1 \define \frac{1}{|\mathcal{\ulineM}|} \sum_{\ulinem}\tr \left\{ \left(I-\mu_{\ulinem}\right)  \theta_{\ulinez^n(\ulinem), \ulined^n(\ulinem)}  \right), \mbox{ and } T_2 \define \frac{1}{|\mathcal{\ulineM}|} \sum_{\ulinem} \norm{\theta_{\ulinez^n(\ulinem), \ulined^n(\ulinem)}-\left(\xi_{\ulinez^n(\ulinem)} \otimes \ketbra{0}\right)}_1. \nonumber 
\end{eqnarray}
We analyze the first term $T_1$, using the Hayashi–Nagaoka inequality \cite{200307TIT_HayNag}. This allows us to decompose $T_1$ as
\begin{eqnarray}
    && T_1 = 2 \: T_{1.1} + 4 \sum_{S \subseteq [4]} T_{1.S}, \mbox{ where} \nonumber \\
    T_{1.1} &\define& \frac{1}{|\mathcal{\ulineM}|} \sum_{\ulinem}\tr \left\{  (I - \gamma^{*}_{\ulinez^n(\ulinem),\ulined^n(\ulinem)}) \theta_{\ulinez^n(\ulinem), \ulined^n(\ulinem)} \right), \mbox{ and}  \nonumber \\
     T_{1.S} &\define& \frac{1}{|\mathcal{\ulineM}|} \sum_{\ulinem} \sum_{\tilde{m}_S \neq m_S}\tr \left\{ \gamma^{*}_{z_S^n(\tilde{m}_S), d^n_S(\tilde{m}_S), z^n_{S^c}(m_{S^{c}}),d_{S^{c}}^n(m_{S^c})}\theta_{\ulinez^n(\ulinem), \ulined^n(\ulinem)} \right).\nonumber
 \end{eqnarray}
The following propositions summarize all the rate constraints that result from bounding these error terms.
\begin{proposition}
\label{Prop:4CQMACFirsttermHay}
For any $\epsilon \in (0,1)$, and for all sufficiently small $\delta, \eta > 0$, and sufficiently large $n$, we have $\mathbb{E}[T_{1.1}] \leq \epsilon$.
\end{proposition}
\begin{proof}
    The proof is provided in Appendix \ref{App:4CQMACFirsttermHay}.
\end{proof}

\begin{proposition}
\label{Prop:4CQMACSectermHay}
For any $\epsilon \in (0,1)$, and for all sufficiently small $\delta, \eta > 0$, and sufficiently large $n$, we have $\mathbb{E}[T_{1.S}] \leq \epsilon,$ if
\begin{eqnarray}
\sum_{s \in S} \tilde{R}_s < I(Y;Z_S|Z_{S^c}), \mbox{ for }  S \subseteq [4]. \nonumber  
\end{eqnarray}
\end{proposition}
\begin{proof}
    The proof is provided in Appendix \ref{App:4CQMACSectermHay}.
\end{proof}
Next, from Proposition \ref{Prop:3CQBCClosnessOFstates}, we have for all $\epsilon \in (0,1)$ and for $\eta = \epsilon^{\frac{1}{5}}$, $\mathbb{E}[T_2] \leq 12 \: \epsilon^{\frac{1}{5}}$. The choice of $\eta$ ensures that $\mathbb{E}[T_{1.1}]$ can be made arbitrarily small, as shown in the proof of Proposition \ref{Prop:4CQMACFirsttermHay}. 
Based on Remark \ref{Rem:BinningandChannelCodingAnalysis} and our exposition in the proof in Sec.~\ref{Sec:3to1CQBC}, specifically the separation of binning and channel coding, this proof is complete.
\end{proof}
\subsection{Step III: Enlarging \texorpdfstring{$\alpha_{S}$}{αS} via Unstructured IID codes to \texorpdfstring{$\alpha_{US}$}{αUS}}
\label{SubSec:Step2}
\begin{figure}
    \centering
    \includegraphics[width=4.5in]{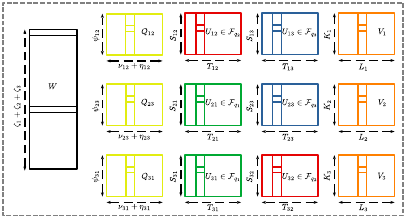}
    \caption{{$13$ codes employed in the proof of Thm.~\ref{Thm:Step2}. The $U_{ji} : ji \in \dbrackthree$ are coset codes built over finite fields. Codes with the same color are built over the same finite field and the smaller of the two is a sub-coset of the larger. The black, yellow and the orange codes are built over auxiliary finite sets  $\CalW : ji \in \dbrackthree, \CalQ_{ji} : ji \in \{12, 23, 31\},$ and $ \CalV_{j} : j \in [3] $ respectively using the conventional IID random code structure.}}
    \label{Fig:3CQBCStepIICodeStructure}
\end{figure}

The inner bound proven in Thm.~\ref{Thm:3CQBCStepIInnerBound} only enables Rxs to decode a bivariate component of the interference. As Marton's common codebook \cite{197905TIT_Mar} strategy proves, there is import to decoding a univariate part of the other user's codewords. In other words, by combining both unstructured codebook based strategy and coset code strategy, we can derive an inner bound that subsumes all current known inner bounds for the $3-$CQBC. The following inner bound is obtained by combining the conventional unstructured IID code based strategy due to Marton \cite{197905TIT_Mar} and the above coset code based strategy (Thm.~\ref{Thm:3CQBCStepIInnerBound}). \textcolor{black}{A proof of this inner bound is involved, but essentially is a generalization of the proof provided for Thm.~\ref{Thm:3CQBCStepIInnerBound}. In regards to the code structure, refer to Fig.~\ref{Fig:3CQBCStepIICodeStructure}.} 
Throughout the statement of Thm.~\ref{Thm:Step2}, we adopt the following notation.

\begin{Notation}
As stated earlier, $i,j,k$ will denote distinct indices in $[3]$, hence $\{i,j,k\}=[3]$ and we let $\dbrackthree \define \{(1,2),(1,3),(2,1),(2,3),(3,1),(3,2)\}$. In addition, $\alpha,\beta,\iota$ will denote distinct indices in $\{ 1,2,3\}$ satisfying $(\alpha,\beta,\iota) \in \{ (1,2,3), (2,3,1), (3,1,2) \}$. $\ulineQ \define (Q_{12},Q_{23},Q_{31})$  and similarly $\ulineCalQ= \CalQ_{12} \times \CalQ_{23} \times \CalQ_{31}$, $\ulineq = (q_{12},q_{23},q_{31})$, and analogously the rest of the objects.
\end{Notation}
 \begin{theorem}
 \label{Thm:Step2}
{ Refer to the above notation. Let $\hat{\alpha}_{US} \in [0,\infty)^{4}$ be the set of all rate-cost quadruples $(R_{1},R_{2},R_{3},\tau) \in
[0,\infty)^{4}$ for which there exists (i) finite sets $\CalW,\CalQ_{12}, \CalQ_{23}, \CalQ_{31}, \CalV_{1}, \CalV_{2}, \CalV_{3}$, (ii) finite fields $\SemiPrivateRVSet_{ij}=\CalF_{\upsilon_{j}}$ for each $ij \in \dbrackthree$, (ii) a PMF $p_{W\ulineQ\ulineU~\!\!\ulineV X}=p_{WQ_{12}Q_{23}Q_{31}U_{1*}U_{2*}U_{3*}V_{1}V_{2}V_{3} X}$ on $\ulineCalU \times \ulineQ \times \ulineCalV\times\CalX$, (iii) nonnegative numbers $\zeta_{j},K_{j},L_{j}: j \in [3]$ $\eta_{\alpha\beta},\nu_{\alpha\beta},\psi_{\alpha\beta}$ for $(\alpha,\beta,\iota) \in \{ (1,2,3), (2,3,1), (3,1,2) \}, S_{ij},T_{ij}:ij \in \dbrackthree$, such that $R_{j}=\zeta_{j}+\sum_{\alpha=1}^{3}\left( \nu_{\alpha\beta}+\eta_{\iota\alpha}\right)\mathds{1}_{\{\alpha=j\}}+T_{ji}+T_{jk}+L_{j}$ for $j \in [3]$, $\mathbb{E}[\kappa(X)] \leq \tau$, }
\begin{eqnarray}
&&S_{A}+M_{B}+K_{C}+\psi_{D} >\Theta(A,B,C,D), \mbox{ where}  \nonumber   \\
\Theta(A,B,C,D)&\define& 
\max_{\substack{(\theta_{j}:j \in B) \in \underset{j \in B}{\prod} \fieldpij}}\left\{
 \begin{array}{c}\sum_{a \in A}\log |\mathcal{U}_{a}| + \sum_{j \in B}\log \upsilon_{j}+\sum_{c \in C} H(V_{c}|W)+\sum_{d \in D}H(Q_{D}|W) - \nonumber \\H(U_{A},U_{ij}\oplus \theta_{j}U_{kj}\!:j \in B,V_{C}|W)\end{array}\right\} \nonumber
\end{eqnarray}
for all $A \subseteq \left\{12,13,21,23,31,32\right\}, B \subseteq \left\{ 1,2,3 \right\}, C \subseteq \left\{ 1,2,3 \right\}, { D \subseteq \left\{ 12, 23, 31\right\}}$,  that satisfy $A \cap A(B) = \phi$, where $A({B}) = \cup_{j \in B}\{ ij,kj\}$, $U_{A} = (U_{a}:a \in A)$, $V_{C}=(V_{c}:c \in C)$, $Q_{D}=(Q_{d}:d \in D)$, $S_{A} = \sum _{a \in A}S_{a}, M_{B}\define \sum_{j \in B} \max\{ S_{ij}+T_{ij},S_{kj}+T_{kj}\}, K_{C} = \sum_{c \in C}K_{c}$, $\psi_{D} = \sum _{d \in D}\psi_{d}$ and
\begin{eqnarray}
\label{Eqn:Chnlbnd1}
S_{\mathcal{A}_{j}} + T_{\mathcal{A}_{j}} + \psi_{\CalC_{j}} + \nu_{\CalC_{j}}+\eta_{\CalC_{j}} &<& \sum_{a \in \mathcal{A}_{j}} \log |\mathcal{U}_{a}| + H(Q_{\CalC_{j}}|Q_{\CalC_{j}^{C}},W) \nonumber \\&& -H(U_{\mathcal{A}_{j}},Q_{\CalC_{j}}|U_{\mathcal{A}_{j}^{c}},Q_{\CalC_{j}^{C}},U_{ij} \oplus U_{kj}, V_{j}, Y_{j}, W),
\\
\label{Eqn:Chnlbnd2}
S_{\mathcal{A}_{j}} + T_{\mathcal{A}_{j}} + \psi_{\CalC_{j}} + \nu_{\CalC_{j}}+\eta_{\CalC_{j}} + S_{ij} + T_{ij} &<& \sum_{a \in \mathcal{A}_{j}} \log |\mathcal{U}_{a}| + \log \upsilon_{j} + H(Q_{\CalC_{j}}|Q_{\CalC_{j}^{C}},W) \nonumber \\ &&- H(U_{\mathcal{A}_{j}}, Q_{\CalC_{j}}, U_{ij} \oplus U_{kj}|U_{\mathcal{A}_{j}^{c}},Q_{\CalC_{j}^{C}},V_{j},Y_{j},W),  \\
\label{Eqn:Chnlbnd4}
S_{\mathcal{A}_{j}}+T_{\mathcal{A}_{j}}+\psi_{\CalC_{j}}+ \nu_{\CalC_{j}}+\eta_{\CalC_{j}}+S_{kj}+T_{kj} &<& \sum_{a \in \mathcal{A}_{j}}\!\!\log |\mathcal{U}_{a}| + \log \upsilon_{j}
+H(Q_{\CalC_{j}}|Q_{\CalC_{j}^{C}},W) \nonumber \\ &&- H(U_{\mathcal{A}_{j}}\!,\!Q_{\CalC_{j}},\!U_{ij}\!\oplus\!
U_{kj}|U_{\mathcal{A}_{j}^{c}},Q_{\CalC_{j}^{C}},V_{j},Y_{j},W),
\\
\label{Eqn:Chnlbnd6}
S_{\mathcal{A}_{j}} +T_{\mathcal{A}_{j}} +\psi_{\CalC_{j}} + \nu_{\CalC_{j}}+\eta_{\CalC_{j}} + K_{j} +L_{j} &<& \sum_{a \in \mathcal{A}_{j}} \log |\mathcal{U}_{a}|+ H(V_{j}|W) + H(Q_{\CalC_{j}}|Q_{\CalC_{j}^{C}},W) \nonumber \\
 &&-H(U_{\mathcal{A}_{j}}, Q_{\CalC_{j}},\!V_{j}|U_{\mathcal{A}_{j}^{c}},Q_{\CalC_{j}^{C}},U_{ij} \oplus
U_{kj},Y_{j},W),
\\
\label{Eqn:Chnlbnd8}
S_{\mathcal{A}_{j}} + T_{\mathcal{A}_{j}} +\psi_{\CalC_{j}} + \nu_{\CalC_{j}}+\eta_{\CalC_{j}}+K_{j}+L_{j}+S_{ij}+T_{ij} &<& \sum_{a \in \mathcal{A}_{j}} \log |\mathcal{U}_{a}| + \log \upsilon_{j} +H(V_{j}|W) +H(Q_{\CalC_{j}}|Q_{\CalC_{j}^{C}},W) \nonumber
\\
\label{Eqn:Chnlbnd9}
&&-H(U_{\mathcal{A}_{j}}, Q_{\CalC_{j}}, V_{j},U_{ij}\oplus U_{kj}|U_{\mathcal{A}_{j}^{c}}, Q_{\CalC_{j}^{C}}, Y_{j},W)
\\
\label{Eqn:Chnlbnd10}
S_{\mathcal{A}_{j}} + T_{\mathcal{A}_{j}} + \psi_{\CalC_{j}} + \nu_{\CalC_{j}}+\eta_{\CalC_{j}} + K_{j}+L_{j}+S_{kj}+T_{kj} &<& \sum_{a \in \mathcal{A}_{j}}\log |\mathcal{U}_{a}| + \log  \upsilon_{j}+H(V_{j}|W)
+H(Q_{\CalC_{j}}|Q_{\CalC_{j}^{C}},W)\nonumber\\
\label{Eqn:Chnlbnd11}
&&- H(U_{\mathcal{A}_{j}}, Q_{\CalC_{j}},V_{j},U_{ij}\oplus
U_{kj}|U_{\mathcal{A}_{j}^{c}},Q_{\CalC_{j}^{C}},Y_{j},W),
\\
\label{Eqn:Chnlbnd12}
\zeta+S_{\CalA_{j}}+T_{\CalA_{j}}+K_{j}+L_{j}+{\psi_{\mathfrak{U}_{j}}}+{\nu_{\mathfrak{U}_{j}}} + {\eta_{\mathfrak{U}_{j}} }&<& \sum_{a \in \mathcal{A}_{j}} \log |\mathcal{U}_{a}|+H(Q_{ji},Q_{jk}|W)+H(V_{j}|W) \nonumber \\
&&-H(U_{\mathcal{A}_{j}},Q_{j*},W,V_{j}|U_{\mathcal{A}_{j}^{c}},Y_{j},U_{ij}\oplus U_{kj}),
\\
\label{Eqn:Chnlbnd13}
\zeta+S_{\CalA_{j}}+T_{\CalA_{j}} + K_{j} + L_{j} + {\psi_{\mathfrak{U}_{j}}}+ {\nu_{\mathfrak{U}_{j}}}+ {\eta_{\mathfrak{U}_{j}}}+ S_{ij} + T_{ij} &<&  \log \upsilon_{j}+\sum_{a \in \mathcal{A}_{j}} \log |\mathcal{U}_{a}|+H(Q_{ji},Q_{jk}|W)+H(V_{j}|W) \nonumber \\
&&-H(U_{\mathcal{A}_{j}}, Q_{j*}, W, V_{j}, U_{ij}\oplus U_{kj}|U_{\mathcal{A}_{j}^{c}},Y_{j}),
\\
\label{Eqn:Chnlbnd14}
\zeta+S_{\CalA_{j}}+T_{\CalA_{j}} + K_{j} + L_{j} + {\psi_{\mathfrak{U}_{j}}} + {\nu_{\mathfrak{U}_{j}}} + {\eta_{\mathfrak{U}_{j}}}+ S_{kj} + T_{kj} &<&  \log \upsilon_{j}+\sum_{a \in \mathcal{A}_{j}} \log |\mathcal{U}_{a}|+H(Q_{ji}, Q_{jk}|W)+H(V_{j}|W) \nonumber \\
&&-H(U_{\mathcal{A}_{j}},Q_{j*}, W, V_{j}, U_{ij}\oplus U_{kj}|U_{\mathcal{A}_{j}^{c}},Y_{j}),
\end{eqnarray}
for every $\mathcal{A}_{j} \subseteq  \left\{ ji,jk\right\},\CalC_{j} \subseteq \mathfrak{U}_{j}\define\left\{ \alpha\beta,\iota\alpha : \alpha=j\right\}$ with
distinct indices $i,j,k$ in $\left\{ 1,2,3 \right\}$, where
$\zeta \define \zeta_{1}+\zeta_{2}+\zeta_{3}$, $S_{\mathcal{A}_{j}} \define \sum_{a \in \mathcal{A}_{j}}S_{a}$,
$T_{\mathcal{A}_{j}} \define \sum_{a \in \mathcal{A}_{j}}T_{a}$, $\psi_{\CalC_{j}} = \sum _{c \in \CalC_{j}}\psi_{c}$, {$\nu_{\CalC_{j}} = \sum _{c \in \CalC_{j}}\nu_{c}$, $\eta_{\CalC_{j}} = \sum _{c \in \CalC_{j}}\eta_{c}$},
$U_{\mathcal{A}_{j}} = (U_{a}:a \in \mathcal{A}_{j})$, $Q_{\mathcal{C}_{j}} = (Q_{c}:c \in \mathcal{C}_{j})$ with all the information quantities are evaluated with respect to state
\begin{eqnarray}
\label{Eqn:StageIITestChnl}
\Phi^{W\ulineQ\!\!~\ulineU\!\!~\ulineU^{\oplus}\!\!~\ulineV X\!\!~\ulineY} \define
\sum_{\substack{w,\ulineq,\ulineu,\ulinev,x\\u_{1}^{\oplus},u_{2}^{\oplus},u_{3}^{\oplus} }}\!\!\!\!\!\!p_{W\ulineQ~\!\!\ulineU~\!\!\ulineV X}(w,\ulineq,\ulineu,\ulinev,x)\mathds{1}{\left\{\substack{u_{ij}\oplus u_{kj}=u_{j}^{\oplus}:j\in[3]}\right\}}\!\ketbra{w~\ulineq~\ulineu~  u_{1}^{\oplus}~ u_{2}^{\oplus}~ u_{3}^{\oplus}~ \ulinev~ x} \!\otimes\! \rho_{x}.\nonumber
\end{eqnarray}
Let $\alpha_{US}$ denote the convex closure of $\hat{\alpha}_{US}$. Then $\alpha_{US} \subseteq \ScrC(\tau)$ is an achievable rate region.
\end{theorem}

\begin{remark}
\label{Rem:afterstep3}
By choosing $U_{ji} = \phi,$ for $(i,j) \in \dbrackthree,$ we recover the best known inner bound achievable using unstructured IID codes.
\end{remark}

{\begin{remark}
\label{Rem:Gen3QBC}
    The above inner bound is characterized for a $3-$CQBC. We can extend this via the standard technique of regularization \cite[Sec.~20.1, Eq.~20.7]{BkWilde_2017} to characterize an inner bound to the bit communicating capacity region of a general $3-$user quantum broadcast channel. We omit a mathematical characterization of this inner bound in the interest of brevity.
\end{remark}}

\section*{Declaration}
This manuscript requires \textbf{NO} supporting or associated data. Every word, phrase and sentence published in this article is the \textbf{sole creation of the authors}. Explicitly stated, \textbf{none} of the generative AI chatbots such as ChatGPT or Deepseek have been utilized in generating \textbf{any} of the words, phrases or sentences.

\appendices
\section{Proof of Proposition \ref{Prop:2CQBC1} and \ref{Prop:2CQBC2}}
\subsection{Proof of Proposition \ref{Prop:2CQBC1}}
\label{App:2CQBC1}
\textit{\underline{Analysis of $T_{j.2.1}$:}} We evaluate the expectation of $T_{j.2.1}$ over the codebook generation distribution. We obtain
\begin{eqnarray}
    \mathbb{E}[T_{j.2.1}] &=& \frac{1}{2^{n(R_1 + R_2)}} \frac{1}{2^{n(K_1+ K_2)}} \sum_{m_1} \sum_{m_2} \sum_{b_1} \sum_{b_2} \sum_{v_1^n} \sum_{v_2^n} q_{V_1}^n(v_1^n) q_{V_2}^n(v_2^n) r_{V_1V_2}^n(v_1^n,v_2^n) \tr\left\{ \left(I^{Y_j} - \pi \: \pi_{v_j^n} \: \pi\right) \rho_{f^n(v_1^n,v_2^n)}^{Y_j}\right\} \nonumber \\ 
    &\overset{(a)}{=}& \frac{1}{2^{n(R_1 + R_2)}} \frac{1}{2^{n(K_1+ K_2)}} \sum_{m_1} \sum_{m_2} \sum_{b_1} \sum_{b_2} \sum_{v_1^n} \sum_{v_2^n} p_{V_1V_2}^n(v_1^n,v_2^n) \tr\left\{ \left(I^{Y_j} - \pi \: \pi_{v_j^n} \: \pi\right) \rho_{f^n(v_1^n,v_2^n)}^{Y_j}\right\} \nonumber \\
    &\overset{(b)}{=}& \frac{1}{2^{n(R_j + K_j)}} \sum_{m_j} \sum_{b_j} \sum_{v_j^n} p_{V_j}^n(v_j^n)  \tr\left\{\left(I^{Y_j} - \pi \: \pi_{v_j^n} \: \pi\right) \rho_{v_j^n}^{Y_j}\right\} \nonumber 
    \\
    &=& \frac{1}{2^{n(R_j + K_j)}} \sum_{m_j} \sum_{b_j} \sum_{v_j^n \notin T_{\delta}^n(p_{V_j})}  p_{V_j}^n(v_j^n)  + \frac{1}{2^{n(R_j + K_j)}} \sum_{m_j}  \sum_{b_j} \sum_{v_j^n \in T_{\delta}^n(p_{V_j})}p_{V_j}^n(v_j^n)  \tr\left\{ \left(I^{Y_j} - \pi \: \pi_{v_j^n} \: \pi\right) \rho_{v_j^n}^{Y_j}\right\}  \nonumber \\
    &\leq& \epsilon + \frac{1}{2^{n(R_j + K_j)}}  \sum_{m_j} \sum_{b_j} \sum_{v_j^n \in T_{\delta}^n(p_{V_j})}  p_{V_j}^n(v_j^n) \tr\left\{ \left(I^{Y_j} - \pi \: \pi_{v_j^n} \: \pi\right) \rho_{v_j^n}^{Y_j}\right\} \nonumber \\
    &\overset{(c)}{\leq}& 2 \epsilon + 2 \sqrt{\epsilon},\nonumber 
\end{eqnarray}
where (a) follows from \eqref{Eqn:Dist}, (b) follows from the definition of $\rho_{v_j^n}^{Y_j}$, and (c) follows from 
\begin{eqnarray}
 \tr\left\{ \pi \pi_{v_j^n} \pi \rho_{v_j^n}^{Y_j}\right\} &=& \tr\left\{\pi_{v_j^n} \pi \rho_{v_j^n}^{Y_j} \pi \right\} \nonumber \\
 &\geq& \tr\left\{\pi_{v_j^n} \rho_{v_j^n}^{Y_j}\right\} - \norm{\pi \rho_{v_j^n}^{Y_j} \pi - \rho_{v_j^n}^{Y_j}}_1 \nonumber \\
 &\geq& 1- \epsilon -2 \sqrt{\epsilon},   \nonumber 
\end{eqnarray}
where the first equality follows from the cyclicity of the trace, first inequality follows from using $\tr(\Delta \rho) \geq \tr(\lambda \sigma ) - \norm{\rho - \sigma}_1$ for $0 \leq \Delta, \rho, \sigma \leq I$ and second inequality follows from (i) the typical property $\tr\left\{ \pi_{v_j^n} \rho_{v_j^n}^{Y_j}\right\} \geq 1- \epsilon$, for $v_j^n \in T_{\delta}^n(p_{V_j})$ and from (ii) the gentle measurement operator \cite{BkWilde_2017}.
\subsection{Proof of Proposition \ref{Prop:2CQBC2}}
\label{App:2CQBC2}
\noindent\textit{\underline{Analysis of $T_{j.2.2}$:}} We evaluate the expectation of $T_{j.2.2}$ over the codebook generation distribution. We obtain 
\begin{eqnarray}
\mathbb{E}[T_{j.2.2}] &=& \frac{1}{2^{n(R_1 + R_2)}} \frac{1}{2^{n(K_1 + K_2)}} \sum_{m_1} \sum_{m_2} \sum_{b_1} \sum_{b_2} \sum_{\tilde{m}_j \neq m_j} \sum_{\tilde{b}_j} \sum_{\tilde{v}_j^n} \sum_{v_1^n} \sum_{v_2^n} 
q_{V_j}^n(\tilde{v}_j^n) \: q_{V_1}^n(v_1^n) \: q_{V_2}^n(v_2^n) \: r_{V_1V_2}^n(v_1^n,v_2^n) \nonumber \\ 
&& \tr\left\{  \pi  \pi_{\tilde{v}_j^n}  \pi \rho_{f^n(v_1^n,v_2^n)}^{Y_j}\right\} \nonumber \\
&\overset{(a)}{=}& \frac{1}{2^{n(R_1 + R_2)}} \frac{1}{2^{n(K_1 + K_2)}} \sum_{m_1} \sum_{m_2} \sum_{b_1} \sum_{b_2} \sum_{\tilde{m}_j \neq m_j} \sum_{\tilde{b}_j}  \sum_{\tilde{v}_j^n} \sum_{v_1^n} \sum_{v_2^n} 
q_{V_j}^n(\tilde{v}_j^n) \: p_{V_1V_2}^n(v_1^n,v_2^n) \: \tr\left\{  \pi  \pi_{\tilde{v}_j^n}  \pi \rho_{f^n(v_1^n,v_2^n)}^{Y_j}\right\} \nonumber \\
&\overset{(b)}{=}& \sum_{\tilde{m}_j \neq m_j} \sum_{\tilde{b}_j} \sum_{\tilde{v}_j^n \in T_{\delta}^n(p_{V_j})} 
q_{V_j}^n(\tilde{v}_j^n) \: \tr\left\{  \pi  \pi_{\tilde{v}_j^n}  \pi \left( \rho^{Y_j} \right) ^{\otimes n }\right\} \nonumber \\
&\overset{(c)}{\leq}& \sum_{\tilde{m}_j } \sum_{\tilde{b}_j}  \sum_{\tilde{v}_j^n \in T_{\delta}^n(p_{V_j})}  2^{-n(D(p_{V_j} || q_{V_j}) - \delta_j)}
\: p_{V_j}^n(\tilde{v}_j^n) \tr\left\{ \pi_{\tilde{v}_j^n} \: \pi \left( \rho^{Y_j} \right) ^{\otimes n } \pi \right\} \label{Eqn:Divergenceterm} \\
&\overset{(d)}{\leq}&  2^{-n\left(I(V_j;Y_j) + D(p_{V_j} || q_{V_j}) - 2 \delta - \delta_j - \left(R_j + K_j \right)\right)}, \nonumber 
\end{eqnarray}
where (a) follows from \eqref{Eqn:Dist}, (b) follows from $\left(\rho^{Y_j}\right) ^{\otimes n } \define \otimes_{t=1}^n \left(\sum_{v_{1_t},v_{2_t}} p_{V_1V_2}(v_{1_t},v_{2_t})  \rho_{f(v_{1_t},v_{2_t})}^{Y_j}\right)$ and from $\pi_{\tilde{v}_j^n}=0,$ for $\tilde{v}_j^n \notin T_{\delta}^n(p_{V_j})$,
(c) follows from (i) the cyclicity of the trace and from (ii) 
\begin{eqnarray}
    q_{V_j}^n(\Tilde{v}_j^n) \leq 2^{-n\left( D(p_{V_j} || q_{V_j}) - \delta_j \right)} \: p_{V_j}^n(\Tilde{v}_j^n), \mbox{ for }  \Tilde{v}_j^n \in \mathcal{T}_{\delta}^n(p_{V_j}) \nonumber 
\end{eqnarray}
(d) follows from  
\begin{eqnarray}
 \tr\left\{ \pi_{\tilde{v}_j^n} \pi \left( \rho^{Y_j}\right)^{\otimes n } \pi \right\} &\leq& 2^{-n(H(Y_j) - \delta) }  
 \tr\left\{ \pi_{\tilde{v}_j^n} \pi \right\} \nonumber \\
 &\leq& 2^{-n(H(Y_j) - \delta) }  
 \tr\left\{ \pi_{\tilde{v}_j^n} \right\} \nonumber \\
 &\leq& 2^{-n(H(Y_j) - \delta) }   2^{n(H(Y_j|V_j) - \delta )}, \mbox{ for } \tilde{v}_j^n \in T_{\delta}^n(p_{V_j}). \nonumber 
\end{eqnarray}
Hence, $T_{j.2.2} \to 0$ as $n \to \infty$, if $R_j + K_j  < I(V_j;Y_j) + D(p_{V_j}||q_{V_j})$.

\noindent \textit{\underline{Analysis of $T_{j.2.3}$:}} Following the identical steps as used in bounding $\mathbb{E}[T_{j.2.2}]$, we obtain 
\begin{eqnarray}
  \mathbb{E}[T_{j.2.3}] \leq 2^{-n\left( I(V_j;Y_j) + D(p_{V_j}||q_{V_j}) - 2\delta - \delta_j - K_j\right)}. \nonumber  
\end{eqnarray}
Hence, $T_{j.2.3} \to 0$ as $n \to \infty$, if $K_j < I(V_j;Y_j) + D(p_{V_j}||q_{V_j})$.

\section{Proof of Proposition \ref{Prop:Dec1Cuffterm} and \ref{Prop:Dec2Cuffterm}}
\label{App:Dec1Cuffterm}
Consider the term $T_{1.1}$. Since $I - \Gamma^{Y_1}_{m_1,a} \leq I$, we have 
\begin{eqnarray}
T_{1.1} &=& \frac{1}{|\mathcal{\ulineM}|} \frac{1}{\upsilon^{s_{\sfb} + t_{\sfb}}} \sum_{\ulinem} \sum_{a} \sum_{a_2,a_3}  \sum_{b_1,b_2,b_3}
  \nonumber \\ 
  &&\left[ \begin{aligned}
&\mathds{1}\left\{A_2(m_{21})=a_2, A_3(m_{31})=a_3,B_1(m_1)=b_1,B_2(m_{22})=b_2,B_3(m_{33})=b_3\right\} -\\ &\frac{r_{U_2U_3\ulineV}^n\left(U_2^n(a_2,m_{21}),U_3^n(a_3,m_{31}),V_1^n(b_1,m_1),V_2^n(b_2,m_{22}),V_3^n(b_3,m_{33})\right)}{2^{n(S_2 + S_3 + K_1 + K_2 + K_3)}} 
\end{aligned} \right] \nonumber \\ 
&&\tr \left\{  \left( I - \Gamma^{Y_1}_{m_1,a} \right)  \left( \rho_{f^n\left(V_1^n(b_1,m_1), U_2^n(a_2,m_{21}), V_2^n(b_2,m_{22}), U_3^n(a_3, m_{31}), V_3^n(b_3,m_{33})\right)}^{Y_1} \otimes \ketbra{0}\right)  \right\} \nonumber \\
&\leq& \frac{1}{|\mathcal{\ulineM}|} \sum_{\ulinem} \sum_{a_2,a_3}  \sum_{b_1,b_2,b_3}
   \left[ \begin{aligned}
&\mathds{1}\left\{A_2(m_{21})=a_2, A_3(m_{31})=a_3,B_1(m_1)=b_1,B_2(m_{22})=b_2,B_3(m_{33})=b_3\right\} -\\ &\frac{r_{U_2U_3\ulineV}^n\left(U_2^n(a_2,m_{21}),U_3^n(a_3,m_{31}),V_1^n(b_1,m_1),V_2^n(b_2,m_{22}),V_3^n(b_3,m_{33})\right)}{2^{n(S_2 + S_3 + K_1 + K_2 + K_3)}} 
\end{aligned} \right] \nonumber
\end{eqnarray}
Now we evaluate the expectation of $T_{1.1}$ with respect to $\left(c_1(m_1),c_2(m_{21}), c_3(m_{31}),c_2(m_{22}),c_3(m_{33})\right)$. We obtain
\begin{eqnarray}
\mathbb{E}\left[T_{1.1}\right] &\leq& 
    \frac{1}{|\mathcal{\ulineM}|} \sum_{\ulinem} \sum_{a_2, a_3}  \sum_{b_1,b_2,b_3} \sum_{c_1(m_1)}  \sum_{c_2(m_{21})} \sum_{c_3(m_{31})} \sum_{c_2(m_{22})} \sum_{c_3(m_{33})}\nonumber \\ && p\left(\begin{aligned}
        &c_1(m_1)=\{v_1^n(b_1,m_1) : b_1 \in [2^{nK_1}]\},c_2(m_{21})=\{u_2^n(a_2,m_{21}) : a_2 \in \mathcal{F}_{\Prime}^{s_2}\}, \\&c_3(m_{31})=\{u_3^n(a_3,m_{31}) : a_3 \in \mathcal{F}_{\Prime}^{s_3}\}, \\ 
        &c_2(m_{22})=\{v_2^n(b_2,m_{22}) : b_2 \in [2^{nK_2}]\}, c_3(m_{33})=\{v_3^n(b_3,m_{33}) : b_3 \in [2^{nK_3}]\}\end{aligned}\right) \nonumber \\ 
 &&\left[ \begin{aligned}
     &\frac{r_{U_2U_3\ulineV}^n\left(u_2^n(a_2,m_{21}), u_3^n(a_3,m_{31}), v_1^n(b_1,m_1), v_2^n(b_2,m_{22}), v_3^n(b_3,m_{33})\right)}{\sum_{\tilde{a}_2, \tilde{a}_3}  \sum_{\tilde{b}_1, \tilde{b}_2, \tilde{b}_3}
r_{U_2U_3\ulineV}^n\left(u_2^n(\tilde{a}_2,m_{21}), u_3^n(\tilde{a}_3,m_{31}), v_1^n(\tilde{b}_1,m_1), v_2^n(\tilde{b}_2,m_{22}), v_3^n(\tilde{b}_3,m_{33}) \right)} \\ & - \frac{ r_{U_2U_3\ulineV}^n\left(u_2^n(a_2,m_{21}), u_3^n(a_3,m_{31}),v_1^n(b_1,m_1), v_2^n(b_2,m_{22}),v_3^n(b_3,m_{33})\right) }{2^{n(S_2 + S_3 + K_1 + K_2 + K_3)}} \end{aligned}
\right] \nonumber \\ 
 &\leq& \frac{1}{|\mathcal{\ulineM}|} \sum_{\ulinem} \sum_{a_2, a_3}    \sum_{b_1,b_2,b_3} \sum_{c_1(m_1)}  \sum_{c_2(m_{21})} \sum_{c_3(m_{31})} \sum_{c_2(m_{22})} \sum_{c_3(m_{33})} \nonumber \\
 &&p\left(\begin{aligned}
        &c_1(m_1)=\{v_1^n(b_1,m_1) : b_1 \in [2^{nK_1}]\},c_2(m_{21})=\{u_2^n(a_2,m_{21}) : a_2 \in \mathcal{F}_{\Prime}^{s_2}\}, \\&c_3(m_{31})=\{u_3^n(a_3,m_{31}) : a_3 \in \mathcal{F}_{\Prime}^{s_3}\}, 
        c_2(m_{22})=\{v_2^n(b_2,m_{22}) : b_2 \in [2^{nK_2}]\}, \\ &c_3(m_{33})=\{v_3^n(b_3,m_{33}) : b_3 \in [2^{nK_3}]\}\end{aligned}\right) \nonumber \\ 
 &&
 \frac{r_{U_2U_3\ulineV}^n\left(u_2^n(a_2,m_{21}), u_3^n(a_3,m_{31}), v_1^n(b_1,m_1), v_2^n(b_2,m_{22}), v_3^n(b_3,m_{33})\right)}{\sum_{\tilde{a}_2, \tilde{a}_3}  \sum_{\tilde{b}_1, \tilde{b}_2, \tilde{b}_3}
r_{U_2U_3\ulineV}^n\left(u_2^n(\tilde{a}_2,m_{21}), u_3^n(\tilde{a}_3,m_{31}), v_1^n(\tilde{b}_1,m_1), v_2^n(\tilde{b}_2,m_{22}), v_3^n(\tilde{b}_3,m_{33}) \right)} \nonumber \\ 
&& \left| \sum_{\tilde{a}_2, \tilde{a}_3} \sum_{\tilde{b}_1, \tilde{b}_2, \tilde{b}_3}  \frac{r_{U_2U_3\ulineV}^n\left(u_2^n(\tilde{a}_2,m_{21}),u_3^n(\tilde{a}_3,m_{31}),v_1^n(\tilde{b}_1,m_1),v_2^n(\tilde{b}_2, m_{22}), v_3^n(\tilde{b}_3, m_{33})\right)}{2^{n(S_2+S_3+K_1+K_2+K_3)}} -1 \right| \nonumber \\ 
 &=& \frac{1}{|\mathcal{\ulineM}|} \sum_{\ulinem} \sum_{c_1(m_1)}  \sum_{c_2(m_{21})} \sum_{c_3(m_{31})} \sum_{c_2(m_{22})} \sum_{c_3(m_{33})} \nonumber \\
 &&p\left(\begin{aligned}
        &c_1(m_1)=\{v_1^n(b_1,m_1) : b_1 \in [2^{nK_1}]\},c_2(m_{21})=\{u_2^n(a_2,m_{21}) : a_2 \in \mathcal{F}_{\Prime}^{s_2}\}, \\&c_3(m_{31})=\{u_3^n(a_3,m_{31}) : a_3 \in \mathcal{F}_{\Prime}^{s_3}\}, \\ 
        &c_2(m_{22})=\{v_2^n(b_2,m_{22}) : b_2 \in [2^{nK_2}]\}, c_3(m_{33})=\{v_3^n(b_3,m_{33}) : b_3 \in [2^{nK_3}]\}\end{aligned}\right) \nonumber \\ 
 && \left|  \sum_{\tilde{a}_2, \tilde{a}_3} \sum_{\tilde{b}_1, \tilde{b}_2, \tilde{b}_3}  \frac{r_{U_2U_3\ulineV}^n\left(u_2^n(\tilde{a}_2,m_{21}),u_3^n(\tilde{a}_3,m_{31}),v_1^n(\tilde{b}_1,m_1),v_2^n(\tilde{b}_2, m_{22}), v_3^n(\tilde{b}_3, m_{33})\right)}{2^{n(S_2+S_3+K_1+K_2+K_3)}} -1 \right| \nonumber \\ 
&=&
 \mathbb{E}\Bigg[\Big|  \sum_{\tilde{a}_2, \tilde{a}_3} \sum_{\tilde{b}_1, \tilde{b}_2, \tilde{b}_3}  
\frac{r_{U_2U_3\ulineV}^n\left(U_2^n(\tilde{a}_2,M_{21}),U_3^n(\tilde{a}_3,M_{31}),V_1^n(\tilde{b}_1,M_1),V_2^n(\tilde{b}_2, M_{22}), V_3^n(\tilde{b}_3, M_{33})\right)}{2^{n(S_2+S_3+K_1+K_2+K_3)}} -1 \Big| \Bigg]. \nonumber 
\end{eqnarray}
Now, using the `cloud mixing' lemma \cite{201311TIT_Cuf}, we have $\mathbb{E}\left[T_{1.1}\right] \leq \epsilon $ if
\begin{eqnarray}
  S_{\mathcal{A}} + K_{\mathcal{B}} &>& |\mathcal{A}|\log\left(\Prime\right) + \sum_{\beta \in \mathcal{B}} H(V_{\beta}) - H(U_{\mathcal{A}},V_{\mathcal{B}}), \nonumber \\
        \max\{S_2 + T_2, S_3 + T_3\} &>& \log(\Prime) - \min_{\theta \in \mathcal{F}_{\Prime} \setminus \{0\}} H(U_2 \oplus \theta U_3), \mbox{ and} \nonumber\\
 K_{\mathcal{B}} + \max\{S_2 + T_2, S_3 + T_3\} &>& \log(\Prime) + \sum_{\beta \in \mathcal{B}} H(V_{\beta}) - \min_{\theta \in \mathcal{F}_{\Prime} \setminus \{0\}} H(U_2 \oplus \theta U_3, V_{\mathcal{B}}), \nonumber
\end{eqnarray}
for $\mathcal{A} \subseteq \{2,3\}$, $ \mathcal{B} \subseteq \{1,2,3\}$, $S_{\mathcal{A}} \define \sum_{\alpha \in \mathcal{A}} S_{\alpha}$,  $U_{\mathcal{A}} \define (U_{\alpha}: \alpha \in \mathcal{A})$, $K_{\mathcal{B}} \define \sum_{\beta \in \mathcal{B}} K_{\beta}$ and $V_{\mathcal{B}} \define (V_{\beta} : \beta \in \mathcal{B})$.

\section{ Proof of Proposition \ref{Prop:3to1CQBCclosnessofstates}: Closeness of the states}
\label{App:3to1CQBCClosenessOfStates}
We mimic the steps in \cite{202103SAD_Sen} in this proof.
For any $\ket{h} \in \boldsymbol{\CalH_{Y_1}^G}$, we have 
\begin{eqnarray}
    \norm{\CalT_{d_1^n,d_2^n,\eta}(\ketbra{h})-\ketbra{h}}_1 \leq 2 \norm{\CalT_{d_1^n,d_2^n,\eta}(\ket{h})-\ket{h}}_2. \nonumber 
\end{eqnarray}
Observe that
\begin{eqnarray}
    \norm{\CalT_{d_1^n,d_2^n,\eta}(\ket{h})-\ket{h}}_2^2 &=& \left( \CalT_{d_1^n,d_2^n,\eta}(\bra{h}) - \bra{h} \right) \left( \CalT_{d_1^n,d_2^n,\eta}(\ket{h}) - \ket{h} \right) \nonumber \\
    &=& <\CalT_{d_1^n,d_2^n,\eta}(h),\CalT_{d_1^n,d_2^n,\eta}(h)>-<\CalT_{d_1^n,d_2^n,\eta}(h),h>-<h,\CalT_{d_1^n,d_2^n,\eta}(h)>+<h,h> \nonumber \\
    &\overset{(a)}{=}& 2-\frac{2}{\sqrt{1 + 2 \eta^2}} \nonumber \\
    & \overset{(b)}{\leq}& 2- 2 e^{-2\eta^2} \nonumber \\
    & \overset{(c)}{\leq}& 4 \eta^2 \nonumber 
\end{eqnarray}
where (a) follows from (i) $\CalT_{d_1^n,d_2^n,\eta}$ being an isometry and (ii) $<\CalT_{d_1^n,d_2^n,\eta}(h),h>=<h,\CalT_{d_1^n,d_2^n,\eta}(h)>= \frac{1}{\sqrt{1 + 2 \eta^2}}$, (b) follows from $\frac{1}{\sqrt{1 + 2 \eta^2}} > e^{-2 \eta^2}$, and (c) follows from $e^{-x} \geq 1-x$, for all $x \geq 0$.
Therefore, we obtain
\begin{eqnarray}
    \norm{\CalT_{d_1^n,d_2^n,\eta}(\ketbra{h})-\ketbra{h}}_1 \leq 2 \norm{\CalT_{d_1^n,d_2^n,\eta}(\ket{h})-\ket{h}}_2  \leq 4 \eta. \nonumber 
\end{eqnarray}

\section{Proof of Proposition \ref{Prop:Dec1firsttermHay}}
\label{App:Dec1FirsttermHay}
We first establish a key property of the decoding POVM elements $G^{\CalJ}_{v_1^n,u^n}$, for $\CalJ\in \{0,1,2\}$, that will be used throughout the proof.
For $\CalJ=1$, we have 
\begin{eqnarray}
     \sum_{v_1^n} \sum_{u^n}  p_{V_1U}^n(v_1^n,u^n) \tr \left\{ G^1_{v_1^n,u^n} \rho^{Y_1}_{v_1^n,u^n}\right\} &\overset{(a)}{=}& \sum_{(v_1^n,u^n) \in \CalT_{\delta}^n(p_{V_1U})} p_{V_1U}^n(v_1^n,u^n) \tr \left\{ \pi_{u^n}^{Y_1}  \pi^{Y_1}_{v_1^n,u^n}  \pi_{u^n}^{Y_1} \rho^{Y_1}_{v_1^n,u^n}  \right\} \nonumber \\
    &\overset{(b)}{=}& \sum_{(v_1^n,u^n) \in \CalT_{\delta}^n(p_{V_1U})} p_{V_1U}^n(v_1^n,u^n)  \tr \left\{  \pi^{Y_1}_{v_1^n,u^n}  \pi_{u^n}^{Y_1}  \rho^{Y_1}_{v_1^n,u^n}  \pi_{u^n}^{Y_1}    \right\} \nonumber \\
    &\overset{(c)}{\geq} & \sum_{(v_1^n,u^n) \in \CalT_{\delta}^n(p_{V_1U})} p_{V_1U}^n(v_1^n,u^n)  \tr \left\{  \pi^{Y_1}_{v_1^n,u^n}  \rho^{Y_1}_{v_1^n,u^n}  \right\}  \nonumber \\
    &-& \sum_{(v_1^n,u^n) \in \CalT_{\delta}^n(p_{V_1U})} p_{V_1U}^n(v_1^n,u^n)  \norm{\pi_{u^n}^{Y_1}  \rho^{Y_1}_{v_1^n,u^n}  \pi_{u^n}^{Y_1}-\rho^{Y_1}_{v_1^n,u^n}}_{1} \nonumber \\
    &\overset{(d)}{\geq}& 1-\epsilon - 2 \sqrt{\epsilon} \nonumber,
\end{eqnarray}
where (a) follows from the definition of $G^1_{v_1^n,u^n}$ in \eqref{Eqn:3to1CQBCtheoriginalpovmelements}, (b) follows from the cyclicity of the trace, (c) follows from the trace inequality  $\tr\left( \Delta \rho\right) \geq \tr\left( \Delta \sigma \right) - \norm{\rho - \sigma}_{1}$, for $0 \leq \Delta, \rho, \sigma \leq I$, and (d) follows from (i) the Gentle Operator Lemma \cite{BkWilde_2017} and from (ii) the typical projector property
\begin{eqnarray}
     \tr \left\{  \pi^{Y_1}_{v_1^n,u^n}  \rho^{Y_1}_{v_1^n,u^n}  \right\}  \geq 1-\epsilon, \mbox{ if } (v_1^n,u^n) \in \CalT_{\delta}^n(p_{V_1U}). \nonumber 
\end{eqnarray}
The same argument applies for $\CalJ \in\{0,2\}$.

Now, we evaluate the expectation of $T_{1.2.1}$ with respect to $\left(D_1^n(b_1,m_1),D_2^n(a)\right)$. We obtain
\begin{eqnarray}
\mathbb{E}[T_{1.2.1}] &=& \frac{1}{|\mathcal{M}_1|} \frac{1}{\upsilon^{s_{\sfb} + t_{\sfb}}} \frac{1}{2^{nK_1}} \frac{1}{|\mathcal{D}_1|^n} \frac{1}{|\mathcal{D}_2|^n} \sum_{m_1} \sum_{a} \sum_{b_1} \sum_{d_1^n, d_2^n}  \sum_{v_1^n} \sum_{u^n} p_{V_1U}^n(v_1^n,u^n) \tr \left\{  \left( I - \gamma^{*}_{v_1^n,d_1^n,u^n,d_2^n} \right) \theta_{v_1^n,d_1^n,u^n,d_2^n} \right\} \nonumber \\
&\overset{(a)}{=}& \frac{1}{|\mathcal{M}_1|} \frac{1}{\upsilon^{s_{\sfb} + t_{\sfb}}} \frac{1}{2^{nK_1}} \frac{1}{|\mathcal{D}_1|^n} \frac{1}{|\mathcal{D}_2|^n} \sum_{m_1} \sum_{a} \sum_{b_1} \sum_{d_1^n, d_2^n}  \sum_{v_1^n} \sum_{u^n} p_{V_1U}^n(v_1^n,u^n) \nonumber \\
   &&\left[\tr \left\{ \theta_{v_1^n,d^n_1,u^n,d^n_2} \right\} - \tr \left\{    \pi_{\boldsymbol{\CalH_{Y_1}^G} }\left(I_{\boldsymbol{\CalH_{Y_1}^{e}}}-  \beta^{*}_{v_1^n,d_1^n,u^n,d_2^n}\right) \theta_{v_1^n,d^n_1,u^n,d^n_2} \left(I_{\boldsymbol{\CalH_{Y_1}^{e}}}-  \beta^{*}_{v_1^n,d_1^n,u^n,d_2^n}\right) \pi_{\boldsymbol{\CalH_{Y_1}^G} } \right\} \right]\nonumber  \\
&\overset{(b)}{\leq}& \frac{4}{|\mathcal{M}_1|} \frac{1}{\upsilon^{s_{\sfb} + t_{\sfb}}} \frac{1}{2^{nK_1}} \frac{1}{|\mathcal{D}_1|^n} \frac{1}{|\mathcal{D}_2|^n} \sum_{m_1} \sum_{a} \sum_{b_1} \sum_{d_1^n, d_2^n}  \sum_{v_1^n} \sum_{u^n} p_{V_1U}^n(v_1^n,u^n)  \nonumber \\
&&\tr \left\{ \left(I_{\boldsymbol{\CalH_{Y_1}^{e}}}-\pi_{\boldsymbol{\CalH_{Y_1}^G} }+\beta^{*}_{v_1^n,d_1^n,u^n,d_2^n} \right)\theta_{v_1^n,d^n_1,u^n,d^n_2} \right\} \nonumber  \\
&\overset{(c)}{=}& \frac{4}{|\mathcal{M}_1|} \frac{1}{\upsilon^{s_{\sfb} + t_{\sfb}}} \frac{1}{2^{nK_1}} \frac{1}{|\mathcal{D}_1|^n} \frac{1}{|\mathcal{D}_2|^n} \sum_{m_1} \sum_{a} \sum_{b_1} \sum_{d_1^n, d_2^n}  \sum_{v_1^n} \sum_{u^n}  p_{V_1U}^n(v_1^n,u^n)  \tr \left\{ \beta^{*}_{v_1^n,d_1^n,u^n,d_2^n}   \: \theta_{v_1^n,d^n_1,u^n,d^n_2} \right\} \nonumber  \\
&\overset{(d)}{\leq}&  \frac{4}{|\mathcal{M}_1|} \frac{1}{\upsilon^{s_{\sfb} + t_{\sfb}}} \frac{1}{2^{nK_1}} \frac{1}{|\mathcal{D}_1|^n} \frac{1}{|\mathcal{D}_2|^n} \sum_{m_1} \sum_{a} \sum_{b_1} \sum_{d_1^n, d_2^n}  \sum_{v_1^n} \sum_{u^n}  p_{V_1U}^n(v_1^n,u^n) \nonumber \\
 &&\left[\tr \left\{ \beta^{*}_{v_1^n,d_1^n,u^n,d_2^n} \left(\rho^{Y_1}_{v_1^n,u^n}\otimes \ketbra{0}\right) \right\} + \norm{\theta_{v_1^n,d^n_1,u^n,d^n_2}- \left(\rho^{Y_1}_{v_1^n,u^n}\otimes \ketbra{0}\right)}_1 \right] \nonumber  \\ 
&\overset{(e)}{\leq}& 16 \eta + \frac{6}{\eta^2} \frac{4}{|\mathcal{M}_1|} \frac{1}{\upsilon^{s_{\sfb} + t_{\sfb}}} \frac{1}{2^{nK_1}} \sum_{m_1} \sum_{a} \sum_{b_1} \sum_{v_1^n} \sum_{u^n}  p_{V_1U}^n(v_1^n,u^n) \nonumber \\ 
&&\left[\tr \left\{ \olineB^1_{v_1^n,u^n} \left(\rho^{Y_1}_{v_1^n,u^n}\otimes \ketbra{0}\right) \right) + \tr \left\{ \olineB^2_{v_1^n,u^n} \left(\rho^{Y_1}_{v_1^n,u^n}\otimes \ketbra{0}\right) \right) + \tr \left\{ \olineB^0_{v_1^n,u^n} \left(\rho^{Y_1}_{v_1^n,u^n}\otimes \ketbra{0}\right) \right) \right] \nonumber  \\
&\overset{(f)}{=}& 16 \eta + \frac{6}{\eta^2} \frac{4}{|\mathcal{M}_1|} \frac{1}{\upsilon^{s_{\sfb} + t_{\sfb}}} \frac{1}{2^{nK_1}}  \sum_{m_1} \sum_{a} \sum_{b_1} \sum_{v_1^n} \sum_{u^n}  p_{V_1U}^n(v_1^n,u^n) \nonumber \\ 
&&\left[1-\tr \left\{ G^1_{v_1^n,u^n} \: \rho^{Y_1}_{v_1^n,u^n} \right) + 1-\tr \left\{ G^2_{v_1^n,u^n} \:\rho^{Y_1}_{v_1^n,u^n}\right) + 1-\tr \left\{ G^0_{v_1^n,u^n} \: \rho^{Y_1}_{v_1^n,u^n} \right) \right] \nonumber  \\
    &\overset{(g)}{\leq}& 16 \eta +  \frac{72}{\eta^2} (\epsilon + 2\sqrt{\epsilon})  \nonumber,
\end{eqnarray}
where (a) follows from (i) the definition of $\gamma^{*}_{v_1^n,d^n_1,u^n,d_2^n}$ in \eqref{Eqn:3to1CQBCPovmelementfordecoder1}, (ii) the cyclicity of the trace and (iii) the property of projectors, (b) follows from applying the non-commutative union bound \cite[Fact.3]{202103SAD_Sen}, (c) follows from $\tr\left\{\left(I_{\boldsymbol{\CalH_{Y_1}^{e}}}-\pi_{\boldsymbol{\CalH_{Y_1}^G} }  \right) \left(\rho^{Y_1}_{v_1^n,u^n}\otimes \ketbra{0}\right) \right\}=0$, since $\rho^{Y_1}_{v_1^n,u^n}\otimes \ketbra{0} $ lives in $\boldsymbol{\CalH_{Y_1}^G}$ and $I_{\boldsymbol{\CalH_{Y_1}^{e}}}-\pi_{\boldsymbol{\CalH_{Y_1}^G}}$ is a projector onto the complement of $\boldsymbol{\CalH_{Y_1}^G}$, (d) follows from the trace inequality $\tr(\Delta \rho ) \leq \tr (\Delta \sigma) + \frac{1}{2} \norm{\rho - \sigma}_1$ for $0 \leq \Delta, \rho, \sigma \leq I$, (e) follows from 
(i) Proposition \ref{Prop:3to1CQBCclosnessofstates} and from (ii) \cite[Corollary 1]{202103SAD_Sen}, (f) follows from (i) the definition of $\olineB^{\CalJ}_{v_1^n,u^n}$ for $\CalJ\in \{0,1,2\}$ in \eqref{Eqn:3to1CQBCcomplementprojector} and from (ii) Gelfand–Naimark’s Theorem \cite[Thm.~3.7]{BkHolevo_2019}, and finally (g) follows from the property established at the beginning of this appendix, for $\CalJ \in \{0,1,2\}$. Setting $\eta = \epsilon^{1/5}$, we obtain
\begin{eqnarray}
    \mathbb{E}[T_{1.2.1}] \leq 16 \epsilon^{\frac{1}{5}} + 72 \epsilon^{\frac{3}{5}} + 144 \epsilon^{\frac{1}{10}}. \nonumber 
\end{eqnarray}
\section{Proof of Proposition \ref{Prop:Dec1SecondtermHay}}
\label{App:Dec1SecondtermHay}
For simplicity, we provide analysis only for terms $ T_{1.2.2}, T_{1.2.3}$ and $T_{1.2.4}$. The other terms can be analyzed analogously.
\textit{\underline{Analysis of $T_{1.2.2}$:}} We evaluate the expectation of $T_{1.2.2}$ over $\left(D_1^n(b_1,m_1), D_2^n(a), V_1^n(\tilde{b}_1,\tilde{m}_1), D_1^n(\tilde{b}_1,\tilde{m}_1) \right)$. We obtain 

\begin{eqnarray}
 \mathbb{E}[T_{1.2.2}] &=& \frac{1}{|\mathcal{M}_1|} \frac{1}{\upsilon^{s_{\sfb} + t_{\sfb}}} \frac{1}{2^{nK_1}} \frac{1}{|\mathcal{D}_1|^{2n}} \frac{1}{|\mathcal{D}_2|^n}  \sum_{m_1} \sum_{a}  \sum_{b_1} \sum_{d_1^n, d_2^n}   \sum_{\tilde{m}_1 \neq m_1} \sum_{\tilde{b}_1} \sum_{\tilde{d}_1^n}   \sum_{v_1^n} \sum_{u^n} \sum_{\tilde{v}_1^n} p_{V_1}^n(\tilde{v}_1^n) p_{V_1U}^n(v_1^n,u^n) \nonumber \\
   && \tr \left\{ \gamma^{*}_{\tilde{v}_1^n,\tilde{d}_1^n,u^n,d_2^n}  \: \theta_{v_1^n,d_1^n,u^n,d_2^n} \right\}\nonumber \\
   &\overset{(a)}{=}& \frac{1}{\upsilon^{s_{\sfb} + t_{\sfb}}}   \frac{1}{|\mathcal{D}_1|^n} \frac{1}{|\mathcal{D}_2|^n}    \sum_{a}  \sum_{\tilde{m}_1 \neq m_1} \sum_{\tilde{b}_1} \sum_{\tilde{d}_1^n,d_2^n}   \sum_{u^n} \sum_{\tilde{v}_1^n} p_{V_1}^n(\tilde{v}_1^n) p_{U}^n(u^n) \tr \left\{ \gamma^{*}_{\tilde{v}_1^n,\tilde{d}_1^n,u^n,d_2^n}  \:  \theta_{u^n,d_2^n} \right\}\nonumber \\
&\overset{(b)}{\leq}& \frac{1}{\upsilon^{s_{\sfb} + t_{\sfb}}} \frac{1}{|\mathcal{D}_1|^n} \frac{1}{|\mathcal{D}_2|^n}  \sum_{a}  \sum_{\tilde{m}_1} \sum_{\tilde{b}_1} \sum_{\tilde{d}_1^n,d_2^n}   \sum_{u^n} \sum_{\tilde{v}_1^n} p_{V_1}^n(\tilde{v}_1^n) p_{U}^n(u^n)
\tr \left\{ \gamma^{*}_{\tilde{v}_1^n,\tilde{d}_1^n,u^n,d_2^n}  \: \CalT_{d_2^n,\eta}\left( \rho_{u^n}^{Y_1} \otimes \ketbra{0}\right) \right\}\nonumber \\
&+&  \frac{1}{\upsilon^{s_{\sfb} + t_{\sfb}}}   \frac{1}{|\mathcal{D}_1|^n} \frac{1}{|\mathcal{D}_2|^n}    \sum_{a}  \sum_{\tilde{m}_1} \sum_{\tilde{b}_1} \sum_{\tilde{d}_1^n,d_2^n}   \sum_{u^n} \sum_{\tilde{v}_1^n} p_{V_1}^n(\tilde{v}_1^n) p_{U}^n(u^n)
 \tr \left\{  \gamma^{*}_{\tilde{v}_1^n,\tilde{d}_1^n,u^n,d_2^n}  \: \CalN_{d_2^n,\eta}\left( \rho_{u^n}^{Y_1} \otimes \ketbra{0} \right) \right\} \label{Eqn:3to1CQBCsecondterminhayashinagaokainequality}
\end{eqnarray}
where (a) follows by using the definition $\theta_{u^n,d_2^n} \define \frac{1}{|\mathcal{D}_1|^n} \sum_{d_1^n} \sum_{v_1^n} p_{V_1|U}^n(v_1^n | u^n) \: \theta_{v_1^n,d_1^n,u^n,d_2^n}$ and (b) follows by using 
\begin{eqnarray}
    \theta_{u^n,d_2^n} = \frac{1 + \eta^2}{1+ 2\eta^2} \mathcal{T}_{d_2^n,\eta}\left( \rho_{u^n} \otimes \ketbra{0} \right) + \mathcal{N}_{d_2^n,\eta} \left( \rho_{u^n} \otimes \ketbra{0} \right), \nonumber 
\end{eqnarray}
as stated in \cite{202103SAD_Sen}. Consider the first term in \eqref{Eqn:3to1CQBCsecondterminhayashinagaokainequality}. We have
\begin{eqnarray}
&&\frac{1}{\upsilon^{s_{\sfb} + t_{\sfb}}}   \frac{1}{|\mathcal{D}_1|^n} \frac{1}{|\mathcal{D}_2|^n}   \sum_{a}  \sum_{\tilde{m}_1} \sum_{\tilde{b}_1} \sum_{\tilde{d}_1^n,d_2^n}   \sum_{u^n} \sum_{\tilde{v}_1^n} p_{V_1}^n(\tilde{v}_1^n) p_{U}^n(u^n)
\tr \left\{ \gamma^{*}_{\tilde{v}_1^n,\tilde{d}_1^n,u^n,d_2^n}  \: \CalT_{d_2^n,\eta}\left( \rho_{u^n}^{Y_1} \otimes \ketbra{0}\right) \right\}\nonumber \\
&\overset{(a)}{\leq}&  \frac{1}{\upsilon^{s_{\sfb} + t_{\sfb}}}   \frac{1}{|\mathcal{D}_1|^n} \frac{1}{|\mathcal{D}_2|^n}  \sum_{a} \sum_{\tilde{m}_1} \sum_{\tilde{b}_1} \sum_{\tilde{d}_1^n,d_2^n}   \sum_{u^n} \sum_{\tilde{v}_1^n} p_{V_1}^n(\tilde{v}_1^n) p_{U}^n(u^n) 
\tr \left\{ 
\left( I_{\boldsymbol{\CalH_{Y_1}^{e}}}-  \beta^{1}_{\tilde{v}_1^n,\tilde{d}^n_1,u^n,d_2^n}\right)  \CalT_{d_2^n,\eta}\left( \rho_{u^n}^{Y_1} \otimes \ketbra{0}\right) \right\}\nonumber \\
&\overset{(b)}{=}& \frac{1}{\upsilon^{s_{\sfb} + t_{\sfb}}}  \frac{1}{|\mathcal{D}_1|^n} \frac{1}{|\mathcal{D}_2|^n}  \sum_{a}\sum_{\tilde{m}_1} \sum_{\tilde{b}_1} \sum_{\tilde{d}_1^n,d_2^n}   \sum_{u^n} \sum_{\tilde{v}_1^n} p_{V_1}^n(\tilde{v}_1^n) p_{U}^n(u^n)  \nonumber \\
&&\tr \left\{ 
\left( I_{\boldsymbol{\CalH_{Y_1}^{e}}}-  \beta^{1}_{\tilde{v}_1^n,\tilde{d}^n_1,u^n,d_2^n}\right)
\pi_{\CalT_{d_2^n,\eta}\left(\boldsymbol{\CalH_{Y_1}^G}\right)} \CalT_{d_2^n,\eta}\left( \rho_{u^n}^{Y_1} \otimes \ketbra{0}\right) \right\}\nonumber \\
&\overset{(c)}{=}& \frac{1}{\upsilon^{s_{\sfb} + t_{\sfb}}}  \frac{1}{|\mathcal{D}_1|^n} \frac{1}{|\mathcal{D}_2|^n}  \sum_{a}\sum_{\tilde{m}_1} \sum_{\tilde{b}_1} \sum_{\tilde{d}_1^n,d_2^n}   \sum_{u^n} \sum_{\tilde{v}_1^n} p_{V_1}^n(\tilde{v}_1^n) p_{U}^n(u^n) \tr \left\{ \left( I_{\CalT_{d_2^n,\eta}\left(\boldsymbol{\CalH_{Y_1}^G}\right)}-\beta^{1}_{\tilde{v}_1^n,\tilde{d}^n_1,u^n,d_2^n}\right)\CalT_{d_2^n,\eta}\left( \rho_{u^n}^{Y_1} \otimes \ketbra{0}\right) \right\}\nonumber \\
&\overset{(d)}{=}&  \frac{1}{\upsilon^{s_{\sfb} + t_{\sfb}}}   \sum_{a}\sum_{\tilde{m}_1} \sum_{\tilde{b}_1} \sum_{u^n} \sum_{\tilde{v}_1^n} p_{V_1}^n(\tilde{v}_1^n) p_{U}^n(u^n)  
\tr \left\{  \left( I-  \olineB^{1}_{\tilde{v}_1^n,u^n}\right)   \left( \rho_{u^n}^{Y_1} \otimes \ketbra{0}\right) \right\} \nonumber \\
&\overset{(e)}{=}&  \frac{1}{\upsilon^{s_{\sfb} + t_{\sfb}}}  \sum_{a}\sum_{\tilde{m}_1} \sum_{\tilde{b}_1} \sum_{u^n} \sum_{\tilde{v}_1^n} p_{V_1}^n(\tilde{v}_1^n) p_{U}^n(u^n)  
\tr \left\{ G^{1}_{\tilde{v}_1^n,u^n} \: \rho_{u^n}^{Y_1} \right\} \nonumber \\
&\overset{(f)}{=}&  \frac{1}{\upsilon^{s_{\sfb} + t_{\sfb}}}  \sum_{a}\sum_{\tilde{m}_1} \sum_{\tilde{b}_1} \sum_{(\tilde{v}_1^n,u^n) \in T_{\delta}^n(p_{V_1U}) }  p_{V_1}^n(\tilde{v}_1^n) p_{U}^n(u^n)  
\tr \left\{ \pi_{u^n}^{Y_1}  \pi^{Y_1}_{\tilde{v}_1^n,u^n}  \pi_{u^n}^{Y_1} \rho_{u^n}^{Y_1} \right\} \nonumber \\
&\overset{(g)}{=}&  \frac{1}{\upsilon^{s_{\sfb} + t_{\sfb}}}  \sum_{a}\sum_{\tilde{m}_1} \sum_{\tilde{b}_1} \sum_{(\tilde{v}_1^n,u^n) \in T_{\delta}^n(p_{V_1U}) }   p_{V_1}^n(\tilde{v}_1^n) p_{U}^n(u^n)  
\tr \left\{ \pi^{Y_1}_{\tilde{v}_1^n,u^n}  \pi_{u^n}^{Y_1} \rho_{u^n}^{Y_1}  \pi_{u^n}^{Y_1}  \right\} \nonumber \\
&\overset{(h)}{\leq}&  \frac{1}{\upsilon^{s_{\sfb} + t_{\sfb}}}  \sum_{a}\sum_{\tilde{m}_1} \sum_{\tilde{b}_1} \sum_{(\tilde{v}_1^n,u^n) \in T_{\delta}^n(p_{V_1U}) }   2^{-n(I(V_1;U) -3 \delta)} p_{V_1U}^n(\tilde{v}_1^n,u^n) \tr \left\{ \pi^{Y_1}_{\tilde{v}_1^n,u^n}  \pi_{u^n}^{Y_1} \rho_{u^n}^{Y_1}  \pi_{u^n}^{Y_1}  \right\} \nonumber \\
&\overset{(i)}{\leq}& 2^{-n(I(V_1;U) -3 \delta -(R_1 + K_1))} 2^{-n(I(V_1;Y_1|U) -2 \delta)},  \nonumber 
\end{eqnarray}
where (a) follows from 
\begin{eqnarray}
    \gamma^*_{\tilde{v}_1^n,\tilde{d}_1^n,u^n,d_2^n} \leq \left( I_{\boldsymbol{\CalH_{Y_1}^{e}}}-  \beta^{*}_{\tilde{v}_1^n,\tilde{d}^n_1,u^n,d_2^n}\right)\leq \left( I_{\boldsymbol{\CalH_{Y_1}^{e}}}-  \beta^{1}_{\tilde{v}_1^n,\tilde{d}^n_1,u^n,d_2^n}\right), \nonumber 
\end{eqnarray}
(b) follows from inserting the projector
$\pi_{\CalT_{d_2^n,\eta}(\boldsymbol{\CalH_{Y_1}^G})}$, (c) follows because the support of $ I_{\boldsymbol{\CalH_{Y_1}^{e}}}-  \beta^{1}_{\tilde{v}_1^n,\tilde{d}^n_1,u^n,d_2^n}$ is contained in $\mathcal{T}_{d_2^n,\eta}\left( \CalH_{Y_1}^{e} \right)$, (d) follows because $\mathcal{T}_{d_2^n, \eta }$ is an isometry, (e) follows from (i) the definition of $\olineB^1_{\tilde{v}_1^n,u^n}$ in \eqref{Eqn:3to1CQBCcomplementprojector} and from (ii) the Gelfand–Naimark’s Theorem \cite[Thm.~3.7]{BkHolevo_2019}, (f) follows from the definition of $G^1_{\tilde{v}_1^n,u^n}$ in \eqref{Eqn:3to1CQBCPovmelementfordecoder1}, 
(g) follows from the cyclicity of the trace, (h) follows from 
\begin{eqnarray}
\sum_{(\tilde{v}_1^n,u^n) \in T_{\delta}^n(p_{V_1U}) }  p_{V_1}^n(\tilde{v}_1^n) p_{U}^n(u^n) \leq \sum_{(\tilde{v}_1^n,u^n) \in T_{\delta}^n(p_{V_1U}) }  2^{-n(I(V_1;U) -3 \delta)} p_{V_1U}^n(\tilde{v}_1^n,u^n), \label{Eqn:Joint}     
\end{eqnarray}
and finally (i) follows from the typical projector property
\begin{eqnarray} 
  &&  \pi_{u^n}^{Y_1} \rho_{u^n}^{Y_1}  \pi_{u^n}^{Y_1} \leq 2^{-n \left(H(Y_1|U) -\delta  \right) }, \mbox{ and }   \tr\left \{\pi^{Y_1}_{\tilde{v}_1^n,u^n}  \right \} \leq 2^{n\left( H(Y_1|V_1,U) + \delta \right)},\mbox{ if }(\tilde{v}_1^n,u^n) \in \CalT_{\delta}^n(p_{V_1U}).\nonumber 
\end{eqnarray}
Now, consider the second term in \eqref{Eqn:3to1CQBCsecondterminhayashinagaokainequality}. We have, 
\begin{eqnarray}
&&\frac{1}{\upsilon^{s_{\sfb} + t_{\sfb}}}   \frac{1}{|\mathcal{D}_1|^n} \frac{1}{|\mathcal{D}_2|^n}    \sum_{a}  \sum_{\tilde{m}_1} \sum_{\tilde{b}_1} \sum_{\tilde{d}_1^n,d_2^n}   \sum_{u^n} \sum_{\tilde{v}_1^n}  p_{V_1}^n(\tilde{v}_1^n) p_{U}^n(u^n)
\tr \left\{  \gamma^{*}_{\tilde{v}_1^n,\tilde{d}_1^n,u^n,d_2^n}  \CalN_{d_2^n,\eta}\left( \rho_{u^n}^{Y_1} \otimes \ketbra{0} \right) \right\} \nonumber \\
&\overset{(a)}{=}& \frac{1}{\upsilon^{s_{\sfb} + t_{\sfb}}}   \frac{1}{|\mathcal{D}_1|^n} \frac{1}{|\mathcal{D}_2|^n}    \sum_{a}  \sum_{\tilde{m}_1} \sum_{\tilde{b}_1} \sum_{\tilde{d}_1^n,d_2^n}  \sum_{(\tilde{v}_1^n,u^n) \in T_{\delta}^n(p_{V_1U})}  p_{V_1}^n(\tilde{v}_1^n) p_{U}^n(u^n) \tr \left\{  \gamma^{*}_{\tilde{v}_1^n,\tilde{d}_1^n,u^n,d_2^n}  \CalN_{d_2^n,\eta}\left( \rho_{u^n}^{Y_1} \otimes \ketbra{0} \right) \right\} \nonumber \\
 &\overset{(b)}{\leq}& \frac{1}{\upsilon^{s_{\sfb} + t_{\sfb}}}   \frac{1}{|\mathcal{D}_1|^n} \frac{1}{|\mathcal{D}_2|^n}    \sum_{a}  \sum_{\tilde{m}_1} \sum_{\tilde{b}_1} \sum_{\tilde{d}_1^n,d_2^n}  \sum_{(\tilde{v}_1^n,u^n) \in T_{\delta}^n(p_{V_1U})} 2^{-n(I(V_1;U) -3 \delta )}p_{V_1U}^n(\tilde{v}_1^n,u^n) \nonumber \\
&&\norm{\gamma^{*}_{\tilde{v}_1^n,\tilde{d}_1^n,u^n,d_2^n}}_1  \norm{\CalN_{d_2^n,\eta}\left( \rho_{u^n}^{Y_1} \otimes \ketbra{0} \right)}_{\infty} \nonumber \\
 &\overset{(c)}{\leq}& 2^{-n(I(V_1;U) - 3 \delta -(R_1+ K_1))} 2^n |\mathcal{H}_{Y_1}|^n   \frac{3 \eta }{\sqrt{|\mathcal{D}_1|^n}} \nonumber 
 \end{eqnarray}
where (a) follows because $\gamma^*_{\tilde{v}_1^n,\tilde{d}_1^n,u^n,d_2^n}=0$, if $(\tilde{v}_1^n,u^n) \notin T_{\delta}^n(p_{V_1U})$, (b) follows from (i) the inequality in \eqref{Eqn:Joint} and from (ii) the trace inequality 
\begin{eqnarray}
     &&|\tr\left( AB\right)| \leq \norm{AB}_1 \leq \min\{\norm{A}_1 \norm{B}_{\infty}, \norm{A}_{\infty} \norm{B}_{1}\}, \label{Eqn:3to1CQBCTraceIneq}
 \end{eqnarray}
 and (c) follows from (i) the inequality 
 \begin{eqnarray}
     \norm{\gamma^{*}_{\tilde{v}_1^n,\tilde{d}^n_1,u^n,d_2^n}}_1 \leq \norm{I_{\boldsymbol{\CalH_{Y_1}^{e}}}- \beta^{*}_{\tilde{v}_1^n,\tilde{d}^n_1,u^n,d_2^n}}_{\infty}^2 \norm{\pi_{\boldsymbol{\CalH_{Y_1}^G}}}_1 \leq 2^n |\CalH_{Y_1}|^n, \label{Eqn:3to1CQBCBndgamma*} 
 \end{eqnarray}
and from (ii) the inequality 
\begin{eqnarray}
   \norm{\CalN_{d_2^n,\eta}\left( \rho_{u^n}^{Y_1} \otimes \ketbra{0} \right)}_{\infty}  \leq \frac{3 \eta }{\sqrt{|\mathcal{D}_1|^n}} \nonumber 
\end{eqnarray}
as stated in \cite{202103SAD_Sen}. Therefore, if we choose $|\CalD_1|\geq \left( 3 \eta 2^n |\CalH_{Y_1}|^n 2^{n \left( I(V_1;Y_1|U) - 2 \delta \right)}\right)^{\frac{2}{n}}$, we have 
\begin{eqnarray}
    \mathbb{E}\left[T_{1.2.2}\right] &\leq& 2^{-n(I(V_1;U) - 3 \delta - (R_1 + K_1))} \left( 2^{-n \left( I(V_1;Y_1|U) -2\delta \right)}+\frac{ 2^n |\CalH_{Y_1}|^n  3 \eta }{\sqrt{|\CalD_1|^n}}\right) \nonumber \\
    &\leq& 2  \: 2^{-n \left( I(V_1;Y_1,U) - 5\delta - (R_1+K_1) \right)}. \nonumber 
\end{eqnarray}
Therefore, for sufficiently large $n$, we have $\mathbb{E}[T_{1.2.2}] \leq \epsilon,$ if
\begin{eqnarray}
R_1 + K_1 &<& I(V_1;Y_1,U) \nonumber \\
&=& H(V_1) - H(V_1|Y_1, U). \nonumber    
\end{eqnarray}

 \textit{\underline{Analysis of $T_{1.2.3}$:}} 
 We evaluate the expectation of $T_{1.2.3}$ with respect to $\left(D_1^n(b_1,m_1), D_2^n(a), U^n(\tilde{a}), D_2^n(\tilde{a}) \right)$. We obtain 

\begin{eqnarray}
   \mathbb{E}[T_{1.2.3}] &=& \frac{1}{|\mathcal{M}_1|} \frac{1}{\upsilon^{s_{\sfb} + t_{\sfb}}}   \frac{1}{2^{nK_1}} \frac{1}{|\mathcal{D}_1|^n} \frac{1}{|\mathcal{D}_2|^{2n}}   \sum_{m_1} \sum_{a} \sum_{b_1} \sum_{d_1^n, d_2^n}  \sum_{\tilde{a} \neq a}  \sum_{\tilde{d}_2^n}  \sum_{v_1^n} \sum_{u^n} \sum_{\tilde{u}^n} q_{U}^n(\tilde{u}^n) p_{V_1U}^n(v_1^n,u^n) \nonumber \\
   &&\tr \left\{ \gamma^{*}_{v_1^n,d_1^n,\tilde{u}^n,\tilde{d}_2^n} \:\theta_{v_1^n,d_1^n,u^n,d_2^n} \right\}\nonumber \\
&\overset{(a)}{=}&
\frac{1}{|\mathcal{M}_1|}   \frac{1}{2^{nK_1}} \frac{1}{|\mathcal{D}_1|^n} \frac{1}{|\mathcal{D}_2|^n}   \sum_{m_1} \sum_{b_1}   \sum_{\tilde{a} \neq a}  \sum_{d_1^n, \tilde{d}_2^n}  \sum_{v_1^n}  \sum_{\tilde{u}^n}
q_{U}^n(\tilde{u}^n) p_{V_1}^n(v_1^n) \tr \left\{ \gamma^{*}_{v_1^n,d_1^n,\tilde{u}^n,\tilde{d}_2^n} \: \theta_{v_1^n,d_1^n} \right\}\nonumber \\
&\overset{(b)}{\leq}&
\frac{1}{|\mathcal{M}_1|}   \frac{1}{2^{nK_1}} \frac{1}{|\mathcal{D}_1|^n} \frac{1}{|\mathcal{D}_2|^n}   \sum_{m_1} \sum_{b_1}   \sum_{\tilde{a}}  \sum_{d_1^n, \tilde{d}_2^n}  \sum_{v_1^n}  \sum_{\tilde{u}^n} q_{U}^n(\tilde{u}^n) p_{V_1}^n(v_1^n) \tr \left\{ \gamma^{*}_{v_1^n,d_1^n,\tilde{u}^n,\tilde{d}_2^n}  \:\mathcal{T}_{d_1^n,\eta}\left(\rho_{v_1^n}^{Y_1} \otimes \ketbra{0}\right) \right\}\nonumber \\
&+& 
\frac{1}{|\mathcal{M}_1|}   \frac{1}{2^{nK_1}} \frac{1}{|\mathcal{D}_1|^n} \frac{1}{|\mathcal{D}_2|^n}   \sum_{m_1} \sum_{b_1}   \sum_{\tilde{a}}  \sum_{d_1^n, \tilde{d}_2^n}  \sum_{v_1^n}  \sum_{\tilde{u}^n} q_{U}^n(\tilde{u}^n) p_{V_1}^n(v_1^n) \nonumber \\
&&\tr \left\{ \gamma^{*}_{v_1^n,d_1^n,\tilde{u}^n,\tilde{d}_2^n}  \:\mathcal{N}_{d_1^n, \eta}\left(\rho_{v_1^n}^{Y_1} \otimes \ketbra{0}\right) \right\} \label{Eqn:3to1CQBCthirdterminhayashinagaokainequality}
\end{eqnarray}
where (a) follows by using the definition $\theta_{v_1^n,d_1^n} \define \frac{1}{|\mathcal{D}_2|^n} \sum_{d_2^n} \sum_{u^n} p_{U|V_1}^n(u^n|v_1^n) \: \theta_{v_1^n,d_1^n,u^n,d_2^n}$, and (b) follows by using
\begin{eqnarray}
\theta_{v_1^n,d_1^n} = \frac{1+ \eta^2}{1+2 \eta^2} \mathcal{T}_{d_1^n,\eta}\left(\rho_{v_1^n}^{Y_1} \otimes \ketbra{0}\right) + \mathcal{N}_{d_1^n, \eta}\left(\rho_{v_1^n}^{Y_1} \otimes \ketbra{0}\right), \nonumber 
\end{eqnarray}
as stated in \cite{202103SAD_Sen}. We begin by bounding the first term in \eqref{Eqn:3to1CQBCthirdterminhayashinagaokainequality}. We have
\begin{eqnarray}
    &&\frac{1}{|\mathcal{M}_1|}   \frac{1}{2^{nK_1}} \frac{1}{|\mathcal{D}_1|^n} \frac{1}{|\mathcal{D}_2|^n}   \sum_{m_1} \sum_{b_1}   \sum_{\tilde{a}}  \sum_{d_1^n, \tilde{d}_2^n}  \sum_{v_1^n}  \sum_{\tilde{u}^n} q_{U}^n(\tilde{u}^n) p_{V_1}^n(v_1^n) \tr \left\{ \gamma^{*}_{v_1^n,d_1^n,\tilde{u}^n,\tilde{d}_2^n}  \mathcal{T}_{d_1^n,\eta}\left(\rho_{v_1^n}^{Y_1} \otimes \ketbra{0}\right) \right\}\nonumber \\
&\overset{(a)}{\leq}& \frac{1}{|\mathcal{M}_1|}   \frac{1}{2^{nK_1}} \frac{1}{|\mathcal{D}_1|^n} \frac{1}{|\mathcal{D}_2|^n}   \sum_{m_1} \sum_{b_1}   \sum_{\tilde{a}}  \sum_{d_1^n, \tilde{d}_2^n}  \sum_{v_1^n}  \sum_{\tilde{u}^n} q_{U}^n(\tilde{u}^n) p_{V_1}^n(v_1^n) \tr \left\{ \left( I_{\boldsymbol{\CalH_{Y_1}^{e}}}-  \beta^{2}_{v_1^n,d^n_1,\tilde{u}^n,\tilde{d}_2^n} \right)  \mathcal{T}_{d_1^n,\eta}\left(\rho_{v_1^n}^{Y_1} \otimes \ketbra{0}\right) \right\}\nonumber \\
&\overset{(b)}{=}& \frac{1}{|\mathcal{M}_1|}   \frac{1}{2^{nK_1}} \frac{1}{|\mathcal{D}_1|^n} \frac{1}{|\mathcal{D}_2|^n}   \sum_{m_1} \sum_{b_1}   \sum_{\tilde{a}}  \sum_{d_1^n, \tilde{d}_2^n}  \sum_{v_1^n}  \sum_{\tilde{u}^n} q_{U}^n(\tilde{u}^n) p_{V_1}^n(v_1^n) \nonumber \\
&&\tr \left\{ \left( I_{\boldsymbol{\CalH_{Y_1}^{e}}}-  \beta^{2}_{v_1^n,d^n_1,\tilde{u}^n,\tilde{d}_2^n} \right)  \pi_{\CalT_{d_1^n,\eta}\left(\boldsymbol{\CalH_{Y_1}^G}\right)}\mathcal{T}_{d_1^n,\eta}\left(\rho_{v_1^n}^{Y_1} \otimes \ketbra{0}\right) \right\}\nonumber \\
&\overset{(c)}{=}& \frac{1}{|\mathcal{M}_1|}   \frac{1}{2^{nK_1}} \frac{1}{|\mathcal{D}_1|^n} \frac{1}{|\mathcal{D}_2|^n}   \sum_{m_1} \sum_{b_1}   \sum_{\tilde{a}}  \sum_{d_1^n, \tilde{d}_2^n}  \sum_{v_1^n}  \sum_{\tilde{u}^n} q_{U}^n(\tilde{u}^n) p_{V_1}^n(v_1^n) \nonumber \\
&&\tr \left\{ \left( I_{\CalT_{d_1^n,\eta}\left(\boldsymbol{\CalH_{Y_1}^G}\right)}-  \beta^{2}_{v_1^n, d^n_1,\tilde{u}^n, \tilde{d}_2^n}\right)  \mathcal{T}_{d_1^n,\eta}\left(\rho_{v_1^n}^{Y_1} \otimes \ketbra{0}\right) \right\}\nonumber \\
&\overset{(d)}{=}& \frac{1}{|\mathcal{M}_1|}   \frac{1}{2^{nK_1}}   \sum_{m_1} \sum_{b_1}   \sum_{\tilde{a}}    \sum_{v_1^n}  \sum_{\tilde{u}^n} q_{U}^n(\tilde{u}^n) p_{V_1}^n(v_1^n) \tr \left\{ \left( I - \olineB^{2}_{v_1^n,\tilde{u}^n}\right)  \left(\rho_{v_1^n}^{Y_1} \otimes \ketbra{0}\right) \right\}\nonumber \\
&\overset{(e)}{=}& \frac{1}{|\mathcal{M}_1|}   \frac{1}{2^{nK_1}}   \sum_{m_1} \sum_{b_1}   \sum_{\tilde{a}}   \sum_{v_1^n}  \sum_{\tilde{u}^n} q_{U}^n(\tilde{u}^n) p_{V_1}^n(v_1^n) \tr \left\{ G^{2}_{v_1^n,\tilde{u}^n} \rho_{v_1^n}^{Y_1}  \right\}\nonumber \\
&\overset{(f)}{\leq}& \frac{1}{|\mathcal{M}_1|}   \frac{1}{2^{nK_1}}  \sum_{m_1} \sum_{b_1}   \sum_{\tilde{a}}     \sum_{(v_1^n, \tilde{u}^n) \in T_{\delta}(p_{V_1U})} 2^{-n \left( D(p_{U}||q_{U}) - \tilde{\delta}\right)} p_{U}^n(\tilde{u}^n) p_{V_1}^n(v_1^n) \tr \left\{ \pi_{v_1^n}^{Y_1} \pi_{v_1^n,\tilde{u}^n}^{Y_1}\pi_{v_1^n}^{Y_1} \rho_{v_1^n}^{Y_1}  \right\}\nonumber \\
&\overset{(g)}{\leq}& \frac{1}{|\mathcal{M}_1|}   \frac{1}{2^{nK_1}}    \sum_{m_1} \sum_{b_1}   \sum_{\tilde{a}}     \sum_{(v_1^n, \tilde{u}^n) \in T_{\delta}(p_{V_1U})} 2^{-n \left( D(p_{U}||q_{U}) - \tilde{\delta}\right)} 2^{-n(I(V_1;U)- 3 \delta)} p_{V_1U}^n(v_1^n,\tilde{u}^n) \tr \left\{  \pi_{v_1^n,\tilde{u}^n}^{Y_1}\pi_{v_1^n}^{Y_1} \rho_{v_1^n}^{Y_1} \pi_{v_1^n}^{Y_1} \right\}\nonumber \\
&\overset{(h)}{\leq}& 2^{-n \left( I(V_1;U) + D(p_{U}||q_{U}) - 3 \delta - \tilde{\delta} - \max\{S_2 + T_2 , S_3 + T_3\} \right)} 2^{-n\left(I(U;Y_1|V_1) - 2 \delta\right)} \nonumber 
\end{eqnarray}
where (a) follows from 
\begin{eqnarray}
    \gamma^*_{v_1^n, d_1^n, \tilde{u}^n, \tilde{d}_2^n} \leq \left( I_{\boldsymbol{\CalH_{Y_1}^{e}}}-  \beta^{*}_{v_1^n, d_1^n, \tilde{u}^n, \tilde{d}_2^n}\right) 
    \leq \left( I_{\boldsymbol{\CalH_{Y_1}^{e}}}-  \beta^{2}_{v_1^n, d_1^n, \tilde{u}^n, \tilde{d}_2^n}\right), \nonumber 
\end{eqnarray}
(b) follows from inserting the projector
$\pi_{\CalT_{d_1^n,\eta}(\boldsymbol{\CalH_{Y_1}^G})}$, (c) follows because the support of $ I_{\boldsymbol{\CalH_{Y_1}^{e}}}-  \beta^{2}_{v_1^n, d_1^n, \tilde{u}^n, \tilde{d}_2^n}$ is contained in $\mathcal{T}_{d_1^n,\eta}\left( \CalH_{Y_1}^{e} \right)$, (d) follows because $\mathcal{T}_{d_1^n, \eta }$ is an isometry, (e) follows from (i) the definition of $\olineB^2_{v_1^n,\tilde{u}^n}$ in \eqref{Eqn:3to1CQBCcomplementprojector} and from (ii) the Gelfand–Naimark’s Theorem \cite[Thm.~3.7]{BkHolevo_2019}, (f) follows from (i)
\begin{eqnarray}
    q_{U}^n(\tilde{u}^n) \leq 2 ^{-n\left(D(p_U || q_U) - \tilde{\delta}\right)} p_U^n(\tilde{u}^n), \mbox{ if } \tilde{u}^n \in T_{\delta}(p_U), \label{Eqn:3to1CQBCDivergence}
\end{eqnarray}
and from (ii) the definition of $G^2_{v_1^n,\tilde{u}^n}$ in \eqref{Eqn:3to1CQBCPovmelementfordecoder1}, 
(g) follows from (i) the cyclicity of the trace and from (ii) the inequality 
\begin{eqnarray}
\sum_{(v_1^n,\tilde{u}^n) \in T_{\delta}^n(p_{V_1U}) }  p_{V_1}^n(v_1^n) p_{U}^n(\tilde{u}^n) \leq \sum_{(v_1^n,\tilde{u}^n) \in T_{\delta}^n(p_{V_1U}) }  2^{-n(I(V_1;U) -3 \delta)} p_{V_1U}^n(v_1^n,\tilde{u}^n), \label{Eqn:Joint1}     
\end{eqnarray}
and finally (h) follows from using the standard typical projector property
\begin{eqnarray} 
  &&  \pi_{v_1^n}^{Y_1} \rho_{v_1^n}^{Y_1}  \pi_{v_1^n}^{Y_1} \leq 2^{-n \left(H(Y_1|V_1) -\delta  \right) },  \mbox{ and }  \tr\left \{\pi_{v_1^n,\tilde{u}^n}^{Y_1} \right \} \leq 2^{n\left( H(Y_1|V_1,U) + \delta \right)}, \mbox{ if }  (v_1^n,\tilde{u}^n) \in \CalT_{\delta}^n(p_{V_1U}).\nonumber 
\end{eqnarray}
Next, consider the second term in \eqref{Eqn:3to1CQBCthirdterminhayashinagaokainequality}. We have
\begin{eqnarray}
&&\frac{1}{|\mathcal{M}_1|}   \frac{1}{2^{nK_1}} \frac{1}{|\mathcal{D}_1|^n} \frac{1}{|\mathcal{D}_2|^n}   \sum_{m_1} \sum_{b_1}   \sum_{\tilde{a}}  \sum_{d_1^n, \tilde{d}_2^n}  \sum_{v_1^n}  \sum_{\tilde{u}^n}  q_{U}^n(\tilde{u}^n) p_{V_1}^n(v_1^n) \tr \left\{ \gamma^{*}_{v_1^n,d_1^n,\tilde{u}^n,\tilde{d}_2^n}  \mathcal{N}_{d_1^n, \eta}\left(\rho_{v_1^n}^{Y_1} \otimes \ketbra{0}\right) \right\} \nonumber \\
&\overset{(a)}{\leq}&
 \frac{1}{|\mathcal{M}_1|}   \frac{1}{2^{nK_1}} \frac{1}{|\mathcal{D}_1|^n} \frac{1}{|\mathcal{D}_2|^n}   \sum_{m_1} \sum_{b_1}   \sum_{\tilde{a}}  \sum_{d_1^n, \tilde{d}_2^n}  
\sum_{(v_1^n, \tilde{u}^n) \in T_{\delta(p_{V_1U})}} 2^{-n(D(p_U||q_U) - \tilde{\delta})}p_{U}^n(\tilde{u}^n) p_{V_1}^n(v_1^n) \nonumber \\
&&\tr \left\{ \gamma^{*}_{v_1^n,d_1^n,\tilde{u}^n,\tilde{d}_2^n}  \mathcal{N}_{d_1^n, \eta}\left(\rho_{v_1^n}^{Y_1} \otimes \ketbra{0}\right) \right\} \nonumber \\
&\overset{(b)}{\leq}&\frac{1}{|\mathcal{M}_1|}   \frac{1}{2^{nK_1}} \frac{1}{|\mathcal{D}_1|^n} \frac{1}{|\mathcal{D}_2|^n}   \sum_{m_1} \sum_{b_1}   \sum_{\tilde{a}}  \sum_{d_1^n, \tilde{d}_2^n}  
\sum_{(v_1^n, \tilde{u}^n) \in T_{\delta(p_{V_1U})}} 2^{-n(D(p_U||q_U) - \tilde{\delta})}   2^{-n(I(V_1;U) - 3 \delta )} p_{V_1U}^n(v_1^n,\tilde{u}^n) \nonumber \\ 
&&\norm{ \gamma^{*}_{v_1^n,d_1^n,\tilde{u}^n,\tilde{d}_2^n}}_1  \norm{\mathcal{N}_{d_1^n, \eta}\left(\rho_{v_1^n}^{Y_1} \otimes \ketbra{0}\right)}_{\infty} \nonumber \\
&\overset{(c)}{\leq}&  2^{-n \left(D(p_{U}||q_{U}) + I(V_1;U) - 3 \delta - \tilde{\delta} - \max\{S_2 + T_2 , S_3 + T_3\} \right)} 2^n |\mathcal{H}_{Y_1}|^n   \frac{3 \eta }{\sqrt{|\mathcal{D}_2|^n}}. \nonumber 
\end{eqnarray}
where (a) follows from (i) the fact that $\gamma^*_{v_1^n,d_1^n,\tilde{u}^n,\tilde{d}_2^n} = 0$, if $(v_1^n,\tilde{u}^n) \notin T_{\delta}^n(p_{V_1U})$, and from (ii) the inequality given in \eqref{Eqn:3to1CQBCDivergence}. (b) follows from (i) the trace inequality defined in \eqref{Eqn:3to1CQBCTraceIneq} and from (ii) the inequality defined in \eqref{Eqn:Joint1}. (c) follows from \eqref{Eqn:3to1CQBCBndgamma*} and from the inequality 
\begin{eqnarray}
   \norm{\CalN_{d_1^n,\eta}\left( \rho_{v_1^n}^{Y_1} \otimes \ketbra{0} \right)}_{\infty}  \leq \frac{3 \eta }{\sqrt{|\mathcal{D}_2|^n}} \nonumber 
\end{eqnarray}
as stated in \cite{202103SAD_Sen}. Therefore, if we choose 
$ |\CalD_2|\geq \left( 3 \eta 2^n   |\CalH_{Y_1}|^n  2^{n \left( D(p_{U}||q_{U}) + I(U;Y_1|V_1) - \tilde{\delta}- 2 \delta \right)}\right)^{\frac{2}{n}}$, we have 
\begin{eqnarray}
    \mathbb{E}[T_{1.2.3}] &\leq& 2^{-n \left( D(p_{U}||q_{U}) + I(V_1;U)  - 3 \delta - \tilde{\delta} - \max\{S_2 + T_2 , S_3 + T_3\} \right)} \left(  2^{-n \left( I(U;Y_1 |V_1) - 2 \delta\right)} + 2^n |\mathcal{H}_{Y_1}|^n   \frac{3 \eta }{\sqrt{|\mathcal{D}_2|^n}}\right) \nonumber \\
    &\leq& 2  \: 2^{-n \left(D(p_{U}||q_{U}) + I(U;Y_1,V_1) - \tilde{\delta} - 5 \delta - \max\{S_2 + T_2 , S_3 + T_3\}\right)} \nonumber 
\end{eqnarray}
Hence, for sufficiently large $n$, we have $\mathbb{E}[T_{1.2.3}]\leq \epsilon,$ if
\begin{eqnarray}
\max\{S_2 + T_2 , S_3 + T_3\} &<& D(p_U || q_U) + I(U;Y_1,V_1) \nonumber \\
&=& \log(\Prime) - H(U | Y_1,V_1). \nonumber 
\end{eqnarray} 

 \textit{\underline{Analysis of $T_{1.2.4}$:}} We evaluate the expectation of $T_{1.2.4}$ with respect to 
 \begin{eqnarray}
 \left(D_1^n(b_1,m_1), D_2^n(a), V_1^n(\tilde{b}_1,\tilde{m}_1), D_1^n(\tilde{b}_1,\tilde{m}_1), U^n(\tilde{a}), D_2^n(\tilde{a}) \right). \nonumber     
 \end{eqnarray}
 We obtain 
\begin{eqnarray}
    \mathbb{E}[T_{1.2.4}] &=& \frac{1}{|\mathcal{M}_1|} \frac{1}{\upsilon^{s_{\sfb} + t_{\sfb}}} \frac{1}{2^{nK_1}} \frac{1}{|\mathcal{D}_1|^{2n}} \frac{1}{|\mathcal{D}_2|^{2n}}  \sum_{m_1} \sum_{a}  \sum_{b_1} \sum_{d_1^n, d_2^n}  \sum_{\tilde{m}_1 \neq m_1} \sum_{\tilde{b}_1}   \sum_{\tilde{a} \neq a} \sum_{\tilde{d}_1^n, \tilde{d}_2^n}  \sum_{v_1^n} \sum_{u^n} \sum_{\tilde{v}_1^n} \sum_{\tilde{u}^n} p_{V_1}^n(\tilde{v}_1^n) q_{U}^n(\tilde{u}^n) \nonumber \\ 
   && p_{V_1U}^n(v_1^n,u^n) \tr \left\{ \gamma^{*}_{\tilde{v}_1^n,\tilde{d}_1^n, \tilde{u}^n, \tilde{d}_2^n} \theta_{v_1^n, d_1^n, u^n, d_2^n} \right\}\nonumber \\
    &\overset{(a)}{=}&  \frac{1}{|\mathcal{D}_1|^{n}} \frac{1}{|\mathcal{D}_2|^{n}}    \sum_{\tilde{m}_1 \neq m_1} \sum_{\tilde{b}_1}   \sum_{\tilde{a} \neq a } \sum_{\tilde{d}_1^n, \tilde{d}_2^n}   \sum_{\tilde{v}_1^n} \sum_{\tilde{u}^n}   p_{V_1}^n(\tilde{v}_1^n) q_{U}^n(\tilde{u}^n) \tr \left\{ \gamma^{*}_{\tilde{v}_1^n,\tilde{d}_1^n, \tilde{u}^n, \tilde{d}_2^n} \theta^{\otimes n } \right\}\nonumber \\
    &\overset{(b)}{\leq}& \frac{1}{|\mathcal{D}_1|^{n}} \frac{1}{|\mathcal{D}_2|^{n}}    \sum_{\tilde{m}_1} \sum_{\tilde{b}_1}   \sum_{\tilde{a}} \sum_{\tilde{d}_1^n, \tilde{d}_2^n}   \sum_{\tilde{v}_1^n} \sum_{\tilde{u}^n}      p_{V_1}^n(\tilde{v}_1^n) q_{U}^n(\tilde{u}^n) \tr \left\{ \gamma^{*}_{\tilde{v}_1^n,\tilde{d}_1^n, \tilde{u}^n, \tilde{d}_2^n} \left(\rho^{Y_1} \otimes \ketbra{0} \right) \right\}\nonumber \\
    &+&\frac{1}{|\mathcal{D}_1|^{n}} \frac{1}{|\mathcal{D}_2|^{n}}    \sum_{\tilde{m}_1} \sum_{\tilde{b}_1}   \sum_{\tilde{a}} \sum_{\tilde{d}_1^n, \tilde{d}_2^n}   \sum_{\tilde{v}_1^n} \sum_{\tilde{u}^n}      p_{V_1}^n(\tilde{v}_1^n) q_{U}^n(\tilde{u}^n) \tr \left\{ \gamma^{*}_{\tilde{v}_1^n,\tilde{d}_1^n, \tilde{u}^n, \tilde{d}_2^n} \mathcal{N}_{\eta}\left(\rho^{Y_1} \otimes \ketbra{0} \right) \right\} \label{Eqn:3to1CQBClastterminhayashinagaokainequality}
\end{eqnarray}
where (a) follows by using the definition $\theta^{\otimes n} \define \frac{1}{|\mathcal{D}_1|^n} \frac{1}{|\mathcal{D}_2|^n} \sum_{d_1^n, d_2^n} \sum_{u^n} \sum_{v_1^n} p_{V_1U}^n(v_1^n,u^n) \theta_{v_1^n,d_1^n,u^n,d_2^n}$, and (b) follows by using 
\begin{eqnarray}
    \theta^{\otimes n } = \frac{1}{1 + 2 \eta^2} \left( \rho^{Y_1} \otimes \ketbra{0}\right) + \mathcal{N}_{\eta}\left( \rho^{Y_1} \otimes \ketbra{0}\right)
\nonumber 
\end{eqnarray}
as stated in \cite{202103SAD_Sen}. Consider the first term in \eqref{Eqn:3to1CQBClastterminhayashinagaokainequality}. We have 
\begin{eqnarray}
&&\frac{1}{|\mathcal{D}_1|^{n}} \frac{1}{|\mathcal{D}_2|^{n}}    \sum_{\tilde{m}_1} \sum_{\tilde{b}_1}   \sum_{\tilde{a}} \sum_{\tilde{d}_1^n, \tilde{d}_2^n}   \sum_{\tilde{v}_1^n} \sum_{\tilde{u}^n}   p_{V_1}^n(\tilde{v}_1^n) q_{U}^n(\tilde{u}^n) \tr \left\{ \gamma^{*}_{\tilde{v}_1^n,\tilde{d}_1^n, \tilde{u}^n, \tilde{d}_2^n} \left(\rho^{Y_1} \otimes \ketbra{0} \right) \right\}\nonumber \\
&\overset{(a)}{\leq}& \frac{1}{|\mathcal{D}_1|^{n}} \frac{1}{|\mathcal{D}_2|^{n}}    \sum_{\tilde{m}_1} \sum_{\tilde{b}_1}   \sum_{\tilde{a}} \sum_{\tilde{d}_1^n, \tilde{d}_2^n}   \sum_{\tilde{v}_1^n} \sum_{\tilde{u}^n}   p_{V_1}^n(\tilde{v}_1^n) q_{U}^n(\tilde{u}^n) \tr \left\{ \left( I_{\boldsymbol{\CalH_{Y_1}^{e}}}-  \beta^0_{\tilde{v}_1^n,\tilde{d}^n_1,\tilde{u}^n,\tilde{d}_2^n} \right) \left(\rho^{Y_1} \otimes \ketbra{0}\right) \right\}\nonumber \\
&\overset{(b)}{=}&  \sum_{\tilde{m}_1} \sum_{\tilde{b}_1}   \sum_{\tilde{a}}  \sum_{\tilde{v}_1^n} \sum_{\tilde{u}^n}   p_{V_1}^n(\tilde{v}_1^n) q_{U}^n(\tilde{u}^n) \tr \left\{ \left( I - \olineB^0_{\tilde{v}_1^n,\tilde{u}^n}\right)  \left(\rho^{Y_1} \otimes \ketbra{0}\right) \right\}\nonumber \\
&\overset{(c)}{=}&  \sum_{\tilde{m}_1} \sum_{\tilde{b}_1}   \sum_{\tilde{a}}  \sum_{\tilde{v}_1^n} \sum_{\tilde{u}^n} p_{V_1}^n(\tilde{v}_1^n) q_{U}^n(\tilde{u}^n) \tr \left\{ G^0_{\tilde{v}_1^n,\tilde{u}^n} \rho^{Y_1}  \right\}\nonumber \\
&\overset{(d)}{\leq}& \sum_{\tilde{m}_1} \sum_{\tilde{b}_1}   \sum_{\tilde{a}}  \sum_{(\tilde{v}_1^n, \tilde{u}^n) \in T_{\delta}^n(p_{V_1U})}   p_{V_1}^n(\tilde{v}_1^n) 2^{-n\left(D(p_U||q_U) - \tilde{\delta} \right)} p_{U}^n(\tilde{u}^n) \tr \left\{ \pi^{Y_1}\pi^{Y_1}_{\tilde{v}_1^n,\tilde{u}^n}\pi^{Y_1}\rho^{Y_1}  \right\}\nonumber \\
&\overset{(e)}{\leq}& \sum_{\tilde{m}_1} \sum_{\tilde{b}_1}   \sum_{\tilde{a}}  \sum_{(\tilde{v}_1^n, \tilde{u}^n) \in T_{\delta}^n(p_{V_1U})} 2^{-n\left(D(p_U||q_U) - \tilde{\delta} \right)} 2^{-n(I(V_1;U) - 3 \delta)} p_{V_1U}^n(\tilde{v}_1^n,\tilde{u}^n)  \tr \left\{  \pi^{Y_1}_{\tilde{v}_1^n,\tilde{u}^n}\pi^{Y_1}\rho^{Y_1} \pi^{Y_1}\right\}\nonumber \\
&\overset{(f)}{\leq}& 2^{-n \left( D(p_U||q_U) + I(V_1;U) - 3 \delta- \tilde{\delta} - (R_1 + K_1 + \max\{S_2 + T_2 , S_3 + T_3\})\right)} 2^{-n\left(I(V_1,U;Y_1) - 2 \delta \right)}, \nonumber 
\end{eqnarray}
where (a) follows from 
\begin{eqnarray}
    \gamma^*_{\tilde{v}_1^n, \tilde{d}_1^n, \tilde{u}^n, \tilde{d}_2^n} \leq \left( I_{\boldsymbol{\CalH_{Y_1}^{e}}}-  \beta^{*}_{\tilde{v}_1^n, \tilde{d}_1^n, \tilde{u}^n, \tilde{d}_2^n}\right) \leq \left( I_{\boldsymbol{\CalH_{Y_1}^{e}}}-  \beta^0_{\tilde{v}_1^n, \tilde{d}_1^n, \tilde{u}^n, \tilde{d}_2^n}\right), \nonumber 
\end{eqnarray}
(b) follows from $\beta^0_{\tilde{v}_1^n, \tilde{d}_1^n, \tilde{u}^n, \tilde{d}_2^n} = \olineB^0_{\tilde{v}_1^n,\tilde{u}^n}$, (c) follows from (i) the definition of $\olineB^0_{\tilde{v}_1^n,\tilde{u}^n}$ in \eqref{Eqn:3to1CQBCcomplementprojector} and from (ii) the Gelfand–Naimark’s Theorem \cite[Thm.~3.7]{BkHolevo_2019}, (d) follows from \eqref{Eqn:3to1CQBCDivergence}
and from the definition of $G^0_{\tilde{v}_1^n,\tilde{u}^n}$ in \eqref{Eqn:3to1CQBCPovmelementfordecoder1}, 
(e) follows from (i) the cyclicity of the trace and from (ii) the inequality 
\begin{eqnarray}
 \sum_{(\tilde{v}_1^n, \tilde{u}^n) \in T_{\delta}^n(p_{V_1U})}   p_{V_1}^n(\tilde{v}_1^n) p_{U}^n(\tilde{u}^n) \leq  \sum_{(\tilde{v}_1^n, \tilde{u}^n) \in T_{\delta}^n(p_{V_1U})}   2^{-n(I(V_1;U) - 3 \delta)} p_{V_1U}^n(\tilde{v}_1^n,\tilde{u}^n), \label{Eqn:Joint2}
\end{eqnarray}
and finally (f) follows from using the standard typical projector property
\begin{eqnarray} 
  &&  \pi^{Y_1}\rho^{Y_1} \pi^{Y_1}\leq 2^{-n \left(H(Y_1) -\delta  \right) },\mbox{ and }  \tr\left \{\pi^{Y_1}_{\tilde{v}_1^n,\tilde{u}^n} \right \} \leq 2^{n\left( H(Y_1|V_1,U) + \delta \right)}, \mbox{ if }(\tilde{v}_1^n,\tilde{u}^n) \in \CalT_{\delta}^n(p_{V_1U}).\nonumber 
\end{eqnarray}
Next, consider the second term in \eqref{Eqn:3to1CQBCthirdterminhayashinagaokainequality}. We have
\begin{eqnarray}
&&\frac{1}{|\mathcal{D}_1|^{n}} \frac{1}{|\mathcal{D}_2|^{n}}    \sum_{\tilde{m}_1} \sum_{\tilde{b}_1}   \sum_{\tilde{a}} \sum_{\tilde{d}_1^n, \tilde{d}_2^n}   \sum_{\tilde{v}_1^n} \sum_{\tilde{u}^n}   p_{V_1}^n(\tilde{v}_1^n) q_{U}^n(\tilde{u}^n) \tr \left\{ \gamma^{*}_{\tilde{v}_1^n,\tilde{d}_1^n, \tilde{u}^n, \tilde{d}_2^n} \mathcal{N}_{\eta}\left(\rho^{Y_1} \otimes \ketbra{0} \right) \right\} \nonumber \\
&\overset{(a)}{\leq}& \frac{1}{|\mathcal{D}_1|^{n}} \frac{1}{|\mathcal{D}_2|^{n}}    \sum_{\tilde{m}_1} \sum_{\tilde{b}_1}   \sum_{\tilde{a}} \sum_{\tilde{d}_1^n, \tilde{d}_2^n} 
\sum_{(\tilde{v}_1^n, \tilde{u}^n)\in T_{\delta}^n(p_{V_1U})}   p_{V_1}^n(\tilde{v}_1^n) 2^{-n\left( D(p_U||q_U) - \tilde{\delta}\right)}p_{U}^n(\tilde{u}^n) \tr \left\{ \gamma^{*}_{\tilde{v}_1^n,\tilde{d}_1^n, \tilde{u}^n, \tilde{d}_2^n} \mathcal{N}_{\eta}\left(\rho^{Y_1} \otimes \ketbra{0} \right) \right\} \nonumber \\
&\overset{(b)}{\leq}& \frac{1}{|\mathcal{D}_1|^{n}} \frac{1}{|\mathcal{D}_2|^{n}}    \sum_{\tilde{m}_1} \sum_{\tilde{b}_1}   \sum_{\tilde{a}} \sum_{\tilde{d}_1^n, \tilde{d}_2^n} 
\sum_{(\tilde{v}_1^n, \tilde{u}^n)\in T_{\delta}^n(p_{V_1U})} 2^{-n\left( D(p_U||q_U) - \tilde{\delta}\right)} 2^{-n(I(V_1;U) - 3 \delta)} p_{V_1U}^n(\tilde{v}_1^n,\tilde{u}^n)   \nonumber \\
&&\norm{ \gamma^{*}_{\tilde{v}_1^n,\tilde{d}_1^n, \tilde{u}^n, \tilde{d}_2^n}}_1 \norm{\mathcal{N}_{\eta}\left(\rho^{Y_1} \otimes \ketbra{0} \right)}_{\infty} \nonumber \\
&\overset{(c)}{\leq}&  2^{-n \left( D(p_U||q_U) + I(V_1;U) - 3 \delta- \tilde{\delta} - (R_1 + K_1 + \max\{S_2 + T_2 , S_3 + T_3\})\right)} 2^n |\mathcal{H}_{Y_1}|^n   \frac{3 \eta }{\sqrt{\max\{|\mathcal{D}_1|^n,|\mathcal{D}_2|^n\}}}. \nonumber 
\end{eqnarray}
where (a) follows from (i) the fact that $\gamma^*_{\tilde{v}_1^n,\tilde{d}_1^n,\tilde{u}^n,\tilde{d}_2^n} = 0$, if $(\tilde{v}_1^n,\tilde{u}^n) \notin T_{\delta}^n(p_{V_1U})$, and from (ii) the inequality given in \eqref{Eqn:3to1CQBCDivergence}. (b) follows from (i) the trace inequality defined in \eqref{Eqn:3to1CQBCTraceIneq} and from (ii) the inequality defined in \eqref{Eqn:Joint2}. (c) follows from \eqref{Eqn:3to1CQBCBndgamma*} and from the inequality
\begin{eqnarray}
   \norm{\CalN_{\eta}\left( \rho^{Y_1} \otimes \ketbra{0} \right)}_{\infty}  \leq \frac{3 \eta }{\sqrt{\max\{|\mathcal{D}_1|^n,|\mathcal{D}_2|^n\}}} \nonumber 
\end{eqnarray}
as stated in \cite{202103SAD_Sen}. Therefore, if we choose 
$\max\{|\mathcal{D}_1|,|\mathcal{D}_2|\}\geq \left( 3 \eta 2^n   |\CalH_{Y_1}|^n  2^{n \left(D(p_{U}||q_{U}) +  I(V_1;U) + I(V_1,U;Y_1) - \tilde{\delta}  - 5 \delta \right)}\right)^{\frac{2}{n}}$, we have 
\begin{eqnarray}
    \mathbb{E}[T_{1.2.4}] &\leq& 2^{-n \left( D(p_U||q_U) + I(V_1;U)  - 3 \delta- \tilde{\delta} - (R_1 + K_1 + \max\{S_2 + T_2 , S_3 + T_3\})\right)} \nonumber \\
    &&\left(  2^{-n \left( I(V_1,U;Y_1) - 2 \delta\right)} + 2^n |\mathcal{H}_{Y_1}|^n   \frac{3 \eta }{\sqrt{\max\{|\mathcal{D}_1|^n,|\mathcal{D}_2|^n\}}}\right) \nonumber \\
    &\leq& 2  \: 2^{-n \left( D(p_U||q_U) +  I(V_1;U) +  I(V_1,U;Y_1)  - \tilde{\delta} - 5 \delta - (R_1 + K_1 + \max\{S_2 + T_2 , S_3 + T_3\})\right)} \nonumber 
\end{eqnarray}
Hence, for sufficiently large $n$, we have $\mathbb{E}[T_{1.2.4}]\leq \epsilon,$ if
\begin{eqnarray}
R_1 + K_1 + \max\{S_2 + T_2 , S_3 + T_3\} &<&  D(p_U || q_U) + I(V_1;U) + I(V_1,U;Y_1)  \nonumber \\
&=& \log(\Prime) + H(V_1) - H(V_1,U | Y_1). \nonumber 
\end{eqnarray}

\section{Proof of Proposition \ref{Prop:Dec2firsttermHay}}
\label{App:Dec2FirsttermHay}
Observe that
\begin{eqnarray}
    \tr \left\{ \Upsilon_{u_2^n,v_2^n} \: \pi^{Y_2}_{v_2^n} \: \rho_{u_2^n, v_2^n}^{Y_2} \: \pi^{Y_2}_{v_2^n}  \right\} &\overset{(a)}{=}& \tr \left\{ \pi^{Y_2} \:  \pi^{Y_2}_{u_2^n} \:  \pi^{Y_2}_{u_2^n, v_2^n} \: \pi^{Y_2}_{u_2^n} \: \pi^{Y_2}  \: \pi^{Y_2}_{v_2^n} \: \rho_{u_2^n, v_2^n}^{Y_2} \: \pi^{Y_2}_{v_2^n}  \right\} \nonumber \\ 
    &\overset{(b)}{=}& \tr \left\{ \pi^{Y_2}_{u_2^n, v_2^n} \: \pi^{Y_2}_{u_2^n} \: \pi^{Y_2}  \: \pi^{Y_2}_{v_2^n} \: \rho_{u_2^n, v_2^n}^{Y_2} \: \pi^{Y_2}_{v_2^n} \: \pi^{Y_2} \: \pi^{Y_2}_{u_2^n}    \right\} \nonumber \\ 
    &\overset{(c)}{\geq}& \tr \left\{ \pi^{Y_2}_{u_2^n, v_2^n} \:  \rho_{u_2^n, v_2^n}^{Y_2}\right\} 
    - \norm{\pi^{Y_2}_{v_2^n} \: \rho_{u_2^n, v_2^n}^{Y_2} \: \pi^{Y_2}_{v_2^n}-\rho_{u_2^n, v_2^n}^{Y_2}}_1 
    - \norm{\pi^{Y_2} \: \rho_{u_2^n, v_2^n}^{Y_2} \: \pi^{Y_2}-\rho_{u_2^n, v_2^n}^{Y_2}}_1 \nonumber \\
    &&- \norm{\pi^{Y_2}_{u_2^n} \: \rho_{u_2^n, v_2^n}^{Y_2} \:  \pi^{Y_2}_{u_2^n}-\rho_{u_2^n, v_2^n}^{Y_2}}_1
    \nonumber \\
    &\overset{(d)}{\geq}& 1- \epsilon - 6 \sqrt{\epsilon}, \nonumber
\end{eqnarray}
where (a) follows from the definition of 
$\Upsilon_{u_2^n,v_2^n}$ in \eqref{Eqn:3to1CQBCPovmelementfordecoder2}, (b) follows from the cyclicity of the trace, (c) follows from applying the inequality $\tr(\Delta \rho) \geq \tr(\Delta \sigma) - \norm{\rho - \sigma}_1$, for $0 \leq \Delta, \rho, \sigma \leq I$, three times, and (d) follows from (i) 
\begin{eqnarray}
    \tr\left\{ \pi^{Y_2}_{u_2^n,v_2^n} \rho^{Y_2}_{u_2^n,v_2^n}\right\} \geq 1 - \epsilon, \nonumber
\end{eqnarray}
and from (ii) the Gentle Operator Lemma \cite{BkWilde_2017}, since each typical projector satisfies
\begin{eqnarray}
    \tr\left \{ \pi^{Y_2}_{u_2^n} \: \rho^{Y_2}_{u_2^n,v_2^n} \right\} \geq 1- \epsilon, \: 
     \tr\left \{ \pi^{Y_2}_{v_2^n} \: \rho^{Y_2}_{u_2^n,v_2^n} \right\} \geq 1- \epsilon, \mbox{ and } \tr\left \{ \pi^{Y_2} \: \rho^{Y_2}_{u_2^n,v_2^n} \right\} \geq 1- \epsilon. \nonumber
\end{eqnarray}
Therefore, we conclude
\begin{eqnarray}
    \mathbb{E}[T_{2.2.1}] &\leq& 1 - (1 - \epsilon - 6 \sqrt{\epsilon}) = \epsilon + 6 \sqrt{\epsilon}. \nonumber
\end{eqnarray}

\section{Proof of Proposition \ref{Prop:Dec2SecondtermHay}}
\label{App:Dec2SecondtermHay}

\textit{\underline{Analysis of $T_{2.2.2}$:}}
We evaluate the expectation of $T_{2.2.2}$ over the random choice of $V^n_2(\tilde{b}_2,\tilde{m}_{22})$. We obtain 
\begin{eqnarray}
    \mathbb{E}[T_{2.2.2}] &=& \frac{1}{2^{n(R_2 + S_2 + K_2)}} \sum_{m_2} \sum_{a_2} \sum_{b_2} \sum_{\tilde{m}_{22} \neq m_{22}} \sum_{\tilde{b}_2} \sum_{u_2^n} \sum_{v_2^n \in T_{\delta}^n(p_{V_2})} \sum_{\tilde{v}_2^n} p_{V_2}^n(\tilde{v}_2^n)  p_{U_2V_2}^n(u_2^n,v_2^n)
    \tr \left\{ \Upsilon_{u_2^n,\tilde{v}_2^n}  \pi^{Y_2}_{v_2^n}  \rho_{u_2^n, v_2^n}^{Y_2}  \pi^{Y_2}_{v_2^n}  \right\} \nonumber \\
&\overset{(a)}{=}&\frac{1}{2^{n(R_2 + S_2+ K_2)}} \sum_{m_2} \sum_{a_2} \sum_{b_2} \sum_{\tilde{m}_{22} \neq m_{22}} \sum_{\tilde{b}_2} \sum_{u_2^n \in T_{\delta}^n(p_{U_2})} \sum_{v_2^n \in T_{\delta}^n(p_{V_2})} \sum_{\tilde{v}_2^n \in T_{\delta}^n(p_{V_2|U_2})} p_{V_2}^n(\tilde{v}_2^n)  p_{U_2V_2}^n(u_2^n,v_2^n) \nonumber \\
&&    \tr \left\{ \pi^{Y_2}  \pi^{Y_2}_{u_2^n}  \pi^{Y_2}_{u_2^n, \tilde{v}_2^n}  \pi^{Y_2}_{u_2^n}  \pi^{Y_2}  \pi^{Y_2}_{v_2^n}  \rho_{u_2^n, v_2^n}^{Y_2}  \pi^{Y_2}_{v_2^n}  \right\} \nonumber \\
&\overset{(b)}{\leq}& \frac{1}{2^{n(R_2 + S_2+ K_2)}} \sum_{m_2} \sum_{a_2} \sum_{b_2} \sum_{\tilde{m}_{22}} \sum_{\tilde{b}_2} \sum_{u_2^n \in T_{\delta}^n(p_{U_2})} \sum_{v_2^n \in T_{\delta}^n(p_{V_2})} \sum_{\tilde{v}_2^n \in T_{\delta}^n(p_{V_2|U_2})} 2^{-n(I(V_2;U_2) - 3 \delta)} 2^{n\left( H(Y_2|U_2, V_2) - \delta \right)}     \nonumber \\ 
    && p_{V_2|U_2}^n(\tilde{v}_2^n|u_2^n)  p_{U_2V_2}^n(u_2^n,v_2^n)
    \tr \left\{ \pi^{Y_2}  \pi^{Y_2}_{u_2^n}  \rho^{Y_2}_{u_2^n, \tilde{v}_2^n}  \pi^{Y_2}_{u_2^n}  \pi^{Y_2}  \pi^{Y_2}_{v_2^n}  \rho_{u_2^n, v_2^n}^{Y_2}  \pi^{Y_2}_{v_2^n}  \right\} \nonumber \\
&\overset{(c)}{\leq}& \frac{1}{2^{n(R_2 + S_2+ K_2)}} \sum_{m_2} \sum_{a_2} \sum_{b_2} \sum_{\tilde{m}_{22}} \sum_{\tilde{b}_2} \sum_{u_2^n \in T_{\delta}^n(p_{U_2})} \sum_{v_2^n \in T_{\delta}^n(p_{V_2})}  2^{-n(I(V_2;U_2) - 3 \delta)} 2^{n\left( H(Y_2|U_2, V_2) - \delta \right)}p_{U_2V_2}^n(u_2^n,v_2^n)
     \nonumber \\ &&\tr \left\{ \pi^{Y_2}  \pi^{Y_2}_{u_2^n}  \rho^{Y_2}_{u_2^n}  \pi^{Y_2}_{u_2^n}  \pi^{Y_2}  \pi^{Y_2}_{v_2^n}  \rho_{u_2^n, v_2^n}^{Y_2}  \pi^{Y_2}_{v_2^n}  \right\} \nonumber \\
&\overset{(d)}{\leq}&\frac{1}{2^{n(R_2 + S_2+ K_2)}} \sum_{m_2} \sum_{a_2} \sum_{b_2} \sum_{\tilde{m}_{22}} \sum_{\tilde{b}_2} \sum_{u_2^n \in T_{\delta}^n(p_{U_2})} \sum_{v_2^n \in T_{\delta}^n(p_{V_2})} 2^{-n(I(V_2;U_2) - 3 \delta)}  2^{n\left( H(Y_2|U_2, V_2) - \delta \right)}  2^{-n\left( H(Y_2|U_2) - \delta \right)} \nonumber \\
    &&p_{U_2V_2}^n(u_2^n,v_2^n) \tr \left\{ \pi^{Y_2}  \pi^{Y_2}_{u_2^n}  \pi^{Y_2}  \pi^{Y_2}_{v_2^n}  \rho_{u_2^n, v_2^n}^{Y_2}  \pi^{Y_2}_{v_2^n}  \right\} \nonumber \\
&\overset{(e)}{\leq}& \frac{1}{2^{n(R_2 + S_2+ K_2)}} \sum_{m_2} \sum_{a_2} \sum_{b_2} \sum_{\tilde{m}_{22}} \sum_{\tilde{b}_2} \sum_{u_2^n \in T_{\delta}^n(p_{U_2})} \sum_{v_2^n \in T_{\delta}^n(p_{V_2})} 2^{-n\left( I(V_2; Y_2,U_2) - 5 \delta \right)} p_{U_2V_2}^n(u_2^n,v_2^n) \tr \left\{ \pi^{Y_2}_{v_2^n}  \rho_{u_2^n, v_2^n}^{Y_2}  \pi^{Y_2}_{v_2^n}  \right\} \nonumber \\
&\overset{(f)}{\leq}& \frac{1}{2^{n(R_2 + S_2+ K_2)}} \sum_{m_2} \sum_{a_2} \sum_{b_2} \sum_{\tilde{m}_{22}} \sum_{\tilde{b}_2} \sum_{v_2^n \in T_{\delta}^n(p_{V_2})} 2^{-n\left( I(V_2; Y_2,U_2) - 5 \delta \right)} p_{V_2}^n(v_2^n) \tr \left\{ \pi^{Y_2}_{v_2^n}  \rho_{v_2^n}^{Y_2}  \pi^{Y_2}_{v_2^n}  \right\} \nonumber \\
&\overset{(g)}{=}& 2^{-n\left( I(V_2; Y_2,U_2) - 5 \delta - (L_2+K_2) \right)} \nonumber 
    \end{eqnarray}
where (a) follows from the definition of $\Upsilon_{u_2^n,\tilde{v}_2^n}$ in \eqref{Eqn:3to1CQBCPovmelementfordecoder2}, (b) follows from (i) the inequality 
\begin{eqnarray}
p_{V_2}^n(\tilde{v}_2^n) \leq 2^{-n(I(V_2;U_2) - 3 \delta)} p_{V_2|U_2}^n(\tilde{v}_2^n|u_2^n), \mbox{ for }  \tilde{v}_2^n \in T_{\delta}^n(p_{V_2|U_2}) \nonumber 
\end{eqnarray}
and from (ii) the fact that if $(u_2^n,\tilde{v}_2^n) \in T_{\delta}^n(p_{U_2V_2})$, then 
\begin{eqnarray}
    \pi^{Y_2}_{u_2^n,\tilde{v}_2^n} &\leq& 2^{n\left( H(Y_2|U_2,V_2) - \delta \right)} \: \pi^{Y_2}_{u_2^n, \tilde{v}_2^n} \: \rho^{Y_2}_{u_2^n, \tilde{v}_2^n} \: \pi^{Y_2}_{u_2^n, \tilde{v}_2^n}  \nonumber \\
    &\leq&  2^{n\left( H(Y_2|U_2,V_2) - \delta \right)} \: \pi^{Y_2}_{u_2^n, \tilde{v}_2^n} \:  \sqrt{\rho^{Y_2}_{u_2^n, \tilde{v}_2^n}} \:  \sqrt{\rho^{Y_2}_{u_2^n, \tilde{v}_2^n}}  \:  \pi^{Y_2}_{u_2^n, \tilde{v}_2^n}  \nonumber \\
    &\leq&  2^{n\left( H(Y_2|U_2,V_2) - \delta \right)} \:  \sqrt{\rho^{Y_2}_{u_2^n, \tilde{v}_2^n}} \:  \pi^{Y_2}_{u_2^n, \tilde{v}_2^n} \:  \sqrt{\rho^{Y_2}_{u_2^n, \tilde{v}_2^n}}    \nonumber \\
    &\leq&  2^{n\left( H(Y_2|U_2,V_2) - \delta \right)} \:  \rho^{Y_2}_{u_2^n, \tilde{v}_2^n},  \nonumber 
\end{eqnarray}
(c) follows from $\rho^{Y_2}_{u_2^n} = \sum_{\tilde{v}_2^n} p_{V_2 |U_2}^n(\tilde{v}_2^n | u_2^n) \rho^{Y_2}_{u_2^n,\tilde{v}_2^n}$, (d) follows from 
\begin{eqnarray}
    \pi^{Y_2}_{u_2^n} \:  \rho^{Y_2}_{u_2^n} 
    \pi^{Y_2}_{u_2^n} \leq 2^{-n \left( H(Y_2|U_2) - \delta \right) } \: \pi^{Y_2}_{u_2^n},  \mbox{ for }  u_2^n \in T_{\delta}^n(p_{U_2}) \nonumber 
\end{eqnarray}
(e) follows from 
\begin{eqnarray}
    \pi^{Y_2} \: \pi^{Y_2}_{u_2^n} \: \pi^{Y_2} \leq \pi^{Y_2} \leq I,  \nonumber 
\end{eqnarray}
(f) follows from $\rho^{Y_2}_{v_2^n} = \sum_{u_2^n} p_{U_2|V_2}^n(u_2^n |v_2^n) \rho^{Y_2}_{u_2^n,v_2^n}$, and finally (g) follows from the cyclicity of the trace and from $\pi^{Y_2}_{v_2^n} \leq I$.
Therefore, for $n$ sufficiently large, we have
\begin{eqnarray}
    \mathbb{E}[T_{2.2.2}] \leq \epsilon,  \mbox{ if }  L_2 + K_2 < I(V_2;Y_2,U_2). \nonumber
\end{eqnarray}

\textit{\underline{Analysis of $T_{2.2.3}$:}} 
We evaluate the expectation of $T_{2.2.3}$ with respect to the random choice of $U^n_2(\tilde{a}_2,\tilde{m}_{21})$. We obtain 
\begin{eqnarray}
 \mathbb{E}[T_{2.2.3}] &=& \frac{1}{2^{n(R_2 + S_2 + K_2)}} \sum_{m_2} \sum_{a_2} \sum_{b_2} \sum_{\tilde{m}_{21} \neq m_{21}} \sum_{\tilde{a}_2} \sum_{u_2^n} \sum_{v_2^n \in T_{\delta}^n(p_{V_2})} \sum_{\tilde{u}_2^n} q_{U_2}^n(\tilde{u}_2^n) p_{U_2V_2}^n(u_2^n,v_2^n)
    \tr \left\{ \Upsilon_{\tilde{u}_2^n,v_2^n} \: \pi^{Y_2}_{v_2^n} \: \rho_{u_2^n, v_2^n}^{Y_2} \: \pi^{Y_2}_{v_2^n}  \right\} \nonumber \\
&\overset{(a)}{=}& \frac{1}{2^{n(R_2 + S_2 + K_2)}} \sum_{m_2} \sum_{a_2} \sum_{b_2} \sum_{\tilde{m}_{21} \neq m_{21}} \sum_{\tilde{a}_2} \sum_{v_2^n \in T_{\delta}^n(p_{V_2})}  \sum_{\tilde{u}_2^n} q_{U_2}^n(\tilde{u}_2^n) p_{V_2}^n(v_2^n)
    \tr \left\{ \Upsilon_{\tilde{u}_2^n,v_2^n} \: \pi^{Y_2}_{v_2^n} \: \rho_{v_2^n}^{Y_2} \: \pi^{Y_2}_{v_2^n}  \right\} \nonumber \\
&\overset{(b)}{\leq}& \frac{1}{2^{n(R_2 + S_2 + K_2)}} \sum_{m_2} \sum_{a_2} \sum_{b_2} \sum_{\tilde{m}_{21}} \sum_{\tilde{a}_2} \sum_{v_2^n \in T_{\delta}^n(p_{V_2})}  \sum_{\tilde{u}_2^n} 2^{-n\left( H(Y_2|V_2) - \delta\right)} q_{U_2}^n(\tilde{u}_2^n) p_{V_2}^n(v_2^n) 
    \tr \left\{ \Upsilon_{\tilde{u}_2^n,v_2^n} \:  \pi^{Y_2}_{v_2^n}  \right\} \nonumber \\
&\overset{(c)}{=}& \frac{1}{2^{n(R_2 + S_2 + K_2)}} \sum_{m_2} \sum_{a_2} \sum_{b_2} \sum_{\tilde{m}_{21}} \sum_{\tilde{a}_2} \sum_{(\tilde{u}_2^n,v_2^n) \in T_{\delta}^n(p_{U_2V_2})} 2^{-n\left( H(Y_2|V_2) - \delta\right)} q_{U_2}^n(\tilde{u}_2^n) p_{V_2}^n(v_2^n) 
  \nonumber \\
  &&\tr \left\{\pi^{Y_2} \: \pi^{Y_2}_{\tilde{u}_2^n} \: \pi^{Y_2}_{\tilde{u}_2^n,v_2^n} \: \pi^{Y_2}_{\tilde{u}_2^n} \: \pi^{Y_2} \:  \pi^{Y_2}_{v_2^n}  \right\} \nonumber \\
&\overset{(d)}{=}& \frac{1}{2^{n(R_2 + S_2 + K_2)}} \sum_{m_2} \sum_{a_2} \sum_{b_2} \sum_{\tilde{m}_{21}} \sum_{\tilde{a}_2} \sum_{(\tilde{u}_2^n,v_2^n) \in T_{\delta}^n(p_{U_2V_2})} 2^{-n\left( H(Y_2|V_2) - \delta\right)}  q_{U_2}^n(\tilde{u}_2^n) p_{V_2}^n(v_2^n) 
   \nonumber \\
  && \tr \left\{\pi^{Y_2}_{\tilde{u}_2^n,v_2^n} \: \pi^{Y_2}_{\tilde{u}_2^n} \: \pi^{Y_2} \:  \pi^{Y_2}_{v_2^n} \: \pi^{Y_2} \: \pi^{Y_2}_{\tilde{u}_2^n}     \right\} \nonumber \\
&\overset{(e)}{\leq}& \frac{1}{2^{n(R_2 + S_2 + K_2)}} \sum_{m_2} \sum_{a_2} \sum_{b_2} \sum_{\tilde{m}_{21}} \sum_{\tilde{a}_2} \sum_{(\tilde{u}_2^n,v_2^n) \in T_{\delta}^n(p_{U_2V_2})} 2^{-n\left( H(Y_2|V_2) - \delta\right)} q_{U_2}^n(\tilde{u}_2^n) p_{V_2}^n(v_2^n) 
    \tr \left\{\pi^{Y_2}_{\tilde{u}_2^n,v_2^n}\right\} \nonumber \\
&\overset{(f)}{\leq}& \frac{1}{2^{n(R_2 + S_2 + K_2)}} \sum_{m_2} \sum_{a_2} \sum_{b_2} \sum_{\tilde{m}_{21}} \sum_{\tilde{a}_2} \sum_{(\tilde{u}_2^n,v_2^n) \in T_{\delta}^n(p_{U_2V_2})} 2^{-n\left( H(Y_2|V_2) - \delta\right)}
    2^{n\left( H(Y_2 |U_2,V_2) + \delta \right)} q_{U_2}^n(\tilde{u}_2^n) p_{V_2}^n(v_2^n) \nonumber \\
&\overset{(g)}{\leq}& \frac{1}{2^{n(R_2 + S_2 + K_2)}} \sum_{m_2} \sum_{a_2} \sum_{b_2} \sum_{\tilde{m}_{21}} \sum_{\tilde{a}_2} \sum_{(\tilde{u}_2^n,v_2^n) \in T_{\delta}^n(p_{U_2V_2})}  \!\!\!2^{-n\left( D(p_{U_2} || q_{U_2}) - \delta_2\right)} 2^{-n\left( I(U_2;Y_2 |V_2) - 2 \delta \right)}p_{U_2}^n(\tilde{u}_2^n) p_{V_2}^n(v_2^n) \nonumber \\
&\overset{(h)}{\leq}& \frac{1}{2^{n(R_2 + S_2 + K_2)}} \sum_{m_2} \sum_{a_2} \sum_{b_2} \sum_{\tilde{m}_{21}} \sum_{\tilde{a}_2} \sum_{(\tilde{u}_2^n,v_2^n) \in T_{\delta}^n(p_{U_2V_2})} \!\!\! 2^{-n\left( D(p_{U_2} || q_{U_2}) - \delta_2\right)} 2^{-n(I(U_2;V_2) - 3 \delta )} \nonumber \\
  &&2^{-n\left( I(U_2;Y_2 |V_2) - 2 \delta \right)}p_{U_2V_2}^n(\tilde{u}_2^n,v_2^n) \nonumber \\
&\leq& 2^{-n \left( D(p_{U_2} ||q_{U_2}) + I(U_2;Y_2,V_2)  - 2 \delta - \delta_2 - (S_2+ T_2)\right)}\nonumber,
\end{eqnarray}
 where (a) follows from $\rho^{Y_2}_{v_2^n} = \sum_{u_2^n} p^n_{U_2|V_2}(u_2^n|v_2^n) \rho^{Y_2}_{u_2^n,v_2^n}$, (b) follows from 
 \begin{eqnarray}
     \pi^{Y_2}_{v_2^n} \: \rho^{Y_2}_{v_2^n} \: \pi^{Y_2}_{v_2^n} \leq 2^{-n \left( H(Y_2|V_2) - \delta \right)} \pi^{Y_2}_{v_2^n}, \mbox{ if }  v_2^n \in T_{\delta}^n(p_{V_2}), \nonumber  
 \end{eqnarray}
 (c) follows from the definition of $\Upsilon_{\tilde{u}_2^n,v_2^n}$ in \eqref{Eqn:3to1CQBCPovmelementfordecoder2}, (d) follows from the cyclicity of the trace, (e) follows from 
 \begin{eqnarray}
      \pi^{Y_2}_{\tilde{u}_2^n} \: \pi^{Y_2} \:  \pi^{Y_2}_{v_2^n} \: \pi^{Y_2} \: \pi^{Y_2}_{\tilde{u}_2^n}      \leq  \pi^{Y_2}_{\tilde{u}_2^n} \:  \pi^{Y_2} \: \pi^{Y_2}_{\tilde{u}_2^n}    \leq \pi^{Y_2}_{\tilde{u}_2^n} \leq I, \nonumber       
 \end{eqnarray}
 (f) follows from 
 \begin{eqnarray}
     \tr\{\pi^{Y_2}_{\tilde{u}_2^n,v_2^n}\} \leq 2^{n\left( H(Y_2 |U_2,V_2) + \delta \right)}, \mbox{ for }  (\tilde{u}_2^n,v_2^n) \in T_{\delta}^n(p_{U_2V_2}), \nonumber 
 \end{eqnarray}
(g) follows from 
 \begin{eqnarray}
     q_{U_2}^n(\tilde{u}_2^n) \leq 2^{-n \left( D(p_{U_2} || q_{U_2}) - \delta_2\right)} p_{U_2}^n(\tilde{u}_2^n), \mbox{ if }  \tilde{u}_2^n \in T_{\delta}^n(p_{U_2}) ,\label{Eqn:3TO1CQBCDec2Divergence} 
 \end{eqnarray}
and finally (h) follows from 
\begin{eqnarray}
\sum_{(\tilde{u}_2^n,v_2^n) \in T_{\delta}^n(p_{U_2V_2})} p_{U_2}^n(\tilde{u}_2^n) p_{V_2}^n(v_2^n) \leq 2^{-n(I(U_2;V_2) - 3 \delta )}p_{U_2V_2}^n(\tilde{u}_2^n,v_2^n).\nonumber 
\end{eqnarray}
Therefore, for $n$ sufficiently large, we have $\mathbb{E}[T_{2.2.3}] \leq \epsilon,$ if
\begin{eqnarray}
S_2 + T_2 &<& D(p_{U_2} || q_{U_2}) + I(U_2;Y_2,V_2)  \nonumber \\
&=&  \log(\Prime) - H(U_2|Y_2,V_2). \nonumber    
\end{eqnarray}

\textit{\underline{Analysis of $T_{2.2.4}$:}}
We evaluate the expectation of $T_{2.2.4}$ with respect to the random choice of $U^n_2(\tilde{a}_2,\tilde{m}_{21})$ and $V_2^n(\tilde{b}_2,\tilde{m}_{21})$. We obtain 
\begin{eqnarray}
 \mathbb{E}[T_{2.2.4}] &=& \frac{1}{2^{n(R_2 + S_2 + K_2)}} \sum_{m_2} \sum_{a_2} \sum_{b_2} \sum_{\tilde{m}_{22} \neq m_{22}} \sum_{\tilde{b}_2}\sum_{\tilde{m}_{21} \neq m_{21}} \sum_{\tilde{a}_2} \sum_{u_2^n} \sum_{v_2^n} \sum_{\tilde{u}_2^n} \sum_{\tilde{v}_2^n}  q_{U_2}^n(\tilde{u}_2^n) p_{V_2}^n(\tilde{v}_2^n) p_{U_2V_2}^n(u_2^n,v_2^n) \nonumber \\ &&\tr\left\{ \Upsilon_{\tilde{u}_2^n, \tilde{v}_2^n} \: \pi^{Y_2}_{v_2^n} \: \rho_{u_2^n, v_2^n}^{Y_2} \: \pi^{Y_2}_{v_2^n}  \right\} \nonumber \\
    &\overset{(a)}{=}& \frac{1}{2^{n(R_2 + S_2 + K_2)}} \sum_{m_2} \sum_{a_2} \sum_{b_2} \sum_{\tilde{m}_{22} \neq m_{22}} \sum_{\tilde{b}_2}\sum_{\tilde{m}_{21} \neq m_{21}} \sum_{\tilde{a}_2} \sum_{v_2^n} \sum_{\tilde{u}_2^n} \sum_{\tilde{v}_2^n}  q_{U_2}^n(\tilde{u}_2^n) p_{V_2}^n(\tilde{v}_2^n) p_{V_2}^n(v_2^n) \nonumber \\&&\tr\left\{ \Upsilon_{\tilde{u}_2^n, \tilde{v}_2^n} \: \pi^{Y_2}_{v_2^n} \: \rho_{v_2^n}^{Y_2} \: \pi^{Y_2}_{v_2^n}  \right\} \nonumber \\
    &\overset{(b)}{\leq}& \frac{1}{2^{n(R_2 + S_2 + K_2)}} \sum_{m_2} \sum_{a_2} \sum_{b_2} \sum_{\tilde{m}_{22}} \sum_{\tilde{b}_2}\sum_{\tilde{m}_{21}} \sum_{\tilde{a}_2} \sum_{v_2^n} \sum_{\tilde{u}_2^n} \sum_{\tilde{v}_2^n}  q_{U_2}^n(\tilde{u}_2^n) p_{V_2}^n(\tilde{v}_2^n) p_{V_2}^n(v_2^n) \tr\left\{ \Upsilon_{\tilde{u}_2^n, \tilde{v}_2^n} \:  \rho_{v_2^n}^{Y_2} \right\} \nonumber \\
    &\overset{(c)}{=}& \frac{1}{2^{n(R_2 + S_2 + K_2)}} \sum_{m_2} \sum_{a_2} \sum_{b_2} \sum_{\tilde{m}_{22}} \sum_{\tilde{b}_2}\sum_{\tilde{m}_{21}} \sum_{\tilde{a}_2}  \sum_{\tilde{u}_2^n} \sum_{\tilde{v}_2^n}  q_{U_2}^n(\tilde{u}_2^n) p_{V_2}^n(\tilde{v}_2^n) \tr\left\{ \Upsilon_{\tilde{u}_2^n, \tilde{v}_2^n} \:  \left(\rho^{Y_2}\right)^{\otimes n } \right\} \nonumber \\
    &\overset{(d)}{=}& \frac{1}{2^{n(R_2 + S_2 + K_2)}} \sum_{m_2} \sum_{a_2} \sum_{b_2} \sum_{\tilde{m}_{22}} \sum_{\tilde{b}_2}\sum_{\tilde{m}_{21}} \sum_{\tilde{a}_2}  \sum_{(\tilde{u}_2^n,\tilde{v}_2^n) \in T_{\delta}^n(p_{U_2V_2})}  q_{U_2}^n(\tilde{u}_2^n) p_{V_2}^n(\tilde{v}_2^n) \nonumber \\&&\tr\left\{ \pi^{Y_2} \: \pi^{Y_2}_{\tilde{u}_2^n} \: \pi^{Y_2}_{\tilde{u}_2^n, \tilde{v}_2^n} \: \pi^{Y_2}_{\tilde{u}_2^n} \: \pi^{Y_2} \:    \left(\rho^{Y_2}\right)^{\otimes n } \right\} \nonumber \\
       &\overset{(e)}{=}& \frac{1}{2^{n(R_2 + S_2 + K_2)}} \sum_{m_2} \sum_{a_2} \sum_{b_2} \sum_{\tilde{m}_{22}} \sum_{\tilde{b}_2}\sum_{\tilde{m}_{21}} \sum_{\tilde{a}_2}  \sum_{(\tilde{u}_2^n,\tilde{v}_2^n) \in T_{\delta}^n(p_{U_2V_2})}  q_{U_2}^n(\tilde{u}_2^n) p_{V_2}^n(\tilde{v}_2^n) \nonumber \\&&\tr\left\{  \pi^{Y_2}_{\tilde{u}_2^n, \tilde{v}_2^n} \: \pi^{Y_2}_{\tilde{u}_2^n} \: \pi^{Y_2} \:    \left(\rho^{Y_2}\right)^{\otimes n } \pi^{Y_2} \: \pi^{Y_2}_{\tilde{u}_2^n} \:\right\} \nonumber \\
    &\overset{(f)}{\leq}& \frac{1}{2^{n(R_2 + S_2 + K_2)}} \sum_{m_2} \sum_{a_2} \sum_{b_2} \sum_{\tilde{m}_{22}} \sum_{\tilde{b}_2}\sum_{\tilde{m}_{21}} \sum_{\tilde{a}_2}  \sum_{(\tilde{u}_2^n,\tilde{v}_2^n) \in T_{\delta}^n(p_{U_2V_2})} 2^{-n \left( H(Y_2) - \delta \right)} q_{U_2}^n(\tilde{u}_2^n) p_{V_2}^n(\tilde{v}_2^n) \nonumber \\&&  \tr\left\{  \pi^{Y_2}_{\tilde{u}_2^n, \tilde{v}_2^n} \: \pi^{Y_2}_{\tilde{u}_2^n} \: \pi^{Y_2} \: \pi^{Y_2}_{\tilde{u}_2^n} \right\} \nonumber \\
    &\overset{(g)}{\leq}& \frac{1}{2^{n(R_2 + S_2 + K_2)}} \sum_{m_2} \sum_{a_2} \sum_{b_2} \sum_{\tilde{m}_{22}} \sum_{\tilde{b}_2}\sum_{\tilde{m}_{21}} \sum_{\tilde{a}_2}  \sum_{(\tilde{u}_2^n,\tilde{v}_2^n) \in T_{\delta}^n(p_{U_2V_2})} 2^{-n \left( H(Y_2) - \delta \right)} q_{U_2}^n(\tilde{u}_2^n) p_{V_2}^n(\tilde{v}_2^n)  \tr\left\{  \pi^{Y_2}_{\tilde{u}_2^n, \tilde{v}_2^n}\right\} \nonumber \\
    &\overset{(h)}{\leq}& \frac{1}{2^{n(R_2 + S_2 + K_2)}} \sum_{m_2} \sum_{a_2} \sum_{b_2} \sum_{\tilde{m}_{22}} \sum_{\tilde{b}_2}\sum_{\tilde{m}_{21}} \sum_{\tilde{a}_2}  \sum_{(\tilde{u}_2^n,\tilde{v}_2^n) \in T_{\delta}^n(p_{U_2V_2})}  2^{-n \left( D(p_{U_2}|| q_{U_2}) - \delta_2 \right)} 2^{-n \left(I(U_2,V_2;Y_2) - 2 \delta\right)} \nonumber \\&& p_{U_2}^n(\tilde{u}_2^n) p_{V_2}^n(\tilde{v}_2^n)  \nonumber \\
    &\overset{(i)}{\leq}& \frac{1}{2^{n(R_2 + S_2 + K_2)}} \sum_{m_2} \sum_{a_2} \sum_{b_2} \sum_{\tilde{m}_{22}} \sum_{\tilde{b}_2}\sum_{\tilde{m}_{21}} \sum_{\tilde{a}_2}  \sum_{(\tilde{u}_2^n,\tilde{v}_2^n) \in T_{\delta}^n(p_{U_2V_2})}  2^{-n \left( D(p_{U_2}|| q_{U_2}) - \delta_2 \right)} 2^{-n(I(U_2;V_2) - 3 \delta )} \nonumber \\&& 2^{-n \left(I(U_2,V_2;Y_2) - 2 \delta\right)}p_{U_2V_2}^n(\tilde{u}_2^n,\tilde{v}_2^n)  \nonumber \\
    &\leq& 2^{-n \left(D(p_{U_2} || q_{U_2}) + I(U_2;V_2) +  I(U_2,V_2;Y_2)   - 5 \delta - \delta_2 -(L_2 + K_2 + S_2 + T_2)\right)} \nonumber  
\end{eqnarray}
where (a) follows from $\rho^{Y_2}_{v_2^n} = \sum_{u_2^n} p_{U_2|V_2}^n(u_2^n|v_2^n) \rho^{Y_2}_{u_2^n,v_2^n}$, (b) follows from
\begin{eqnarray}
    \pi^{Y_2}_{v_2^n} \: \rho^{Y_2}_{v_2^n} \: \pi^{Y_2}_{v_2^n} = \pi^{Y_2}_{v_2^n} \: \sqrt{\rho^{Y_2}_{v_2^n}} \: \sqrt{\rho^{Y_2}_{v_2^n}} \: \pi^{Y_2}_{v_2^n} = \sqrt{\rho^{Y_2}_{v_2^n}} \: \pi^{Y_2}_{v_2^n} \: \sqrt{\rho^{Y_2}_{v_2^n}}   \leq \rho^{Y_2}_{v_2^n}, \nonumber 
\end{eqnarray}
(c) follows from $\left( \rho^{Y_2} \right)^{\otimes n} = \sum_{v_2^n} p_{V_2}^n(v_2^n) \rho^{Y_2}_{v_2^n}$, (d) follows from the definition of $\Upsilon_{\tilde{u}_2^n,\tilde{v}_2^n}$ in \eqref{Eqn:3to1CQBCPovmelementfordecoder2}, (e) follows from the cyclicity of the trace, (f) follows from
\begin{eqnarray}
    \pi^{Y_2} \left(\rho^{Y_2} \right)^{\otimes n} \pi^{Y_2} \leq 2^{-n\left( H(Y_2) - \delta \right) } \pi^{Y_2}, \nonumber 
\end{eqnarray}
(g) follows from 
\begin{eqnarray}
    \pi^{Y_2}_{\tilde{u}_2^n} \: \pi^{Y_2} \: \pi^{Y_2}_{\tilde{u}_2^n} \leq \pi^{Y_2}_{\tilde{u}_2^n} \leq I, \nonumber
\end{eqnarray}
(h) follows from \eqref{Eqn:3TO1CQBCDec2Divergence} and from 
\begin{eqnarray}
    \tr\{\pi^{Y_2}_{\tilde{u}_2^n,\tilde{v}_2^n}\} \leq 2^{n\left( H(Y_2|U_2,V_2) - \delta \right)},  \mbox{ for } (\tilde{u}_2^n,\tilde{v}_2^n) \in T_{\delta}^n(p_{U_2V_2}), \nonumber  
\end{eqnarray}
and finally (i) follows from 
\begin{eqnarray}
\sum_{(\tilde{u}_2^n,\tilde{v}_2^n) \in T_{\delta}^n(p_{U_2V_2})}
p_{U_2}^n(\tilde{u}_2^n) p_{V_2}^n(\tilde{v}_2^n)
\leq \sum_{(\tilde{u}_2^n,\tilde{v}_2^n) \in T_{\delta}^n(p_{U_2V_2})} 2^{-n(I(U_2;V_2) - 3 \delta )}p_{U_2V_2}^n(\tilde{u}_2^n,\tilde{v}_2^n), \nonumber     
\end{eqnarray}
Therefore, for $n$ sufficiently large, we have $
\mathbb{E}[T_{2.2.4}] \leq \epsilon,$ if
\begin{eqnarray}
L_2 + K_2 + S_2 + T_2 &<& D(p_{U_2} || q_{U_2}) + I(U_2;V_2) + I(U_2,V_2;Y_2)  \nonumber \\ &=&  \log(\Prime) + H(V_2) -H(U_2;V_2|Y_2) . \nonumber    
\end{eqnarray}

\section{Proof of Proposition \ref{Prop:4CQMACFirsttermHay}}
\label{App:4CQMACFirsttermHay}
We begin by establishing a key property of the decoding POVM elements $G_{\ulinez^n}^S$ for $S \subseteq [4]$, which will be used throughout this appendix.
\begin{eqnarray}
        &&\sum_{\ulinez^n} p_{\ulineZ}^n(\ulinez^n) \tr\left( G^S_{\ulinez^n} \xi_{\ulinez^n} \right) \geq 1- \epsilon - 2 \sqrt{\epsilon}. \nonumber     \end{eqnarray}
The proof follows the same arguments as in Appendix~\ref{App:Dec1FirsttermHay}. Now, we evaluate the expectation of $T_{1.1}$ over the codebook generation distribution. We obtain
\begin{eqnarray}
    \mathbb{E}[T_{1.1}]&=&\frac{1}{|\mathcal{\ulineM}|} \frac{1}{|\mathcal{\ulineD}|^n} \sum_{\ulinem} \sum_{\ulined^n} \sum_{\ulinez^n} p^n_{\ulineZ}(\ulinez^n) \tr \left\{ \left(I-\gamma^*_{\ulinez^n, \ulined^n}\right) \theta_{\ulinez^n, \ulined^n} \right\} \nonumber \\
    &\overset{(a)}{=}&\frac{1}{|\mathcal{\ulineM}|} \frac{1}{|\mathcal{\ulineD}|^n} \sum_{\ulinem} \sum_{\ulined^n} \sum_{\ulinez^n} p^n_{\ulineZ}(\ulinez^n) \left[\tr \left\{\theta_{\ulinez^n, \ulined^n} \right\} - \tr \left\{ \pi_{\boldsymbol{\CalH_{Y_G}}} \left(I_{\boldsymbol{\CalH_{Y}^{e}}}-  \beta^{*}_{\ulinez^{n},\ulined^{n}}\right) \theta_{\ulinez^n, \ulined^n} \left(I_{\boldsymbol{\CalH_{Y}^{e}}}-  \beta^{*}_{\ulinez^{n},\ulined^{n}}\right) \pi_{\boldsymbol{\CalH_{Y_G}}} \right\} \right],\nonumber \\
    &\overset{(b)}{\leq}& \frac{4}{|\mathcal{\ulineM}|} \frac{1}{|\mathcal{\ulineD}|^n} \sum_{\ulinem} \sum_{\ulined^n} \sum_{\ulinez^n} p^n_{\ulineZ}(\ulinez^n) \tr \left\{ \left(I-\pi_{\boldsymbol{\CalH_{Y_G}}} + \beta^*_{\ulinez^n, \ulined^n}\right) \theta_{\ulinez^n, \ulined^n} \right\} \nonumber \\
    &\overset{(c)}{\leq}& \frac{4}{|\mathcal{\ulineM}|} \frac{1}{|\mathcal{\ulineD}|^n} \sum_{\ulinem} \sum_{\ulined^n} \sum_{\ulinez^n} p^n_{\ulineZ}(\ulinez^n) \tr \left\{ \left(I-\pi_{\boldsymbol{\CalH_{Y_G}}} + \beta^*_{\ulinez^n, \ulined^n}\right) \left(\xi_{\ulinez^n} \otimes \ketbra{0}\right) \right\} \nonumber \\
&+& \frac{4}{|\mathcal{\ulineM}|} \frac{1}{|\mathcal{\ulineD}|^n} \sum_{\ulinem} \sum_{\ulined^n} \sum_{\ulinez^n} p^n_{\ulineZ}(\ulinez^n) \norm{\theta_{\ulinez^n, \ulined^n}-\left(\xi_{\ulinez^n} \otimes \ketbra{0}\right)}_1  \label{Eqn:goingbacktooriginalstate},
    \end{eqnarray}
    where (a) follows from (i) the definition of $\gamma^*_{\ulinez^n, \ulined^n}$ in \eqref{Eqn:3CQBCgammadef}, (ii) the cyclicity of the trace and (iii) the property of the projector i.e., $\pi_{\boldsymbol{\CalH_{Y_G}}}^2=\pi_{\boldsymbol{\CalH_{Y_G}}}$, (b) follows from the non-commutative union bound \cite[Fact.3]{202103SAD_Sen} and (c) follows from the trace inequality $\tr(\Lambda \rho) \leq \tr(\Lambda \sigma) + \frac{1}{2} \norm{\rho - \sigma}_1$, with $0 \leq \Lambda, \rho, \sigma \leq I$.
Using Proposition \ref{Prop:3CQBCClosnessOFstates}, we bound the second term by $12 \eta$ and since the state $\xi_{\ulinez^n} \otimes \ketbra{0} $ lives in $\boldsymbol{\CalH_{Y_G}}$ and $I-\pi_{\boldsymbol{\CalH_{Y_G}}}$ is a projector onto its complement, we are left with
\begin{eqnarray}
  \mathbb{E}[T_{1.1}] &\leq&  \frac{4}{|\mathcal{\ulineM}|} \frac{1}{|\mathcal{\ulineD}|^n} \sum_{\ulinem} \sum_{\ulined^n} \sum_{\ulinez^n} p^n_{\ulineZ}(\ulinez^n) \tr \left\{ \beta^*_{\ulinez^n, \ulined^n} \left(\xi_{\ulinez^n} \otimes \ketbra{0}\right) \right\}
+  12 \eta \nonumber \\
&\leq& \frac{42}{\eta^2}\frac{4}{|\mathcal{\ulineM}|}  \sum_{\ulinem} \sum_{\ulinez^n} p^n_{\ulineZ}(\ulinez^n) \left[ \sum_{S \subseteq [4]} \tr \left\{ \olineB^S_{\ulinez^n} \left(\xi_{\ulinez^n} \otimes \ketbra{0}\right) \right\} \right]
+  12 \eta \nonumber \\
&=& \frac{42}{\eta^2}\frac{4}{|\mathcal{\ulineM}|}  \sum_{\ulinem}  \left[ 15- \sum_{S \subseteq [4]} \sum_{\ulinez^n} p^n_{\ulineZ}(\ulinez^n) \tr \left\{ G^S_{\ulinez^n} \xi_{\ulinez^n}  \right\} \right]
+  12 \eta \nonumber.
\end{eqnarray}
where the second inequality follows from \cite[Corollary.1]{202103SAD_Sen} and the equality follows from (i) the definition of $\olineB_{\ulinez^n}^S$ in \eqref{Eqn:3CQBCComplementprojector} and (ii) from Gelfand–Naimark’s Thm.~\cite[Thm.~3.7]{BkHolevo_2019}. Hence, by applying the property established at the beginning of this appendix, for $S \subseteq [4]$, we obtain 
\begin{eqnarray}
    && \mathbb{E}[T_{1.1}] \leq \frac{168}{\eta^2} (15\epsilon + 30 \sqrt{\epsilon}) + 12 \eta.\nonumber 
\end{eqnarray}
With $\eta=\epsilon^{\frac{1}{5}}$, we have
\begin{eqnarray}
    && \mathbb{E}[T_{1.1}] \leq 168 (15\epsilon^{\frac{3}{5}} + 30 \epsilon^{\frac{1}{10}}) + 12 \epsilon^{\frac{1}{5}} \nonumber 
\end{eqnarray}

\section{Proof of Proposition \ref{Prop:4CQMACSectermHay}}
\label{App:4CQMACSectermHay}

 We elaborate on the analysis of $\sum_{S \subseteq [4]} T_{1S}$, and in doing so we leverage the \textit{smoothing} and \textit{augmentation} \cite{202103SAD_Sen} properties of the tilting maps we have defined in \eqref{Eqn:3CQICTiltingMaps}. As is standard, this is broken into fifteen terms—four singleton errors, six pair errors, four triple errors, and one quadruple error. The analysis of the quadruple error is slightly different from the analysis of the remaining fourteen terms. To analyze the quadruple error, we first prove that the difference between the average of the original state and the average of the tilted state over $\mathcal{\ulineZ}^n$ is small. This is done analogously to the steps used in \cite{202103SAD_Sen}. Having done this, the rest of the analysis is straightforward, since the corresponding projector has not been tilted, and we are left with it operating on the average of the original state. 
 
 We evaluate the expectation of $T_{1.S}$ for $S=4$ over the codebook generation distribution. We obtain
 \begin{eqnarray}
     \mathbb{E}[T_{1.4}]&=& \frac{1}{|\mathcal{\ulineM}|}  \frac{1}{|\mathcal{\ulineD}|^{2n}}  \sum_{\ulinem}  \sum_{\tilde{\ulinem} \neq \ulinem}  \sum_{\uline{\tilde{d}}^n}  \sum_{\uline{\tilde{z}}^n}  \sum_{\ulined^n}  \sum_{\ulinez^n}  p_{\ulineZ}^n(\uline{\tilde{z}}^n)  p_{\ulineZ}^n(\ulinez^n) \tr \left\{  \gamma^*_{\uline{\tilde{z}}^n, \uline{\tilde{d}}^n}  \theta_{\ulinez^n, \ulined^n}  \right\} \nonumber \\
     &\overset{(a)}{=}& \frac{1}{|\mathcal{\ulineM}|}  \frac{1}{|\mathcal{\ulineD}|^{n}}  \sum_{\ulinem}  \sum_{\tilde{\ulinem} \neq \ulinem}  \sum_{\uline{\tilde{d}}^n}  \sum_{\uline{\tilde{z}}^n}  p_{\ulineZ}^n(\uline{\tilde{z}}^n) \tr \left\{  \gamma^*_{\uline{\tilde{z}}^n, \uline{\tilde{d}}^n}  \theta^{\otimes n}  \right\} \nonumber \\
    &\overset{(b)}{\leq}& \frac{1}{\Omega(\eta)}  \frac{1}{|\mathcal{\ulineM}|}  \frac{1}{|\mathcal{\ulineD}|^{n}}  \sum_{\ulinem}  \sum_{\tilde{\ulinem}}  \sum_{\uline{\tilde{d}}^n}  \sum_{\uline{\tilde{z}}^n}  p_{\ulineZ}^n(\uline{\tilde{z}}^n) \tr \left\{  \gamma^*_{\uline{\tilde{z}}^n, \uline{\tilde{d}}^n}  \left(\xi^{\otimes n} \otimes \ketbra{0}\right)  \right\} \nonumber \\
     &+& \frac{1}{|\mathcal{\ulineM}|}  \frac{1}{|\mathcal{\ulineD}|^{n}}  \sum_{\ulinem}  \sum_{\tilde{\ulinem}}  \sum_{\uline{\tilde{d}}^n}  \sum_{\uline{\tilde{z}}^n}  p_{\ulineZ}^n(\uline{\tilde{z}}^n) \tr \left\{  \gamma^*_{\uline{\tilde{z}}^n, \uline{\tilde{d}}^n}  \CalN_{\eta}\left(\xi^{\otimes n} \otimes \ketbra{0}\right)  \right\}, \label{Eqn:lasttermofhayashinagaokainequality}
 \end{eqnarray} 
where (a) follows by using the definition $\theta^{\otimes n} \define \frac{1}{|\mathcal{\ulineD}|^n} \sum_{\ulined^n} \sum_{\ulinez^n} p_{\ulineZ}^n(\ulinez^n) \theta_{\ulinez^n, \ulined^n}$,
and (b) follows by using  
    \begin{eqnarray}
&&\theta^{\otimes n} = \frac{1}{\Omega(\eta)} \left( \xi^{\otimes n } \otimes \ketbra{0} \right) + \mathcal{N}_{\eta}\left( \xi^{\otimes n } \otimes \ketbra{0} \right),\nonumber 
\end{eqnarray}
as stated in \cite{202103SAD_Sen}. Consider the first term in \eqref{Eqn:lasttermofhayashinagaokainequality}, we have 
 \begin{eqnarray}
      &&\frac{1}{\Omega(\eta)}  \frac{1}{|\mathcal{\ulineM}|}  \frac{1}{|\mathcal{\ulineD}|^{n}}  \sum_{\ulinem}  \sum_{\tilde{\ulinem}}  \sum_{\uline{\tilde{d}}^n}  \sum_{\uline{\tilde{z}}^n}  p_{\ulineZ}^n(\uline{\tilde{z}}^n) \tr \left\{  \gamma^*_{\uline{\tilde{z}}^n, \uline{\tilde{d}}^n}  \left(\xi^{\otimes n} \otimes \ketbra{0}\right)  \right\} \nonumber \\
      &\overset{(a)}{\leq}&   \frac{1}{|\mathcal{\ulineM}|}  \frac{1}{|\mathcal{\ulineD}|^{n}}  \sum_{\ulinem}  \sum_{\tilde{\ulinem}}  \sum_{\uline{\tilde{d}}^n}  \sum_{\uline{\tilde{z}}^n}  p_{\ulineZ}^n(\uline{\tilde{z}}^n) \tr \left\{  \left( I - \beta^{[4]}_{\uline{\tildez}^n,\uline{\tilde{d}}^n} \right) \left(\xi^{\otimes n} \otimes \ketbra{0}\right)  \right\} \nonumber \\
      &\overset{(b)}{=}& \frac{1}{|\mathcal{\ulineM}|}  \sum_{\ulinem}  \sum_{\tilde{\ulinem}}    \sum_{\uline{\tilde{z}}^n}  p_{\ulineZ}^n(\uline{\tilde{z}}^n) \tr \left\{  \left( I - \olineB^{[4]}_{\uline{\tildez}^n} \right) \left(\xi^{\otimes n} \otimes \ketbra{0}\right)  \right\} \nonumber \\
      &\overset{(c)}{=}& \frac{1}{|\mathcal{\ulineM}|}    \sum_{\ulinem}  \sum_{\tilde{\ulinem}}    \sum_{\uline{\tilde{z}}^n}  p_{\ulineZ}^n(\uline{\tilde{z}}^n) \tr \left\{   G^{[4]}_{\uline{\tildez}^n} \xi^{\otimes n}   \right\} \nonumber \\
      &\overset{(d)}{=}& \frac{1}{|\mathcal{\ulineM}|}   \sum_{\ulinem}  \sum_{\tilde{\ulinem}}    \sum_{\uline{\tilde{z}}^n \in T_{\delta}^n(p_{\ulineZ})}  p_{\ulineZ}^n(\uline{\tilde{z}}^n) \tr \left\{  \pi \pi_{\uline{\tildez}^n} \pi \xi^{\otimes n}   \right\} \nonumber \\
      &\overset{(e)}{=}& \frac{1}{|\mathcal{\ulineM}|}  \sum_{\ulinem}  \sum_{\tilde{\ulinem}}   \sum_{\uline{\tilde{z}}^n \in T_{\delta}^n(p_{\ulineZ})}p_{\ulineZ}^n(\uline{\tilde{z}}^n) \tr \left\{  \pi_{\uline{\tildez}^n} \pi \xi^{\otimes n} \pi  \right\} \nonumber \\
      &\overset{(f)}{\leq}& \frac{1}{|\mathcal{\ulineM}|}    \sum_{\ulinem}  \sum_{\tilde{\ulinem}}  \sum_{\uline{\tilde{z}}^n \in T_{\delta}^n(p_{\ulineZ})}  2^{-n\left(H(Y) - \delta\right)} p_{\ulineZ}^n(\uline{\tilde{z}}^n) \tr \left\{  \pi_{\uline{\tildez}^n}  \right\} \nonumber \\
      &\overset{(g)}{\leq}& 2^{n(\tilde{R}_1+\tilde{R}_2 + \tilde{R}_3 + \tilde{R}_4)}  2^{-n(I(Y; \ulineZ)-2\delta)},\nonumber 
 \end{eqnarray}
where (a) follows from $\gamma^*_{\uline{\tilde{z}}^n, \uline{\tilde{d}}^n} \leq I-\beta^*_{\uline{\tilde{z}}^n, \uline{\tilde{d}}^n} \leq I-\beta^{[4]}_{\uline{\tilde{z}}^n, \uline{\tilde{d}}^n}$, (b) follows from $\beta^{[4]}_{\uline{\tilde{z}}^n, \uline{\tilde{d}}^n} \define \olineB_{\uline{\tilde{z}}^n}^{[4]}$, (c) follows from (i)  $\olineB_{\uline{\tilde{z}}^n}^{[4]} \define I - \olineG^{[4]}_{\tilde{\ulinez}^n}$ and (ii) from the Gelfand–Naimark’s Thm.~\cite[Thm.~3.7]{BkHolevo_2019}, (d) follows from the definition of $G^{[4]}_{\tilde{\ulinez}^n}$
in \eqref{Eqn:3CQBCOriginalPOVM}, (e) follows from the cyclicity of the trace, and finally (f) and (g) follows from the  standard typical projector property
\begin{eqnarray}
    \pi \xi^{\otimes n} \pi \leq 2^{-n\left(H(Y)-\delta \right)}, \mbox{ and }  \tr\{\pi_{\uline{\tildez}^n}\} \leq 2^{n \left( H(Y|\ulineZ) + \delta\right)}, \mbox{ for } \uline{\tildez}^n \in T_{\delta}^n(p_{\ulineZ}). \nonumber 
\end{eqnarray}

\med Next, consider the second term in \eqref{Eqn:lasttermofhayashinagaokainequality}, 
 \begin{eqnarray}
    &&\frac{1}{|\mathcal{\ulineM}|}  \frac{1}{|\mathcal{\ulineD}|^{n}}  \sum_{\ulinem}  \sum_{\tilde{\ulinem}}  \sum_{\uline{\tilde{d}}^n}  \sum_{\uline{\tilde{z}}^n}  p_{\ulineZ}^n(\uline{\tilde{z}}^n) \tr \left\{  \gamma^*_{\uline{\tilde{z}}^n, \uline{\tilde{d}}^n}  \CalN_{\eta}\left(\xi^{\otimes n} \otimes \ketbra{0}\right)  \right\} \nonumber \\ 
    &\leq&  \frac{1}{|\mathcal{\ulineM}|}  \frac{1}{|\mathcal{\ulineD}|^{n}}  \sum_{\ulinem}  \sum_{\tilde{\ulinem} }  \sum_{\uline{\tilde{d}}^n}  \sum_{\uline{\tilde{z}}^n}  p_{\ulineZ}^n(\uline{\tilde{z}}^n)  \norm{ \gamma^*_{\uline{\tilde{z}}^n, \uline{\tilde{d}}^n}}_1 \norm{ \CalN_{\eta}\left(\xi^{\otimes n} \otimes \ketbra{0}\right) }_{\infty} \nonumber \\ 
    &\leq&  2^{n(\tilde{R}_1+\tilde{R}_2 + \tilde{R}_3 + \tilde{R}_4)}  \: 2 |\CalH_Y|^n \: \frac{21\eta}{\sqrt{|\mathcal{\ulineD}|^n}}, \nonumber 
 \end{eqnarray}
The first inequality follows from the standard trace-norm inequality
\begin{eqnarray}
    \label{Eqn:3CQBCTraceNormineq}
|\tr(AB)| \leq \norm{AB}_{1} \leq \norm{A}_{1} \norm{B}_{\infty}.
\end{eqnarray}
The last inequality follows from (i)
 \begin{eqnarray}
 &&  \norm{\gamma^*_{\uline{\tilde{z}}^n, \uline{\tilde{d}}^n}}_1 \leq \norm{I-\beta^*_{\uline{\tilde{z}}^n, \uline{\tilde{d}}^n}}_{\infty} \norm{\pi_{\boldsymbol{\CalH_{Y_G}} }\left(I_{\boldsymbol{\CalH_{Y}^{e}}}-  \beta^{*}_{\ulinez^{n},\ulined^{n}}\right)}_1 \leq \norm{I-\beta^*_{\uline{\tilde{z}}^n, \uline{\tilde{d}}^n}}_{\infty}^2 \norm{\pi_{\boldsymbol{\CalH_{Y_G}} }}_1 \leq 2|\CalH_{Y}|^n, \nonumber 
 \end{eqnarray}
 and (ii) from the inequality stated in \cite{202103SAD_Sen}
 \begin{eqnarray}
     &&\norm{ \CalN_{\eta}\left(\xi^{\otimes n} \otimes \ketbra{0}\right) }_{\infty} \leq \frac{21\eta}{\sqrt{|\mathcal{\ulineD}|^n}}, \nonumber 
 \end{eqnarray}
 which implies that the difference between $\theta^{\otimes n}$ and $\left(\xi^{\otimes n} \otimes \ketbra{0}\right)$ is small. This establishes the smoothing property. Therefore, if we choose $|\ulineD| \geq (42 \eta^2)^{\frac{1}{n}} |\CalH_Y|^2 2^{2(I(Y;\ulineZ)+2\delta)}$, we have  
\begin{eqnarray}
      &&\mathbb{E}[T_{1.4}] \leq  2^{n(\tilde{R}_1 + \tilde{R}_2 + \tilde{R}_3 + \tilde{R}_4)} \left( 2^{-n\left(I\left(Y; \ulineZ \right)-2\delta\right)} + \frac{42 |\CalH_Y|^n \eta}{\sqrt{|\mathcal{\ulineD}|^n}} \right) \leq    2 \: 2^{-n\left(I\left(Y; \ulineZ \right)-2\delta-(\tilde{R}_1 + \tilde{R}_2 + \tilde{R}_3 + \tilde{R}_4)\right)}. \nonumber 
\end{eqnarray}
Hence, if
\begin{eqnarray}
\tilde{R}_1 + \tilde{R}_2 + \tilde{R}_3 + \tilde{R}_4 < I\left(Y; \ulineZ \right), \mbox{ then }  \mathbb{E}[T_{1.4}] \leq \epsilon. \nonumber   
\end{eqnarray}
Returning to the original notation (see Fig.~\ref{TabNotataion4CQMAC}), we have 
$\mathbb{E}[T_{1.4}] \leq \epsilon,$ if
\begin{eqnarray}
S_{ji} + T_{ji} + S_{jk} + T_{jk} + K_j + L_j + \max\{S_{ij} + T_{ij},S_{kj} + T_{kj}\} < I(Y_j;U_{ji},U_{jk},V_j,U_{ij}\oplus U_{kj}). \nonumber 
\end{eqnarray}
The above bound is obtained when binning is ignored. However, since our analysis includes binning and, moreover, each of the codebooks is picked according to the corresponding marginal distribution, the right-hand side of the bound we obtain will contain an additional divergence term. Specifically, we obtain
\begin{eqnarray}
&&S_{ji} + T_{ji} + S_{jk} + T_{jk} + K_j + L_j + \max\{S_{ij} + T_{ij},S_{kj} + T_{kj}\} \nonumber \\
&<& I(Y_j;U_{ji},U_{jk},V_j,U_{ij}\oplus U_{kj}) + D\left( p_{U_{ji}U_{jk}V_jU_{ij}\oplus U_{kj}} \Big|\Big| \frac{p_{V_j}}{|\mathcal{U}_{ji}| |\mathcal{U}_{jk}| \Prime_j}\right) \nonumber \\
&=& \log \left(|\mathcal{U}_{ji}| \right) + \log \left(|\mathcal{U}_{jk}| \right) + \log \left(\Prime_j \right) + H(V_j) - H(U_{ji}, U_{jk}, V_j,U_{ij}\oplus U_{kj} |Y_j).\nonumber 
\end{eqnarray}

Note that the above bound is identical to the one obtained in  \eqref{Eqn:CQBCChannelCodingStep1Bounds6}, \eqref{Eqn:CQBCChannelCodingStep1Bounds7}
of the Thm.~\ref{Thm:3CQBCStepIInnerBound}, with $\mathcal{A}_j = \{ji,jk\}$.

Having described the analysis of the quadruple error term, we are left with fourteen terms. The analysis of each of these fourteen terms is essentially identical. Therefore, we describe the analysis for a general $S \subsetneq [4]$. 
Consider
\begin{eqnarray}
    T_{1.S}=\frac{1}{|\mathcal{\ulineM}|} \sum_{\ulinem} \sum_{\tilde{m}_S \neq m_S}\tr \left\{ \gamma^*_{Z^n_S(\tilde{m}_S),D^n_S(\tilde{m}_S),Z^n_{S^c}(m_{S^c}), D^n_{S^c}(m_{S^c})} \theta_{\ulineZ^n(\ulinem),\ulineD^n(\ulinem)}\right\}. \nonumber
\end{eqnarray}
We evaluate the expectation over the codebook generation distribution. We obtain
\begin{eqnarray}
    \mathbb{E}[T_{1.S}] &=& \frac{1}{|\mathcal{\ulineM}|} \frac{1}{|\mathcal{\ulineD}|^n} \frac{1}{|\mathcal{D}_S|^n} \sum_{\ulinem} \sum_{\tilde{m}_S \neq m_S} \sum_{\tilde{d}^n_S} \sum_{\tilde{z}^n_S} \sum_{\ulined^n} \sum_{\ulinez^n} p_{\ulineZ}^n(\ulinez^n) p_{Z_S}^n(\tilde{z}_S^n)\tr \left\{ \gamma^*_{\tilde{z}^n_S, \tilde{d}^n_S,z^n_{S^c}, d^n_{S^c}} \theta_{\ulinez^n,\ulined^n} \right\} \nonumber\\
    &\overset{(a)}{=}&\frac{1}{|\mathcal{\ulineM}|} \frac{1}{|\mathcal{\ulineD}|^n} \sum_{\ulinem} \sum_{\tilde{m}_S \neq m_S} \sum_{\tilde{d}^n_S} \sum_{\tilde{z}^n_S} \sum_{d_{S^c}^n} \sum_{z_{S^c}^n} p_{Z_{S^c}}^n(z_{S^c}^n) p_{Z_S}^n(\tilde{z}_S^n)\tr \left\{ \gamma^*_{\tilde{z}^n_S, \tilde{d}^n_S,z^n_{S^c}, d^n_{S^c}} \theta_{z^n_{S^c},d^n_{S^c}} \right\} \nonumber \\
    &\overset{(b)}{\leq}&  \frac{1}{|\mathcal{\ulineM}|} \frac{1}{|\mathcal{\ulineD}|^n} \sum_{\ulinem} \sum_{\tilde{m}_S} \sum_{\tilde{d}^n_S} \sum_{\tilde{z}^n_S} \sum_{d_{S^c}^n}  \sum_{z_{S^c}^n} p_{Z_{S^c}}^n(z_{S^c}^n) p_{Z_S}^n(\tilde{z}_S^n)\tr \left\{ \gamma^*_{\tilde{z}^n_S, \tilde{d}^n_S,z^n_{S^c}, d^n_{S^c}} \CalT^{S^c}_{d^n_{S^c}, \eta}\left(\xi_{z^n_{S^c}} \otimes \ketbra{0}\right) \right\} \nonumber \\
    &+& \frac{1}{|\mathcal{\ulineM}|} \frac{1}{|\mathcal{\ulineD}|^n} \sum_{\ulinem} \sum_{\tilde{m}_S} \sum_{\tilde{d}^n_S} \sum_{\tilde{z}^n_S}  \sum_{d_{S^c}^n}  \sum_{z_{S^c}^n} p_{Z_{S^c}}^n(z_{S^c}^n) p_{Z_S}^n(\tilde{z}_S^n)\tr \left\{ \gamma^*_{\tilde{z}^n_S, \tilde{d}^n_S,z^n_{S^c}, d^n_{S^c}} \CalN_{d^n_{S^c}}\left(\xi_{z^n_{S^c}} \otimes \ketbra{0}\right) \right\}, \label{Eqn:Oneofthefourteentermaftersplitting}
\end{eqnarray}
where (a) follows by using the definition $\theta_{z^n_S,d^n_S} \define \frac{1}{|\mathcal{D}_{S^c}|^n} \sum_{d^n_{S^c}} \sum_{z^n_{S^c}} p_{Z_{S^c}|Z_S}^n(z^n_{S^c}|z^n_S) \theta_{\ulinez^n, \ulined^n}$, and (b) follows by using 
\begin{eqnarray}
        &&\theta{z^n_S, d^n_S}= \frac{\Omega(S, \eta)}{\Omega(\eta)} \mathcal{T}_{d^n_S, \eta} \left( \xi_{z^n_S} \otimes \ketbra{0}\right) + \mathcal{N}_{d^n_S, \eta}\left( \xi_{z^n_S} \otimes \ketbra{0}\right). \nonumber 
\end{eqnarray}
Consider the first term in \eqref{Eqn:Oneofthefourteentermaftersplitting}. We have
\begin{eqnarray}
         &&\frac{1}{|\mathcal{\ulineM}|} \frac{1}{|\mathcal{\ulineD}|^n} \sum_{\ulinem} \sum_{\tilde{m}_S} \sum_{\tilde{d}^n_S} \sum_{\tilde{z}^n_S}  \sum_{d_{S^c}^n}  \sum_{z_{S^c}^n} p_{Z_{S^c}}^n(z_{S^c}^n) p_{Z_S}^n(\tilde{z}_S^n)\tr \left\{ \gamma^*_{\tilde{z}^n_S, \tilde{d}^n_S,z^n_{S^c}, d^n_{S^c}} \CalT^{S^c}_{d^n_{S^c}, \eta}\left(\xi_{z^n_{S^c}} \otimes \ketbra{0}\right) \right\} \nonumber \\
         &\overset{(a)}{\leq}& \frac{1}{|\mathcal{\ulineM}|} \frac{1}{|\mathcal{\ulineD}|^n} \sum_{\ulinem} \sum_{\tilde{m}_S} \sum_{\tilde{d}^n_S} \sum_{\tilde{z}^n_S}  \sum_{d_{S^c}^n}  \sum_{z_{S^c}^n} p_{Z_{S^c}}^n(z_{S^c}^n) p_{Z_S}^n(\tilde{z}_S^n)\tr \left\{ \left(I-\beta^S_{\tilde{z}^n_S, \tilde{d}^n_S,z^n_{S^c}, d^n_{S^c}}\right) \CalT^{S^c}_{d^n_{S^c}, \eta}\left(\xi_{z^n_{S^c}} \otimes \ketbra{0}\right) \right\} \nonumber \\
         &=& \frac{1}{|\mathcal{\ulineM}|} \frac{1}{|\mathcal{\ulineD}|^n} \sum_{\ulinem} \sum_{\tilde{m}_S} \sum_{\tilde{d}^n_S} \sum_{\tilde{z}^n_S} \sum_{d_{S^c}^n}  \sum_{z_{S^c}^n} p_{Z_{S^c}}^n(z_{S^c}^n) p_{Z_S}^n(\tilde{z}_S^n)\tr \left\{ \left(I-\beta^S_{\tilde{z}^n_S, \tilde{d}^n_S,z^n_{S^c}, d^n_{S^c}}\right) \pi_{\CalT^{S^c}_{d^n_{S^c},\eta}(\boldsymbol{\CalH_{Y_G}})} \CalT^{S^c}_{d^n_{S^c}, \eta}\left(\xi_{z^n_{S^c}} \otimes \ketbra{0}\right) \right\} \nonumber \\
         &\overset{(b)}{=}& \frac{1}{|\mathcal{\ulineM}|} \frac{1}{|\mathcal{\ulineD}|^n} \sum_{\ulinem} \sum_{\tilde{m}_S} \sum_{\tilde{d}^n_S} \sum_{\tilde{z}^n_S} \sum_{d_{S^c}^n}   \sum_{z_{S^c}^n} p_{Z_{S^c}}^n(z_{S^c}^n) p_{Z_S}^n(\tilde{z}_S^n)\tr \left\{ \left(I_{\CalT_{d^n_{S^c},\eta}^{S^c}(\boldsymbol{\CalH_{Y_G}})}-\beta^S_{\tilde{z}^n_S, \tilde{d}^n_S,z^n_{S^c}, d^n_{S^c}}\right)  \CalT^{S^c}_{d^n_{S^c}, \eta}\left(\xi_{z^n_{S^c}} \otimes \ketbra{0}\right) \right\} \nonumber \\
         &\overset{(c)}{=}& \frac{1}{|\mathcal{\ulineM}|}  \sum_{\ulinem} \sum_{\tilde{m}_S}  \sum_{\tilde{z}^n_S}  \sum_{z_{S^c}^n} p_{Z_{S^c}}^n(z_{S^c}^n) p_{Z_S}^n(\tilde{z}_S^n)\tr \left\{ \left(I_{\boldsymbol{\CalH_{Y_G}}}-\olineB^S_{\tilde{z}^n_S,z^n_{S^c}}\right) \left(\xi_{z^n_{S^c}} \otimes \ketbra{0}\right) \right\} \nonumber \\
         &\overset{(d)}{=}& \frac{1}{|\mathcal{\ulineM}|} \sum_{\ulinem} \sum_{\tilde{m}_S}  \sum_{\tilde{z}^n_S}  \sum_{z_{S^c}^n} p_{Z_{S^c}}^n(z_{S^c}^n) p_{Z_S}^n(\tilde{z}_S^n)\tr \left\{ G^S_{\tilde{z}^n_S,z^n_{S^c}} \xi_{z^n_{S^c}} \right\} \nonumber \\
         &\overset{(e)}{=}& \frac{1}{|\mathcal{\ulineM}|} \sum_{\ulinem} \sum_{\tilde{m}_S} \sum_{(\tilde{z}^n_S, z_{S^c}^n) \in T_{\delta}^n(p_{\ulineZ})} p_{Z_{S^c}}^n(z_{S^c}^n) p_{Z_S}^n(\tilde{z}_S^n)\tr \left\{
         \pi_{z_{S^{c}}^{n}} \pi_{\tilde{z}^n_S,z^n_{S^c}}\pi_{z_{S^{c}}^{n}}
          \xi_{z^n_{S^c}} \right\} \nonumber \\
         &\overset{(f)}{=}&\frac{1}{|\mathcal{\ulineM}|}  \sum_{\ulinem} \sum_{\tilde{m}_S}  \sum_{(\tilde{z}^n_S, z_{S^c}^n) \in T_{\delta}^n(p_{\ulineZ})}p_{Z_{S^c}}^n(z_{S^c}^n) p_{Z_S}^n(\tilde{z}_S^n)\tr \left\{ \pi_{\tilde{z}^n_S,z^n_{S^c}} \: \pi_{z_{S^{c}}^{n}}
          \xi_{z^n_{S^c}} \pi_{z_{S^{c}}^{n}} \right\} \nonumber \\
        &\overset{(g)}{\leq}& 2^{n \left(\sum_{s \in S} \tilde{R}_s\right)} 2^{-n(I(Y; Z_S|Z_{S^c})-2\delta)} , \nonumber
\end{eqnarray}
where (a) follows from $\gamma^*_{\tilde{z}^n_S, \tilde{d}^n_S,z^n_{S^c}, d^n_{S^c}} \leq I-\beta^*_{\tilde{z}^n_S, \tilde{d}^n_S,z^n_{S^c}, d^n_{S^c}} \leq I- \beta^S_{\tilde{z}^n_S, \tilde{d}^n_S,z^n_{S^c}, d^n_{S^c}}$,
(b) follows from $\beta^S_{\tilde{z}^n_S, \tilde{d}^n_S,z^n_{S^c}, d^n_{S^c}}$ having support contained in $\CalT^{S^c}_{d^n_{S^c},\eta}(\boldsymbol{\CalH_{Y_G}})$, (c) follows from extracting the isometry $\mathcal{T}_{d^n_{S^c}, \eta}$, (d) follows from (i) $\olineB^S_{\tilde{z}_S^n,z^n_{S^c}} \define I_{\boldsymbol{\CalH_{Y_G}}} - \olineG^S_{\tilde{z}_S^n,z^n_{S^c}}$ and (ii) from Gelfand–Naimark’s Thm.~\cite[Thm.~3.7]{BkHolevo_2019}, (e) follows from the definition of $G^S_{\tilde{z}^n_S,z^n_{S^c}}$ in \eqref{Eqn:3CQBCOriginalPOVM}, (f) follows from the cyclicity of the trace, and finally (g) follows from 
\begin{eqnarray}
    \pi_{z_{S^{c}}^{n}}
          \xi_{z^n_{S^c}} \pi_{z_{S^{c}}^{n}} \leq 2^{-n\left(H(Y|Z_{S^c}) - \delta \right)}, \mbox{ and } \tr\{\pi_{\tilde{z}^n_S,z^n_{S^c}}\} \leq 2^{-n \left( H(Y|\ulineZ) + \delta \right)}, \mbox{ if }  (\tilde{z}^n_S, z_{S^c}^n) \in T_{\delta}^n(p_{\ulineZ}). \nonumber 
\end{eqnarray}
Now consider the second term in \eqref{Eqn:Oneofthefourteentermaftersplitting}.
\begin{eqnarray}
    &&\frac{1}{|\mathcal{\ulineM}|} \frac{1}{|\mathcal{\ulineD}|^n} \sum_{\ulinem} \sum_{\tilde{m}_S} \sum_{\tilde{d}^n_S} \sum_{\tilde{z}^n_S} \sum_{d_{S^c}^n}  \sum_{z_{S^c}^n} p_{Z_{S^c}}^n(z_{S^c}^n) p_{Z_S}^n(\tilde{z}_S^n)\tr \left\{ \gamma^*_{\tilde{z}^n_S, \tilde{d}^n_S,z^n_{S^c}, d^n_{S^c}} \CalN_{d^n_{S^c}}\left(\xi_{z^n_{S^c}} \otimes \ketbra{0}\right) \right\} \nonumber \\
    &\overset{(a)}{\leq}&\frac{1}{|\mathcal{\ulineM}|} \frac{1}{|\mathcal{\ulineD}|^n} \sum_{\ulinem} \sum_{\tilde{m}_S} \sum_{\tilde{d}^n_S} \sum_{\tilde{z}^n_S} \sum_{d_{S^c}^n}   \sum_{z_{S^c}^n} p_{Z_{S^c}}^n(z_{S^c}^n) p_{Z_S}^n(\tilde{z}_S^n) \norm{ \gamma^*_{\tilde{z}^n_S, \tilde{d}^n_S,z^n_{S^c}, d^n_{S^c}}}_{1} \norm{\CalN_{d^n_{S^c}}\left(\xi_{z^n_{S^c}} \otimes \ketbra{0}\right)}_{\infty} \nonumber \\
    &\overset{(b)}{\leq}&  2^{n\left(\sum_{s \in S} \tilde{R}_s\right)} \: 2 |\CalH_Y|^n \: \frac{21\eta}{\sqrt{|\CalD_S|^n}}. \nonumber 
\end{eqnarray}
where (a) follows from the trace-norm inequality given in \eqref{Eqn:3CQBCTraceNormineq} and (b) follows from (i) 
 \begin{eqnarray}
 \norm{\gamma^*_{\tilde{z}^n_S, \tilde{d}^n_S,z^n_{S^c}, d^n_{S^c}}}_{1}  &\leq& \norm{I-\beta^*_{\tilde{z}^n_S, \tilde{d}^n_S,z^n_{S^c}, d^n_{S^c}}}_{\infty} \norm{\pi_{\boldsymbol{\CalH_{Y_G}} }\left(I_{\boldsymbol{\CalH_{Y}^{e}}}-  \beta^{*}_{\tilde{z}^n_S, \tilde{d}^n_S,z^n_{S^c}, d^n_{S^c}}\right)}_1 \nonumber \\
 &\leq& \norm{I-\beta^*_{\tilde{z}^n_S, \tilde{d}^n_S,z^n_{S^c}, d^n_{S^c}}}_{\infty}^2 \norm{\pi_{\boldsymbol{\CalH_{Y_G}} }}_1 \nonumber \\
 &\leq& 2|\CalH_{Y}|^n, \nonumber 
 \end{eqnarray}
and from (ii) the inequality 
\begin{eqnarray}
     \norm{\mathcal{N}_{d^n_S, \eta}\left( \xi_{z^n_S} \otimes \ketbra{0}\right)}_{\infty} \leq \frac{21 \eta}{\sqrt{|\mathcal{D_S}|^n}}, \nonumber 
\end{eqnarray}
as stated in \cite{202103SAD_Sen}. Therefore, if we choose $|\CalD_S| \geq (42 \eta)^{\frac{2}{n}}|\CalH_Y|^2 2^{2\left(I(Y;Z_S|Z_{S^c})-2 \delta\right)}$, we have 
\begin{eqnarray}
    \mathbb{E}[T_{1.S}] &\leq&  2^{n\left(\sum_{s \in S} \tilde{R}_s\right)} \left( 2^{-n(I(Y;Z_S|Z_{S^c})-2\delta)}+ \frac{42 |\CalH_Y|^n \eta}{\sqrt{|\CalD_S|^n}}\right) \nonumber \\ &\leq& 2 2^{-n\left(I(Y;Z_S|Z_{S^c})-2\delta -\left(\sum_{s \in S} \tilde{R}_s\right)\right)} \nonumber 
    \end{eqnarray}
Hence, $\mathbb{E}[T_{1.S}]\leq \epsilon,$ if
\begin{eqnarray}
  \sum_{s \in S} \tilde{R}_s < I(Y;Z_S|Z_{S^c}). \nonumber 
\end{eqnarray}
As an example, we take $S=\{1,2\}$, i.e., $\mathbb{E}[T_{1.S}]\leq \epsilon,$ if
\begin{eqnarray}
\tilde{R}_1+\tilde{R}_2 < I(Y;Z_1,Z_2|Z_3,Z_4). \nonumber     
\end{eqnarray}
Returning to the original notation (see Fig.~\ref{TabNotataion4CQMAC}), we have $ \mathbb{E}[T_{1.S}]\leq \epsilon,$ if
\begin{eqnarray}
S_{ji} + T_{ji}+ S_{jk} + T_{jk}  < I(Y_j;U_{ji},U_{jk}|V_j,U_{ij}\oplus U_{kj}). \nonumber     
\end{eqnarray}
Now if the analysis is carried out with binning, we obtain the following bound
\begin{eqnarray}
S_{ji} + T_{ji}+ S_{jk} + T_{jk}  &<& I(Y_j;U_{ji},U_{jk}|V_j,U_{ij}\oplus U_{kj})  + D\left( p_{U_{ji}U_{jk}} \Big|\Big| \frac{1}{|\mathcal{U}_{ji}||\mathcal{U}_{jk}|} \right) \nonumber \\
&=& \log \left( |\mathcal{U}_{ji}|\right) + \log \left( |\mathcal{U}_{jk}|\right) - H\left(U_{ji}, U_{jk} | U_{ij} \oplus U_{kj}, V_j, Y_j \right), \nonumber
\end{eqnarray}
which corresponds to the bound \eqref{Eqn:CQBCChannelCodingStep1Bounds2} in Thm.~\ref{Thm:3CQBCStepIInnerBound}, with $\mathcal{A}_j=\{ji,jk\}$.

\bibliographystyle{IEEEtran}
{
\bibliography{CosetCdsFor3CQChnls}
\end{document}